\documentclass[aip,pof,amsmath,amssymb]{revtex4-1}

\usepackage[dvips]{graphicx,color}

\begin{document}

\newcommand\etal{\mbox{\textit{et al.}}}
\newcommand\Real{\mbox{Re}} 
\newcommand\Imag{\mbox{Im}} 
\newcommand\Rey{\mbox{\textit{Re}}} % Reynolds number
\newcommand\Pra{\mbox{\textit{Pr}}} % Prandtl number
\newcommand\Pec{\mbox{\textit{Pe}}} % Peclet number

\title{Linear stability analysis of ice growth under supercooled water film driven by a laminar airflow}

\author{Kazuto Ueno}

\email{k.ueno@kyudai.jp}

\author{Masoud Farzaneh}

\affiliation
{NSERC/Hydro-Quebec/UQAC Industrial Chair on Atmospheric Icing of Power Network Equipment (CIGELE)
and Canada Research Chair on Atmospheric Icing Engineering of Power Networks (INGIVRE), 
Universit$\acute{e}$ du Qu$\acute{e}$bec $\grave{a}$ Chicoutimi, 
Chicoutimi, Qu$\acute{e}$bec G7H 2B1, Canada}

%\date{\today}

\begin{abstract}
We propose a theoretical model for ice growth under a wind-driven supercooled water film. The thickness and surface velocity of the water layer are variable by changing the air stream velocity. For a given water supply rate, linear stability analysis is carried out to study the morphological instability of the ice-water interface. In this model, water and air boundary layers are simultaneously disturbed due to the change in ice shape, and the effect of the interaction between air and water flows on the growth condition of the ice-water interface disturbance is taken into account. It is shown that as wind speed increases, the amplification rate of the disturbance is significantly affected by variable stresses exerted on the water-air interface by the air flow as well as restoring forces due to gravity and surface tension. We predict that an ice pattern of a centimeter scale in wavelength appears and the wavelength decreases as wind speed increases, and that the ice pattern moves in the direction opposite to the water flow. The effect of the air stress disturbance on the heat transfer coefficient at the water-air interface is also investigated for various wind speeds. 
\end{abstract}


\maketitle

%%%%%%%%%%%%%%%%%%%%%%%
\section{Introduction}
%%%%%%%%%%%%%%%%%%%%%%%

Thin liquid films are ubiquitous entities in a variety of settings and display interesting dynamics depending on various forces acting on the liquid film and the surface geometries on which the fluid moves. \cite{Oron97}
A number of studies for the stability of a gravity-driven viscous liquid flowing on an inclined flat plane have been done, beginning from the pioneering works of Benjamin \cite{Benjamin57} and Yih. \cite{Yih63} The stability of a wind-driven liquid film flowing over a horizontal flat plate \cite{Craik66} and airfoils \cite{Yih90, Tsao97} was investigated for small disturbances. In these works an interaction between air and liquid flows was considered, and it was shown that the thin liquid film becomes unstable to small disturbances, and that waves arise due to the variable stresses exerted on the liquid-air interface by the airflow. 

On the other hand, glaze (wet) ice formation and icicle growth are the problem of ice growth from a liquid film flow accompanying a phase change. \cite{Farzaneh08} Glaze ice forms when water is collected from the impingement of supercooled water droplets, whereas icicles grow from the water of melting snow and ice at the root of the icicles. 
Typically, icicles also make up an important part of the total ice load in freezing rain. \cite{Makkonen98}
The glaze ice and icicle surfaces are covered with a supercooled water film, and ice grows from a part of the water film by releasing latent heat into the ambient air below 0 $^{\circ}$C.
It is well known that a solid surface under a supercooled liquid film is morphologically unstable, resulting in dendritic growth and a material is a microscopic mixture of solid and liquid. When a film of water is supplied by impinging droplets and cooled from its surface by cold air, the growing ice always initially entraps a considerable amount of liquid water. This is called spongy ice. \cite{Makkonen87, Farzaneh08} 
It was recently shown that sponginess is a material parameter (70 $\%$ of ice), and is independent of the growth conditions. \cite{Makkonen10}
The remaining unfrozen water flows on the ice surface. It should be noted that water flow is significantly different over an accreting ice layer than a non-accreting substrate and hence the problem of ice accretion in the presence of a flowing water film is highly complex. 

Ringlike ripples of a centimeter-scale in wavelength are formed. \cite{Knight80, Maeno94, Farzaneh08} 
Although the basic mechanism of icicle growth is well known, \cite{Makkonen88, Maeno94} the mechanism of icicle ripple formation remained unsolved. 
Ogawa and Furukawa first attempted a theoretical explanation for the icicle ripple formation in the absence of airflow around icicles, where the icicle surface was covered with a gravity driven supercooled water film with a free surface. \cite{Ogawa02} In an improved model proposed by Ueno, \cite{Ueno03, Ueno04, Ueno07} the influence of the shape of the water-air interface due to the action of gravity and surface tension on the ice growth conditions was newly taken into account, and a quite different mechanism on the origin of ripples on icicles was proposed. The theoretical results obtained by Ueno have recently been compared very favorably with experimental results.\cite{UFYT10} 
We extended the above theoretical framework to include natural convection airflow around icicles. \cite{Ueno10} It was found that the enhancement of the rate of latent heat loss from the water-air interface to the surrounding air due to airflow caused the amplification rate of the ice-water interface disturbance to increase. However, the wavelength of ice ripples was not significantly affected by the airflow.

Since the natural convection airflow around icicles was less than 1 m/s in velocity, the free shear stress condition at the water-air interface was still satisfied, and hence driving force of the water film over the ice surface was gravity only. \cite{Ueno10} However, when wind speed is large, the wind drag at the water-air interface also drives the water film.  
For example, the combination of two driving mechanism due to gravity and wind drag produces a variety of aufeis (also referred to as icings) morphologies with various surface features. \cite{Streitz02} 
Aufeis are spreading and thickening ice accretions that form in cold air when a thin sheet of water flows or trickles over a cold surface. According to the aufeis formation experiments,
an initial morphologies of aufeis appeared essentially wavelike (or terraced) on a planar aluminum and a smooth ice surface, and their spacing and height varied with slope of an inclined plane and wind speed. \cite{Streitz02}
The morphological instability of the ice-water interface under the water film flow due to the two driving forces is  also relevant to the surface roughness characteristics associated with
glaze icing formation around aircraft wings and structures. \cite{Shin96, Kollar10}
Furthermore, the morphological instability of the surface of growing ice is closely related to various natural phenomena where a thin layer of moving fluid separates the developing solid from the surrounding air. \cite{Meakin10} 

The physical model commonly used in ice accretion codes is mainly based on conservation of energy and mass
within numerical cells along the ice accreting surface. \cite{Poots96, Farzaneh08} However, in glaze icing conditions the numerical results are poor agreement with experimental results. Therefore, some investigators developed theoretical and numerical understanding of the dynamic effect of water film flow on the glaze ice accretion. 
Bourgault $\etal$ applied a simple film flow model to the problem of aircraft icing. \cite{Bourgault00}
Myers $\etal$ introduced a mathematical model for ice accretion with water flow driven by air shear, gravity, and surface tension, employing the lubrication approximation to describe the water film flow. \cite{Myers02_1, Myers02_2, Myers04}
A version of the Myers $\etal$ model is used in the ICECREMO commercial aircraft icing code.
The aerodynamic forces, as modified by the accreted ice, are significant in determining the wind drag and lift on iced structures. However, most analyses in current glaze ice accretion models lack the physical motivation for the effects of either surface roughness or profile change of ice on the heat transfer coefficient, and the roughness is treated as input to the code. \cite{Gent00} 
The lack of roughness formation in standard icing models indicates that the related surface instabilities must originate from more localized structures in which air flow can interact with the water film. \cite{Tsao98, Tsao00}

A more microscopic, rather than global, energy balance and detailed analysis of the interaction between the air and water flows are required to predict fine details of localized roughness. \cite{Tsao98, Tsao00, Gent00} Therefore, herein we perform a linear stability analysis for ice growth under a supercooled water film driven by a laminar airflow, taking into account the effect of interaction between the air and water flows on the ice growth conditions. 
There is a significant difference between the current and previous works, \cite{Ueno03, Ueno04, Ueno07, UFYT10, Ueno10} as follows: Our previous works featured a gravity driven water flow, and the shape of the water-air interface was determined by the action of gravity and surface tension only. In the current model, water flow driven by air shear stress is considered. We will show that when the air and water flows are coupled, tangential and normal air shear stress disturbances as well as gravity and surface tension play an important role in determining the shape of the water-air interface as air stream velocity increases. Without employing any of the empirical methods used in standard icing models, the heat transfer coefficient at the water-air interface is determined explicitly by solving the governing equations for the air and water flows and the air temperature field.
It will be shown that the growth conditions of the ice-water interface disturbance as well as the heat transfer coefficient at the water-air interface are strongly affected by the air shear stress disturbances, which is particularly a new effect not found in gravity driven water flow.

%%%%%%%%%%%%%%%%%%%%%%%%%%%%%%%%%
\section{\label{sec:model}Model}
%%%%%%%%%%%%%%%%%%%%%%%%%%%%%%%%%

%%%%%% Fig.1 %%%%%%%%%%%%%%%%%%%%%%%%%%%%%%%%%%%%%%%%%%%%%%%%%%%%%%%%%%%%%%%%%%%%%%%%%%%%%%%%%%%%%%%%%%%%%%%%%%%%%%%
\begin{figure}
\begin{center}
\includegraphics[width=12cm,height=12cm,keepaspectratio,clip]{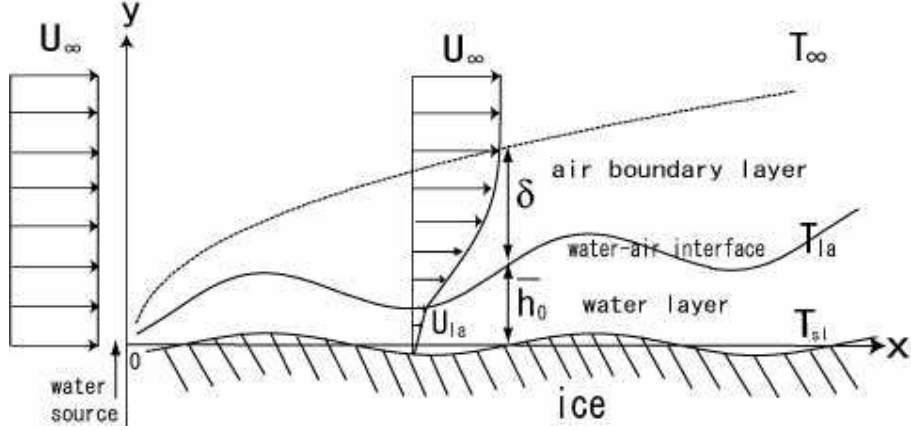}
\end{center}
\caption{Physical model of air-water-ice multi-phase system. Vertical height is not to scale.}
\label{fig:ice-water-air}
\end{figure}
%%%%%%%%%%%%%%%%%%%%%%%%%%%%%%%%%%%%%%%%%%%%%%%%%%%%%%%%%%%%%%%%%%%%%%%%%%%%%%%%%%%%%%%%%%%%%%%%%%%%%%%%%%%%%%%%%%%%

The model configuration is shown in Fig. \ref{fig:ice-water-air}, which is based on the experiments of aufeis formation on a smooth ice substrate in a wind tunnel by Streitz and Ettema. \cite{Streitz02} In that experiment, water was supplied through a row of holes located at the upstream end of the wind tunnel in a refrigerated laboratory, and the water film was driven by gravity and wind drag for various plane slopes and wind speeds. The following analysis is restricted to a two-dimensional vertical cross-section. The position $x$ is measured from the leading edge where water is supplied at a constant rate, and the $y$ axis is normal to it. The ice is covered with a thin water film of $\bar{h}_{0}$. A cold air stream flows over the thin water layer, and the airflow is assumed to be laminar. The surface of the water layer moves at a velocity of $u_{la}$ under the influence of wind drag. The air velocity approaches the free stream velocity $u_{\infty}$ at a distance $\delta$ from the water-air interface. Simultaneous water and air boundary layers occur. Since the air temperature at $T_{\infty}$ is lower than the ice-water interface temperature, $T_{sl}$, ice grows from a part of the water film by releasing the latent heat to the air through the water-air interface at temperature $T_{la}$.

In this model, water is supplied from only the leading edge and there is no impingement of supercooled water droplets on the water film surface. The water and air boundary layers start at $x=0$. On the other hand, aircraft icing is primarily due to the impact of supercooled water droplets on a cold surface. One could see ice buildup at the leading edge of the airfoil, taken as $x=0$, where unfrozen water flow is slowest, but the water layer is thickest. In this sense, the model herein is not yet truly relevant to the aircraft icing problems. \cite{Bourgault00, Myers02_1, Myers02_2, Myers04, Gent00, Tsao98, Tsao00} 
In addition, the following assumptions are used in the current model:
(1) The water film is driven by wind drag only. Hence the analysis is only valid on a horizontal surface, and the free stream velocity $u_{\infty}$ is constant in space. 
(2) Density remains constant through the phase change. 
(3) Change in ice shape disturbs the water-air interface, and the flow and temperature fields in the water film and air. A quasi-stationary approximation is used for the disturbed fields and unsteadiness only enters through the Stephan condition, due to the long time scale of the ice-water interface motion.
(4) Heat conduction into a substrate beneath ice sheet is not included. The ice sheet is assumed to be thick and the undisturbed part of temperature gradient in the ice does not exist.
(5) The presence of waves on the water film is ignored because the waves did not interact with the forming ice in any observable manner in the experiments, except for enhancing the spreading of the water over the aufeis surface. \cite{Streitz02}

%%%%%%%%%%%%%%%%%%%%%%%%%%%%%%%%%
\subsection{Governing equations}
%%%%%%%%%%%%%%%%%%%%%%%%%%%%%%%%%

The velocity components in the $x$ and $y$ directions in the air, $u_{a}$ and $v_{a}$, are governed by \cite{Landau59} 
\begin{equation}
\frac{\partial u_{a}}{\partial t}
+u_{a}\frac{\partial u_{a}}{\partial x}
+v_{a}\frac{\partial u_{a}}{\partial y} 
=-\frac{1}{\rho_{a}}\frac{\partial p_{a}}{\partial x}
+\nu_{a}\left(\frac{\partial^{2}u_{a}}{\partial x^{2}}
+\frac{\partial^{2}u_{a}}{\partial y^{2}}\right),
\label{eq:geq-ua} 
\end{equation}

\begin{equation}
\frac{\partial v_{a}}{\partial t}
+u_{a}\frac{\partial v_{a}}{\partial x}
+v_{a}\frac{\partial v_{a}}{\partial y}
=-\frac{1}{\rho_{a}}\frac{\partial p_{a}}{\partial y}
+\nu_{a}\left(\frac{\partial^{2}v_{a}}{\partial x^{2}}
+\frac{\partial^{2}v_{a}}{\partial y^{2}}\right), 
\label{eq:geq-va}
\end{equation}

\begin{equation}
\frac{\partial u_{a}}{\partial x}+\frac{\partial v_{a}}{\partial y}=0,
\label{eq:continuity-air}
\end{equation}
where $p_{a}$ is the air pressure, $\rho_{a}=1.3$ ${\rm kg/m^{3}}$, the density of air, and $\nu_{a}=1.3 \times 10^{-5}$ ${\rm m^{2}/s}$, the kinematic viscosity of air. 
The velocity components in the $x$ and $y$ directions in the water layer, $u_{l}$ and $v_{l}$, are governed by \cite{Landau59} 
\begin{equation}
\frac{\partial u_{l}}{\partial t}
+u_{l}\frac{\partial u_{l}}{\partial x}
+v_{l}\frac{\partial u_{l}}{\partial y} 
=-\frac{1}{\rho_{l}}\frac{\partial p_{l}}{\partial x}
+\nu_{l}\left(\frac{\partial^{2}u_{l}}{\partial x^{2}}
+\frac{\partial^{2}u_{l}}{\partial y^{2}}\right),
\label{eq:geq-ul} 
\end{equation}

\begin{equation}
\frac{\partial v_{l}}{\partial t}
+u_{l}\frac{\partial v_{l}}{\partial x}
+v_{l}\frac{\partial v_{l}}{\partial y}
=-\frac{1}{\rho_{l}}\frac{\partial p_{l}}{\partial y}
+\nu_{l}\left(\frac{\partial^{2}v_{l}}{\partial x^{2}}
+\frac{\partial^{2}v_{l}}{\partial y^{2}}\right)
-g, 
\label{eq:geq-vl}
\end{equation}

\begin{equation}
\frac{\partial u_{l}}{\partial x}+\frac{\partial v_{l}}{\partial y}=0,
\label{eq:continuity-water}
\end{equation}
where $\nu_{l}=1.8 \times 10^{-6}$ ${\rm m^{2}/s}$ and $\rho_{l}=1.0 \times 10^{3}$ ${\rm kg/m^{3}}$ are the kinematic viscosity and density of water, respectively, $p_{l}$ is the water pressure and $g$ the gravitational acceleration. 
The continuity equations (\ref{eq:continuity-air}) and (\ref{eq:continuity-water}) can be satisfied by introducing the stream functions $\psi_{a}$ and $\psi_{l}$ such that
$u_{a}=\partial \psi_{a}/\partial y$, 
$v_{a}=-\partial \psi_{a}/\partial x$,
$u_{l}=\partial \psi_{l}/\partial y$, and  
$v_{l}=-\partial \psi_{l}/\partial x$.

Neglecting viscous dissipation in the energy equation, the equations for the temperatures in the air $T_{a}$, water $T_{l}$ and ice $T_{s}$ are \cite{Landau59} 
\begin{equation}
\frac{\partial T_{a}}{\partial t}
+u_{a}\frac{\partial T_{a}}{\partial x}
+v_{a}\frac{\partial T_{a}}{\partial y}
=\kappa_{a}\left(\frac{\partial^{2} T_{a}}{\partial x^{2}}
+\frac{\partial^{2} T_{a}}{\partial y^{2}}\right),
\label{eq:geq-Ta}
\end{equation}

\begin{equation}
\frac{\partial T_{l}}{\partial t}
+u_{l}\frac{\partial T_{l}}{\partial x}
+v_{l}\frac{\partial T_{l}}{\partial y}
=\kappa_{l}\left(\frac{\partial^{2} T_{l}}{\partial x^{2}}+\frac{\partial^{2} T_{l}}{\partial y^{2}}\right),
\label{eq:geq-Tl}
\end{equation}

\begin{equation}
\frac{\partial T_{s}}{\partial t}
=\kappa_{s}\left(\frac{\partial^{2} T_{s}}{\partial x^{2}}
+\frac{\partial^{2} T_{s}}{\partial y^{2}}\right),
\label{eq:geq-Ts}
\end{equation}
where $\kappa_{a}=1.87 \times 10^{-5}$ ${\rm m^{2}/s}$, $\kappa_{l}=1.33 \times 10^{-7}$ ${\rm m^{2}/s}$ and  $\kappa_{s}=1.15 \times 10^{-6}$ ${\rm m^{2}/s}$ are the thermal diffusivities of air, water and ice, respectively. 

%%%%%%%%%%%%%%%%%%%%%%%%%%%%%%%%%%%%%%%%%%%%%%%%%
\subsection{\label{sec:BC}Boundary conditions}
%%%%%%%%%%%%%%%%%%%%%%%%%%%%%%%%%%%%%%%%%%%%%%%%%

The following boundary conditions are the same as those used in a previous paper \cite{Ueno10} except for the first condition in Eq. (\ref{eq:bc-infinity}) herein.
Ignoring the density difference between ice and water, there is no normal fluid motion at the ice-water interface. Then
both velocity components $u_{l}$ and $v_{l}$ at a disturbed ice-water interface, $y=\zeta(t,x)$, must satisfy the no-slip condition:\cite{Myers02_1, Myers02_2, Myers04}
\begin{equation}
u_{l}|_{y=\zeta}=0,
\hspace{1cm}
v_{l}|_{y=\zeta}=0.
\label{eq:bc-ul-vl-zeta}
\end{equation}
Since there is no impingement of supercooled water droplets on the water film,
the kinematic condition at a disturbed water-air interface, $y=\xi(t,x)$, is \cite{Benjamin57,Oron97}
\begin{equation}
\frac{\partial \xi}{\partial t}+u_{l}|_{y=\xi}\frac{\partial \xi}{\partial x}=v_{l}|_{y=\xi}.
\label{eq:bc-kinematic-xi}
\end{equation}
The continuity of velocities of water film flow and airflow at the water-air interface is \cite{Yih90} 
\begin{equation}
u_{l}|_{y=\xi}=u_{a}|_{y=\xi},
\qquad
v_{l}|_{y=\xi}=v_{a}|_{y=\xi}.
\label{eq:bc-ul-vl-ua-va-xi}
\end{equation}
The continuity of tangential and normal stresses at the water-air interface leads to \cite{Landau59, Craik66, Yih90}
\begin{equation}
\mu_{l}\left(\frac{\partial u_{l}}{\partial y}\Big|_{y=\xi}
+\frac{\partial v_{l}}{\partial x}\Big|_{y=\xi}\right)
=\mu_{a}\left(\frac{\partial u_{a}}{\partial y}\Big|_{y=\xi}
+\frac{\partial v_{a}}{\partial x}\Big|_{y=\xi}\right),
\label{eq:bc-shear-stress-xi}
\end{equation}
\begin{equation}
-p_{a}|_{y=\xi}+2\mu_{a}\frac{\partial v_{a}}{\partial y}\Big|_{y=\xi}
-\left(-p_{l}|_{y=\xi}+2\mu_{l}\frac{\partial v_{l}}{\partial y}\Big|_{y=\xi}\right)
=-\gamma\frac{\partial^{2}\xi}{\partial x^{2}}\left[1+\left(\frac{\partial \xi}{\partial x}\right)^{2}\right]^{-3/2},
\label{eq:bc-normal-stress-xi}
\end{equation}
where $\mu_{l}=\rho_{l}\nu_{l}$ and $\mu_{a}=\rho_{a}\nu_{a}$ are the viscosities of water and air, respectively, and $\gamma=7.6 \times 10^{-2}$ N/m is the surface tension. 
The curvature term on the right hand side in Eq. (\ref{eq:bc-normal-stress-xi}) determines the magnitude of the surface tension induced stress. Hence, the condition expressed by Eq. (\ref{eq:bc-normal-stress-xi}) is that the capillary force resisting displacement and the normal stress on either side of the water-air interface should be in equilibrium. \cite{Craik66}

The continuity condition of temperature at the ice-water interface is 
\begin{equation}
T_{l}|_{y=\zeta}=T_{s}|_{y=\zeta}=T_{i},
\label{eq:Tsl}
\end{equation}
in which the interfacial temperature $T_{i}$ is an unknown to be determined.
The Stephan condition is 
\begin{equation}
L\left(\bar{V}+\frac{\partial \zeta}{\partial t} \right)
=K_{s}\frac{\partial T_{s}}{\partial y}\Big|_{y=\zeta}
      -K_{l}\frac{\partial T_{l}}{\partial y}\Big|_{y=\zeta},
\label{eq:heatflux-zeta}
\end{equation}
which is based on the assumption that ice grows in proportion to the heat flux difference across the ice-water interface.
Here $L=3.3 \times 10^{8}$ ${\rm J/m^{3}}$ is the latent heat per unit volume, $\bar{V}$ is the undisturbed ice growth rate, and $K_{s}=2.22$ ${\rm J/(m\,K\,s)}$ and $K_{l}=0.56$ ${\rm J/(m\,K\,s)}$ are thermal conductivities of ice and water, respectively. 

The continuity condition of temperature at the water-air interface is
\begin{equation}
T_{l}|_{y=\xi}=T_{a}|_{y=\xi}=T_{la},
\label{eq:bc-Tla}
\end{equation}
where $T_{la}$ is the temperature at the water-air interface and will be determined later.
The continuity of heat flux at the water-air interface is
\begin{equation}
-K_{l}\frac{\partial T_{l}}{\partial y}\Big|_{y=\xi}
=-K_{a}\frac{\partial T_{a}}{\partial y}\Big|_{y=\xi},
\label{eq:heatflux-xi}
\end{equation}
where $K_{a}=0.024$ ${\rm J/(m\,K\,s)}$ is the thermal conductivity of air. 
Far away from the air boundary layer, the velocities and temperature asymptote to their far-field values:
\begin{equation}
u_{a}|_{y=\infty}=u_{\infty}, \qquad
v_{a}|_{y=\infty}=0, \qquad
T_{a}|_{y=\infty}=T_{\infty}.
\label{eq:bc-infinity}
\end{equation}

%%%%%%%%%%%%%%%%%%%%%%%%%%%%%%%%%%%%%%%%%%%%%%%%%%%%%%%
\subsection{\label{sec:LSA}Linear stability analysis}
%%%%%%%%%%%%%%%%%%%%%%%%%%%%%%%%%%%%%%%%%%%%%%%%%%%%%%%

In this paper, the stability analysis will be limited to one-dimensional disturbances. A simple normal-mode analysis is applied to the field variables. Suppose an ice-water interface disturbance with a small amplitude $\zeta_{k}$ resulting in $\zeta(t,x)=\zeta_{k}{\rm exp}[\sigma t+i kx]$,
where $k$ is the wave number and $\sigma=\sigma^{(r)}+i \sigma^{(i)}$, $\sigma^{(r)}$ and $v_{p} \equiv -\sigma^{(i)}/k$ are the amplification rate and phase velocity of the disturbance, respectively.
We separate $\xi$, $\psi_{a}$, $\psi_{l}$, $p_{a}$, $p_{l}$, $T_{a}$, $T_{l}$ and $T_{s}$ into undisturbed steady fields with bar and disturbed parts with prime as follows:
$\xi=\bar{h}_{0}+\xi'$,
$\psi_{a}=\bar{\psi}_{a}+\psi'_{a}$,
$\psi_{l}=\bar{\psi}_{l}+\psi'_{l}$,
$p_{a}=\bar{p}_{a}+p'_{a}$,
$p_{l}=\bar{p}_{l}+p'_{l}$,
$T_{a}=\bar{T}_{a}+T'_{a}$,
$T_{l}=\bar{T}_{l}+T'_{l}$
and
$T_{s}=\bar{T}_{s}+T'_{s}$.
The undisturbed velocities in the air and water are derived from 
$\bar{u}_{a}= \partial \bar{\psi}_{a}/\partial y$, 
$\bar{v}_{a}=-\partial \bar{\psi}_{a}/\partial x$,
$\bar{u}_{l}= \partial \bar{\psi}_{l}/\partial y$, and  
$\bar{v}_{l}=-\partial \bar{\psi}_{l}/\partial x$.
We define 
$\bar{G}_{a} \equiv -\partial \bar{T}_{a}/\partial y|_{y=\bar{h}_{0}}$ and  
$\bar{G}_{l} \equiv -\partial \bar{T}_{l}/\partial y|_{y=0}$
as temperature gradients at the undisturbed water-air interface and ice-water interface, respectively. 
The disturbed field variables are assumed to be expanded in normal mode form, as follows:
\begin{equation}
\left(
\begin{array}{c}
\xi' \\ 
\psi'_{a} \\ 
\psi'_{l} \\ 
p'_{a} \\ 
p'_{l} \\ 
T'_{a} \\ 
T'_{l} \\ 
T'_{s}  
\end{array}
\right)
=
\left(
\begin{array}{c}
\xi_{k} \\ 
u_{\infty}f_{a}(\eta)\xi_{k} \\ 
u_{la}f_{l}(y_{*})\zeta_{k} \\
(\rho_{a}u_{\infty}^{2}/\delta_{0})g_{a}(\eta)\xi_{k} \\ 
(\rho_{l}u_{la}^{2}/\bar{h}_{0})g_{l}(y_{*})\zeta_{k} \\ 
H_{a}(\eta)\bar{G}_{a}\xi_{k} \\ 
H_{l}(y_{*})\bar{G}_{l}\zeta_{k} \\ 
H_{s}(y_{*})\bar{G}_{l}\zeta_{k} 
\end{array}
\right)
\exp[\sigma t+i kx].
\label{eq:pertset}
\end{equation}

We introduce the following two dimensionless variables $\eta=(y-\bar{h}_{0})/\delta_{0}$ in the air and $y_{*}=y/\bar{h}_{0}$ in the water layer. 
Here
$\delta_{0}=(2\nu_{a}x/u_{\infty})^{1/2}$ is a scaled measure of the air boundary layer thickness, \cite{Schlichting99} 
$u_{\infty}$ is the free stream velocity and 
$x$ is the distance from the leading edge where water is supplied.
$u_{la}$ in Eq. (\ref{eq:pertset}) is the surface velocity of the water film driven by wind drag.
As shown in Eq. (\ref{eq:h0-ula}) herein, $\bar{h}_{0}$ and $u_{la}$ are functions of $x$.
$\xi_{k}$ is the amplitude of the water-air interface disturbance, and $f_{a}$, $f_{l}$, $g_{a}$, $g_{l}$, $H_{a}$, $H_{l}$ and $H_{s}$ are dimensionless amplitudes of disturbed parts of the stream function $\psi$, pressure $p$ and temperature $T$. 
In the following, quasi-stationary approximation is used for the disturbed fields as in previous papers, \cite{Ueno03, Ueno04, Ueno07, UFYT10, Ueno10} and we assume that the undisturbed part of temperature gradient within the ice does not exist, hence $\bar{T}_{s}=T_{sl}$ ($T_{sl}=$0 $^{\circ}$C for pure water).

%%%%%%%%%%%%%%%%%%%%%%%%%%%%%%%%%%%%%%%%%%%%%%%%%%%%%%%%%%%%%%%%%%%%%%%%%%%%%%%%%%%%%%%%%%%%%%%%%%%%%%%%%%%%%%%%%%%
\subsubsection{Equations and boundary conditions for undisturbed flows and temperatures in the air and water film}
%%%%%%%%%%%%%%%%%%%%%%%%%%%%%%%%%%%%%%%%%%%%%%%%%%%%%%%%%%%%%%%%%%%%%%%%%%%%%%%%%%%%%%%%%%%%%%%%%%%%%%%%%%%%%%%%%%%

Using the Blasius-type similarity transformations
and substituting
$\bar{\psi}_{a}=u_{\infty}\delta_{0}\bar{F}_{a}(\eta)$ and 
$\bar{T}_{a*}=(\bar{T}_{a}-T_{\infty})/(T_{la}-T_{\infty})$
into the partial differential equations (\ref{eq:geq-ua}), (\ref{eq:geq-va}) and (\ref{eq:geq-Ta}),
a set of ordinary differential equations for the dimensionless functions $\bar{F}_{a}$ and $\bar{T}_{a*}$
and their boundary conditions are obtained: \cite{Schlichting99} 
\begin{equation}
\frac{d^{3}\bar{F}_{a}}{d\eta^{3}}
=-\bar{F}_{a}\frac{d^{2}\bar{F}_{a}}{d\eta^{2}},
\label{eq:geq-basicFa}
\end{equation}
\begin{equation}
\frac{d^{2}\bar{T}_{a*}}{d\eta^{2}}
=-Pr_{a}\bar{F}_{a}\frac{d\bar{T}_{a*}}{d\eta},
\label{eq:geq-basicTa}
\end{equation}
\begin{equation}
\frac{d\bar{F}_{a}}{d\eta}\Big|_{\eta=0}=0, \qquad
\bar{F}_{a}|_{\eta=0}=0, \qquad
\frac{d\bar{F}_{a}}{d\eta}\Big|_{\eta=\infty}=1, \qquad
\bar{T}_{a*}|_{\eta=0}=1, \qquad
\bar{T}_{a*}|_{\eta=\infty}=0,
\label{eq:bc-basic_air}
\end{equation}
where $Pr_{a}=\nu_{a}/\kappa_{a}$ is the Prandtl number of air. 
The first and second equations in Eq. (\ref{eq:bc-basic_air}) are derived from the undisturbed parts in Eq. (\ref{eq:bc-ul-vl-ua-va-xi}) by using the fact that the free stream velocity $u_{\infty}$ is much larger than the water surface velocity $u_{la}$,\cite{Ueno10} as shown in Table \ref{tab:tableI}.  
The third equation in Eq. (\ref{eq:bc-basic_air}) is the result of the condition $\bar{u}_{a}|_{y=\infty}=\partial\bar{\psi}_{a}/\partial y|_{y=\infty}=u_{\infty}$. 
The boundary conditions $\bar{T}_{a}|_{y=\bar{h}_{0}}=T_{la}$ and $\bar{T}_{a}|_{y=\infty}=T_{\infty}$ yield the fourth and fifth equations in Eq. (\ref{eq:bc-basic_air}).

For water film, we assume the following scaling $\bar{h}_{0}=C_{1}x^{a}$, $u_{la}=C_{2}x^{b}$ and $T_{sl}-T_{la}=C_{3}x^{c}$, and 
$\bar{\psi}_{l}=u_{la}\bar{h}_{0}\bar{F}_{l}(y_{*})$. 
Here the constants $C_{1}$, $C_{2}$, $C_{3}$, $a$, $b$ and $c$ are determined from the boundary conditions, as follows.
First, if the volumetric water flow rate per width, 
\begin{equation}
Q/l_{w}=\int_{0}^{\bar{h}_{0}}\bar{u}_{l}dy=C_{1}C_{2}x^{a+b}\int_{0}^{1}\bar{u}_{l*}dy_{*},
\label{eq:Qoverl}
\end{equation}
is constant, $a+b=0$ must hold, where $\bar{u}_{l*}\equiv\bar{u}_{l}/u_{la}=d\bar{F}_{l}/dy_{*}$. 
This is also derived by substituting 
$\bar{\psi}_{l}=u_{la}\bar{h}_{0}\bar{F}_{l}(y_{*})$ into the undisturbed part of Eq. (\ref{eq:bc-kinematic-xi}), $\bar{u}_{l}|_{y=\bar{h}_{0}}d\bar{h}_{0}/dx=\bar{v}_{l}|_{y=\bar{h}_{0}}$. 
Second, the undisturbed part of Eq. (\ref{eq:bc-shear-stress-xi}) yields
\begin{equation}
\frac{C_{2}}{C_{1}}\mu_{l}\frac{d\bar{u}_{l*}}{dy_{*}}\Big|_{y_{*}=1}x^{b-a}
=\left(\frac{u_{\infty}^{3}}{2\nu_{a}}\right)^{1/2}\mu_{a}\frac{d^{2}\bar{F}_{a}}{d\eta^{2}}\Big|_{\eta=0}x^{-1/2},
\label{eq:C1-C2}
\end{equation}
from which $b-a=-1/2$ must hold. 
Finally, the undisturbed part of Eq. (\ref{eq:heatflux-xi}) yields 
\begin{equation}
K_{l}\frac{C_{3}}{C_{1}}\frac{d\bar{T}_{l*}}{dy_{*}}\Big|_{y_{*}=1}x^{c-a}
=-K_{a}\left(\frac{u_{\infty}}{2\nu_{a}}\right)^{1/2}T_{\infty}\frac{d\bar{T}_{a*}}{d\eta}\Big|_{\eta=0}x^{-1/2},
\label{eq:C1-C3}
\end{equation}
where $T_{la}-T_{\infty}\approx -T_{\infty}$ is used because we assume $|T_{la}| \ll |T_{\infty}|$.
From Eq. (\ref{eq:C1-C3}) $c-a=-1/2$ must hold. 

Substituting 
$\bar{\psi}_{l}=u_{la}\bar{h}_{0}\bar{F}_{l}(y_{*})$ and
$\bar{T}_{l*}=(\bar{T}_{l}-T_{sl})/(T_{sl}-T_{la})$ 
into the partial differential equations (\ref{eq:geq-ul}) and (\ref{eq:geq-Tl}), a set of differential equations for the dimensionless functions $\bar{u}_{l*}$ and $\bar{T}_{l*}$ is obtained:
\begin{equation}
\frac{d^{2}\bar{u}_{l*}}{dy_{*}^{2}}
=2b\frac{\Rey_{l}\bar{h}_{0}}{\Rey_{a}\delta_{0}}\bar{u}_{l*}^{2},
\label{eq:geq-basicFl}
\end{equation}
\begin{equation}
\frac{d^{2}\bar{T}_{l*}}{dy_{*}^{2}}
=2c\frac{\Rey_{l}\bar{h}_{0}}{\Rey_{a}\delta_{0}}Pr_{l}\bar{u}_{l*}\bar{T}_{l*},
\label{eq:geq-basicTl}
\end{equation}
where $\Rey_{l}=u_{la}\bar{h}_{0}/\nu_{l}$ and $\Rey_{a}=u_{\infty}\delta_{0}/\nu_{a}$ are the Reynolds numbers of the water and air, and $Pr_{l}=\nu_{l}/\kappa_{l}$ is the Prandtl number of water.
Since $\Rey_{l}\bar{h}_{0}/(\Rey_{a}\delta_{0}) \ll 1$ for values shown in Table \ref{tab:tableI}, 
Eqs. (\ref{eq:geq-basicFl}) and (\ref{eq:geq-basicTl}) can be approximated as 
$d^{2}\bar{u}_{l*}/dy_{*}^{2}=0$ and $d^{2}\bar{T}_{l*}/dy_{*}^{2}=0$. 
The boundary conditions 
$\bar{u}_{l}|_{y=0}=0$, 
$\mu_{l}\partial\bar{u}_{l}/\partial y|_{y=\bar{h}_{0}}
=\mu_{a}\partial\bar{u}_{a}/\partial y|_{y=\bar{h}_{0}}$, 
$\bar{T}_{l}|_{y=0}=T_{sl}$ and
$\bar{T}_{l}|_{y=\bar{h}_{0}}=T_{la}$
can be written as
$\bar{u}_{l*}|_{y_{*}=0}=0$,
$d\bar{u}_{l*}/dy_{*}|_{y_{*}=1}=1$, 
$\bar{T}_{l*}|_{y_{*}=0}=0$
and 
$\bar{T}_{l*}|_{y_{*}=1}=-1$, respectively, by defining $u_{la}=\mu_{a}u_{\infty}\bar{h}_{0}d^{2}\bar{F}_{a}/d\eta^{2}|_{\eta=0}/(\mu_{l}\delta_{0})$.
Therefore, the solutions of the undisturbed velocity and temperature profiles in the water film are linear in $y_{*}$,
that is,
$\bar{u}_{l*}=y_{*}$ and $\bar{T}_{l*}=-y_{*}$.
This is in agreement with the more usual lubrication approach for describing icing with shear.
\cite{Oron97, Myers02_1, Myers02_2, Myers04}

From $a+b=0$, $b-a=-1/2$ and $c-a=-1/2$, we obtain $a=1/4$, $b=-1/4$, $c=-1/4$. The value of $a$ coincides with that stated in previous papers. \cite{Nelson95,Tsao97}
$C_{1}$, $C_{2}$ and $C_{3}$ are determined from Eqs. (\ref{eq:Qoverl}), (\ref{eq:C1-C2}) and (\ref{eq:C1-C3}). 
Hence
$\bar{h}_{0}$ and $u_{la}$ 
can be expressed as follows:
\begin{equation}
\bar{h}_{0}=\left[
\frac{2\mu_{l}(2\nu_{a})^{1/2}}{\mu_{a}\frac{d^{2}\bar{F}_{a}}{d\eta^{2}}\Big|_{\eta=0}}
\right]^{1/2}(Q/l_{w})^{1/2}u_{\infty}^{-3/4}x^{1/4}, \qquad
u_{la}=\left[\frac{2\mu_{a}\frac{d^{2}\bar{F}_{a}}{d\eta^{2}}\Big|_{\eta=0}}{\mu_{l}(2\nu_{a})^{1/2}}
\right]^{1/2}(Q/l_{w})^{1/2}u_{\infty}^{3/4}x^{-1/4}. 
\label{eq:h0-ula}
\end{equation}
It is found that $\bar{h}_{0}$ and $u_{la}$ depend on $Q/l_{w}$ and $u_{\infty}$, and vary slowly with $x$. 
We assume that the scaling of $\bar{h}_{0}$ and $u_{la}$ for $x$ holds except for the very vicinity of water source. 
Equations (\ref{eq:geq-basicFa}) and (\ref{eq:bc-basic_air}) yield $d^{2}\bar{F}_{a}/d\eta^{2}|_{\eta=0}=0.47$.
The shear rate for the undisturbed water layer is then given by
\begin{equation}
\frac{\partial\bar{u}_{l}}{\partial y}\Big|_{y=0}
=\frac{u_{la}}{\bar{h}_{0}}\frac{d\bar{u}_{l*}}{dy_{*}}\Big|_{y_{*}=0}
=\left(\frac{1}{2\nu_{a}x}\right)^{1/2}
\frac{\mu_{a}}{\mu_{l}}\frac{d^{2}\bar{F}_{a}}{d\eta^{2}}\Big|_{\eta=0}u_{\infty}^{3/2},
\label{eq:ul0overh0}
\end{equation}
and its value is in the range 30.6 to 449.3 ${\rm s}^{-1}$ for the range of $u_{\infty}=5$ to 30 m/s at $x=0.1$ m. Hence the value of the time defined by the inverse of shear rate lies within the range of $2.2\times 10^{-3}$ to $3.3 \times 10^{-2}$ s for the above parameters.

Using Eq. (\ref{eq:C1-C3}), $C_{1}=\bar{h}_{0}x^{-1/4}$ and $\delta_{0}=(2\nu_{a}x/u_{\infty})^{1/2}$, we can express  
\begin{equation}
T_{sl}-T_{la}
=-C_{1}\frac{K_{a}}{K_{l}}\left(\frac{u_{\infty}}{2\nu_{a}}\right)^{1/2}\bar{G}_{a*}T_{\infty}x^{-1/4}
=-\frac{K_{a}}{K_{l}}\frac{\bar{h}_{0}}{\delta_{0}/\bar{G}_{a*}}T_{\infty},
\label{eq:Tla}
\end{equation}
where $\bar{G}_{a*}\equiv -d\bar{T}_{a*}/d\eta|_{\eta=0}$.
Since the undisturbed part of temperature gradient within the ice does not exist in the model herein,
the undisturbed part of Eq. (\ref{eq:heatflux-zeta}) yields
$L\bar{V}=K_{l}(T_{sl}-T_{la})/\bar{h}_{0}$,
into which Eq. (\ref{eq:Tla}) is substituted to obtain the undisturbed ice growth rate
\begin{equation}
\bar{V}=-\frac{K_{a}T_{\infty}}{L(\delta_{0}/\bar{G}_{a*})}.
\label{eq:V}
\end{equation}
Equations (\ref{eq:Tla}) and (\ref{eq:V}) are the same form as those in a previous paper, \cite{Ueno10}
but the value of $\bar{G}_{a*}$ is different. From Eqs. (\ref{eq:geq-basicFa}), (\ref{eq:geq-basicTa}) and (\ref{eq:bc-basic_air}), we obtain $\bar{G}_{a*}=0.413$ for $Pr_{a}=0.7$. The variation of $T_{la}$ and $\bar{V}$ against $u_{\infty}$ is shown in Table \ref{tab:tableII}. 

%%%%%%%%%%%%%%%%%%%%%%%%%%%%%%%%%%%%%%%%%%%%%%%%%%%%%%%%%%%%%%%%%%%%%%%%%%%%%%%%%%%%%%%%%%%%%%%%%%%%%%%%%%%%%%%%%
\subsubsection{Equations and boundary conditions for disturbed flows and temperatures in the air and water film}
%%%%%%%%%%%%%%%%%%%%%%%%%%%%%%%%%%%%%%%%%%%%%%%%%%%%%%%%%%%%%%%%%%%%%%%%%%%%%%%%%%%%%%%%%%%%%%%%%%%%%%%%%%%%%%%%%

When the assumed forms of $\psi_{a}$ and $T_{a}$ are substituted into the complete equations (\ref{eq:geq-ua}), (\ref{eq:geq-va}) and (\ref{eq:geq-Ta}), the differential equations for the amplitudes $f_{a}$ and $H_{a}$ are obtained:
\begin{eqnarray}
\frac{d^{4}f_{a}}{d\eta^4}
&=&-\bar{F}_{a}\frac{d^{3}f_{a}}{d\eta^{3}}
+\left\{2k_{a*}^{2}-(2-i k_{a*}\Rey_{a})\frac{d\bar{F}_{a}}{d\eta}\right\}\frac{d^{2}f_{a}}{d\eta^{2}} \nonumber \\
&& +\left\{k_{a*}^{2}\left(\bar{F}_{a}+2\eta\frac{d\bar{F}_{a}}{d\eta}\right)
-\frac{d^{2}\bar{F}_{a}}{d\eta^{2}}\right\}\frac{df_{a}}{d\eta}
-\left\{k_{a*}^{4}+i k_{a*}\Rey_{a}\left(k_{a*}^{2}\frac{d\bar{F}_{a}}{d\eta}+\frac{d^{3}\bar{F}_{a}}{d\eta^{3}}\right)\right\}f_{a},\nonumber \\
\label{eq:geq-fa} 
\end{eqnarray}
\begin{eqnarray}
\frac{d^{2}(\bar{G}_{a*}H_{a})}{d\eta^{2}}
&=&-Pr_{a}\bar{F}_{a}\frac{d(\bar{G}_{a*}H_{a})}{d\eta}
+\left\{k_{a*}^{2}+Pr_{a}(-1+i k_{a*}\Rey_{a})\frac{d\bar{F}_{a}}{d\eta}\right\}(\bar{G}_{a*}H_{a}) \nonumber \\ 
&& -i k_{a*}Pr_{a}\Rey_{a}\frac{d\bar{T}_{a*}}{d\eta}f_{a},
\label{eq:geq-Ha}
\end{eqnarray}
where $k_{a*}=k\delta_{0}$ is the dimensionless wave number normalized by the length $\delta_{0}$. 
The disturbed part of Eq. (\ref{eq:bc-ul-vl-ua-va-xi}) and the boundary conditions 
$u'_{a}|_{y=\infty}=\partial \psi'_{a}/\partial y|_{y=\infty}=0$ and 
$v'_{a}|_{y=\infty}=-\partial \psi'_{a}/\partial x|_{y=\infty}=0$
yield
\begin{equation}
\frac{df_{a}}{d\eta}\Big|_{\eta=0}=-\frac{d^{2}\bar{F}_{a}}{d\eta^{2}}\Big|_{\eta=0}, \qquad
f_{a}|_{\eta=0}=0, \qquad
\frac{df_{a}}{d\eta}\Big|_{\eta=\infty}=0, \qquad
f_{a}|_{\eta=\infty}=0. 
\label{eq:bc-fa}
\end{equation}
We note that, as shown in Table \ref{tab:tableI}, since the water surface velocity $u_{la}$ is significantly lower than the free stream velocity $u_{\infty}$, $u_{a}|_{y=\xi}=0$ and $v_{a}|_{y=\xi}=0$ are good approximation, from which the first and second equations in Eq. (\ref{eq:bc-fa}) are obtained. 
Furthermore, the disturbed part of Eq. (\ref{eq:bc-Tla}) and the boundary condition $T_{a}'|_{y=\infty}=0$ give
\begin{equation}
H_{a}|_{\eta=0}=1, \qquad
H_{a}|_{\eta=\infty}=0.
\label{eq:bc-Ha}
\end{equation}

On the other hand, when the assumed forms of $\psi_{l}$ and $T_{l}$ are substituted into Eqs. (\ref{eq:geq-ul}) (\ref{eq:geq-vl}) and (\ref{eq:geq-Tl}) and neglecting the terms with $\Rey_{l}\bar{h}_{0}/(\Rey_{a}\delta_{0}) \ll 1$, the disturbed parts yield the equation for the amplitudes $f_{l}$ and $H_{l}$: 
\begin{equation}
\frac{d^{4}f_{l}}{dy_{*}^{4}}
=\left(2k_{l*}^{2}+ik_{l*} \Rey_{l}\bar{u}_{l*}\right)\frac{d^{2}f_{l}}{dy_{*}^{2}}
-\left\{k_{l*}^{4}+ik_{l*} \Rey_{l}\left(k_{l*}^{2}\bar{u}_{l*}+\frac{d^{2}\bar{u}_{l*}}{dy_{*}^{2}}\right)\right\}f_{l},
\label{eq:geq-fl}
\end{equation}
\begin{equation}
\frac{d^{2}H_{l}}{dy_{*}^{2}}
=\left(k_{l*}^{2}+ik_{l*} \Pec_{l}\bar{u}_{l*}\right)H_{l} 
-ik_{l*} \Pec_{l}\frac{d\bar{T}_{l*}}{d y_{*}}f_{l},
\label{eq:geq-Hl}
\end{equation}
where $k_{l*}=k\bar{h}_{0}$ is the dimensionless wave number normalized by the length $\bar{h}_{0}$,
$\bar{T}_{l*}(y_{*})=-y_{*}$ and $\Pec_{l}\equiv u_{la}\bar{h}_{0}/\kappa_{l}$ is the Peclet number.
We note that Eqs. (\ref{eq:geq-fl}) and (\ref{eq:geq-Hl}) are in the same form as found in previous papers, \cite{Ueno03, Ueno04, Ueno07, UFYT10,Ueno10}
but the form of $\bar{u}_{l*}$ is different. In this paper, $\bar{u}_{l*}=y_{*}$ is used. 
Using Eq. (\ref{eq:h0-ula}), the Reynolds number and the Peclet number of the water
can be written as $\Rey_{l}=(2/\nu_{l})Q/l_{w}$ and $\Pec_{l}=(2/\kappa_{l})Q/l_{w}$, which are independent of $u_{\infty}$ and $x$.

Linearization of the disturbed parts of Eq. (\ref{eq:bc-ul-vl-zeta}) at $y=0$ and Eqs. (\ref{eq:bc-shear-stress-xi}) and (\ref{eq:bc-normal-stress-xi}) at $y=\bar{h}_{0}$ yield the boundary conditions for $f_{l}$:
\begin{equation}
\frac{df_{l}}{dy_{*}}\Big|_{y_{*}=0}+1=0, \qquad
f_{l}|_{y_{*}=0}=0,
\label{eq:bc-fl0-dfl0}
\end{equation}
\begin{equation}
\frac{d^{2}f_{l}}{dy_{*}^{2}}\Big|_{y_{*}=1}
+\left(k_{l*}^{2}+\Sigma_{a}\right)f_{l}|_{y_{*}=1}=0, 
\label{eq:bc-shearstress}
\end{equation}
\begin{eqnarray}
\frac{d^{3}f_{l}}{dy_{*}^{3}}\Big|_{y_{*}=1}
-\left(3k_{l*}^{2}+ik_{l*} \Rey_{l}\right)\frac{df_{l}}{dy_{*}}\Big|_{y_{*}=1} \nonumber \\
+ik_{l*}\Rey_{l}
\left(1+\frac{1}{Fr^{2}}+Wek_{l*}^{2}+\Pi_{a}
\right)f_{l}|_{y_{*}=1}=0,
\label{eq:bc-normalstress}
\end{eqnarray}
where
\begin{equation}
\Sigma_{a}=\frac{\mu_{a}}{\mu_{l}}\frac{u_{\infty}}{u_{la}}\left(\frac{\bar{h}_{0}}{\delta_{0}}\right)^{2}
\frac{d^{2}f_{a}}{d\eta^{2}}\Big|_{\eta=0},
\label{eq:shearstress_a}
\end{equation}
\begin{eqnarray}
\Pi_{a}=-\frac{\rho_{a}}{\rho_{l}}\left(\frac{u_{\infty}}{u_{la}}\right)^{2}\frac{\bar{h}_{0}}{\delta_{0}}
\frac{1+ik_{a*}\Rey_{a}}{1+(k_{a*}\Rey_{a})^{2}}
\left\{
\frac{d^{3}f_{a}}{d\eta^{3}}\Big|_{\eta=0}
+3k_{a*}^{2}\frac{d^{2}\bar{F}_{a}}{d\eta^{2}}\Big|_{\eta=0}
\right\}, \nonumber \\
\label{eq:normalstress_a}
\end{eqnarray}
$Fr=u_{la}/(g\bar{h}_{0})^{1/2}$ is the Froude number, and $We=\gamma/(\rho_{l}u_{la}^{2}\bar{h}_{0})$ is the Weber number.

Linearization of the disturbed part of Eqs. (\ref{eq:bc-Tla}) and (\ref{eq:heatflux-xi}) at $y=\bar{h}_{0}$ yields
\begin{equation}
H_{l}|_{y_{*}=1}+f_{l}|_{y_{*}=1}=0,
\label{eq:Tla-xi}
\end{equation}
\begin{equation}
\frac{dH_{l}}{dy_{*}}\Big|_{y_{*}=1}
-\frac{\bar{h}_{0}}{\delta_{0}}\left(-\frac{dH_{a}}{d\eta}\Big|_{\eta=0}\right)f_{l}|_{y_{*}=1}=0.
\label{eq:heatflux-xi-h0}
\end{equation}
In deriving Eqs. (\ref{eq:bc-shearstress}), (\ref{eq:bc-normalstress}), (\ref{eq:Tla-xi}) and (\ref{eq:heatflux-xi-h0}), 
the relation between the amplitude of the water-air interface and that of the ice-water interface,
$\xi_{k}=-f_{l}|_{y_{*}=1}\zeta_{k}$, 
is used, which is derived from the linearization of Eq. (\ref{eq:bc-kinematic-xi}) at $y=\bar{h}_{0}$. \cite{Ueno03, Ueno04, Ueno07, UFYT10, Ueno10}
It should be noted that the water flow is affected by the terms $\Sigma_{a}$ in (\ref{eq:bc-shearstress}) and $\Pi_{a}$ in (\ref{eq:bc-normalstress}) due to the tangential and normal air shear stress disturbances, respectively. 
The coupling between the air and water flows affects the disturbed part of temperature distribution in the water layer through the boundary conditions found in Eqs. (\ref{eq:Tla-xi}) and (\ref{eq:heatflux-xi-h0}). 

%%%%%%%%%%%%%%%%%%%%%%%%%%%%%%%%%%%%
\subsubsection{Dispersion relation}
%%%%%%%%%%%%%%%%%%%%%%%%%%%%%%%%%%%%

The disturbed parts of Eqs. (\ref{eq:Tsl}) and (\ref{eq:heatflux-zeta}) give
the dimensionless amplification rate
$\sigma_{*}^{(r)}\equiv \sigma^{(r)}/(K_{a}\bar{G}_{a}/L\bar{h}_{0})$, 
and the dimensionless phase velocity 
$v_{p*}\equiv -\sigma^{(i)}/(kK_{a}\bar{G}_{a}/L)$, \cite{Ueno03, Ueno04, Ueno07, UFYT10, Ueno10}
\begin{equation}
\sigma_{*}^{(r)}=-\frac{dH_{l}^{(r)}}{dy_{*}}\Big|_{y_{*}=0}+K^{s}_{l}k_{l*}(H_{l}^{(r)}|_{y_{*}=0}-1),
\label{eq:amp}
\end{equation}
\begin{equation}
v_{p*}=-\frac{1}{k_{l*}}\left(-\frac{dH_{l}^{(i)}}{dy_{*}}\Big|_{y_{*}=0}+K^{s}_{l}k_{l*}H_{l}^{(i)}|_{y_{*}=0}\right),
\label{eq:phasevel}
\end{equation}
where $H_{l}^{(r)}$ and $H_{l}^{(i)}$ are the real and imaginary parts of $H_{l}$, and $K^{s}_{l}=K_{s}/K_{l}=3.96$ is the ratio of the thermal conductivity of ice to that of water. 

The numerical procedure for calculating Eqs. (\ref{eq:amp}) and (\ref{eq:phasevel}) is as follows: 
First, Eqs. (\ref{eq:geq-basicFa}), (\ref{eq:geq-basicTa}), (\ref{eq:geq-fa}) and (\ref{eq:geq-Ha}) are simultaneously solved for a given $u_{\infty}$ and $x$ with boundary conditions (\ref{eq:bc-basic_air}), (\ref{eq:bc-fa}) and (\ref{eq:bc-Ha}). The derived solutions $\bar{F}_{a}$ and $f_{a}$ are substituted into Eqs. (\ref{eq:shearstress_a}) and (\ref{eq:normalstress_a}). Using the boundary conditions (\ref{eq:bc-fl0-dfl0}), (\ref{eq:bc-shearstress}) and (\ref{eq:bc-normalstress}), Eq. (\ref{eq:geq-fl}) is solved. 
Then Eq. (\ref{eq:geq-Hl}) is solved with the boundary conditions (\ref{eq:Tla-xi}) and (\ref{eq:heatflux-xi-h0}), using solutions $f_{l}$ and $H_{a}$.   
Finally, substituting solution $H_{l}$ into Eqs. (\ref{eq:amp}) and (\ref{eq:phasevel}) and replacing $k_{l*}$ with $(\bar{h}_{0}/\delta_{0})k_{a*}$, Eqs. (\ref{eq:amp}) and (\ref{eq:phasevel}) are obtained with respect to $k_{a*}$.

%%%%%% Table.1 %%%%%%%%%%%%%%%%%%%%%%%%%%%%%%%%%%%%%%%%%%%%%%%%%%%%%%%%%%%%%%%%%%%%%%%%%%%%%%%%%%%%%%%%%%%
\begin{table}[ht]
\caption{\label{tab:tableI}  
Variation of a length, $\delta_{0}$,
thickness of water film, $\bar{h}_{0}$, 
water-air surface velocity, $u_{la}$,
inverse of square of the Froude number, $1/Fr^{2}=g\bar{h}_{0}/u_{la}^{2}$,
the Weber number, $We=\gamma/(\rho_{l}u_{la}^{2}\bar{h}_{0})$,
the Reynolds number of air, $\Rey_{a}=u_{\infty}\delta_{0}/\nu_{a}$,
against the free stream velocity, $u_{\infty}$,
for $Q/l_{w}=1000$ $[{\rm (ml/h)/cm}]$ and $x=0.1$ m. 
The corresponding values of the Reynolds number and the Peclet number of water are 
$\Rey_{l}=u_{la}\bar{h}_{0}/\nu_{l}=31$ and 
$\Pec_{l}=u_{la}\bar{h}_{0}/\kappa_{l}=418$, respectively.}
\begin{ruledtabular}
\begin{tabular}{ccccccc}
$u_{\infty}$ (m/s) & $\delta_{0}$ ($\mu$m) & $\bar{h}_{0}$ ($\mu$m) &  $u_{la}$ (cm/s) 
& $1/Fr^{2}$ & $We$ & $\Rey_{a}$ \\ 
  5 & 721 & 1348 &  4.1 & 7.78 & 33.19 &  277 \\
 10 & 510 &  802 &  6.9 & 1.64 & 19.74 &  392 \\
 15 & 416 &  591 &  9.4 & 0.66 & 14.56 &  480 \\
 20 & 361 &  477 & 11.7 & 0.34 & 11.74 &  555 \\ 
 25 & 322 &  403 & 13.8 & 0.21 &  9.93 &  620 \\
 30 & 294 &  352 & 15.8 & 0.14 &  8.66 &  679 \\
100 & 161 &  143 & 39.0 & 0.01 &  3.51 & 1240 \\  
\end{tabular}
\end{ruledtabular}
\end{table}
%%%%%%%%%%%%%%%%%%%%%%%%%%%%%%%%%%%%%%%%%%%%%%%%%%%%%%%%%%%%%%%%%%%%%%%%%%%%%%%%%%%%%%%%%%%%%%%%%%%%%%%%%
 
%%%%%%%%%%%%%%%%%%%%%%%%%%%%%%%%%%%%%%
\section{\label{sec:results}Results}
%%%%%%%%%%%%%%%%%%%%%%%%%%%%%%%%%%%%%%

\begin{figure}[ht]
\begin{center}
\includegraphics[width=7cm,height=7cm,keepaspectratio,clip]{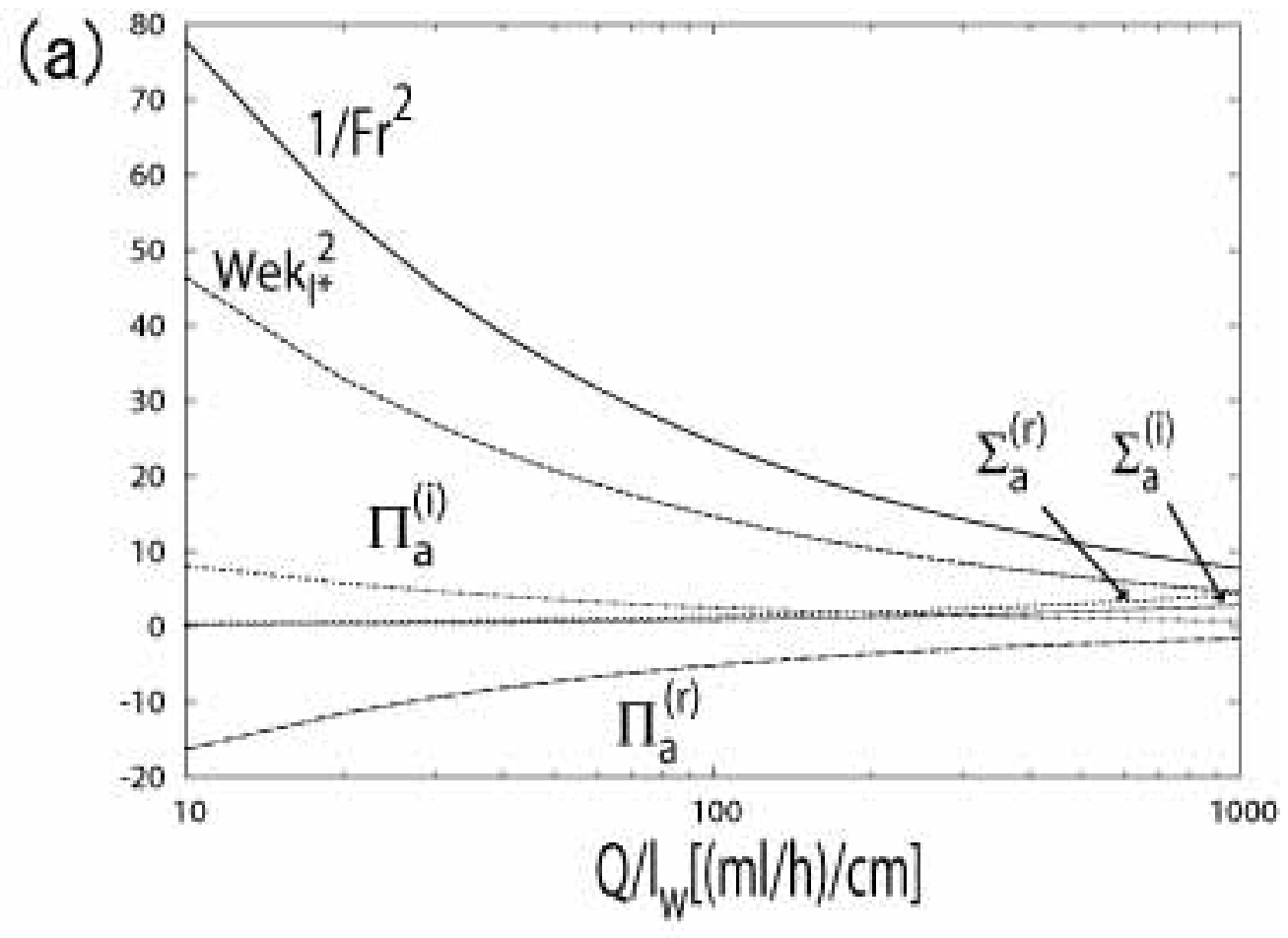}\qquad
\includegraphics[width=7cm,height=7cm,keepaspectratio,clip]{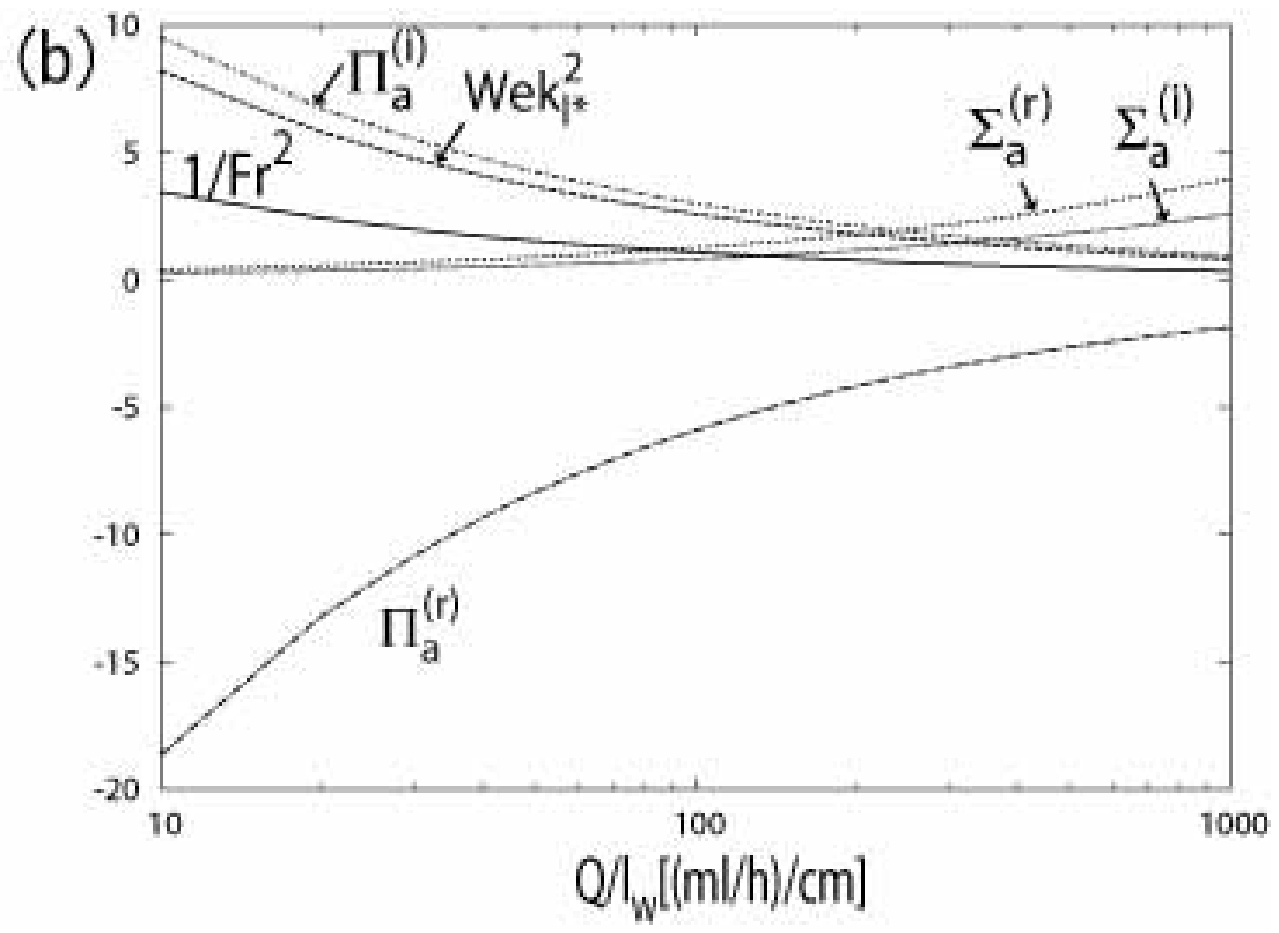}\\[5mm]
\includegraphics[width=7cm,height=7cm,keepaspectratio,clip]{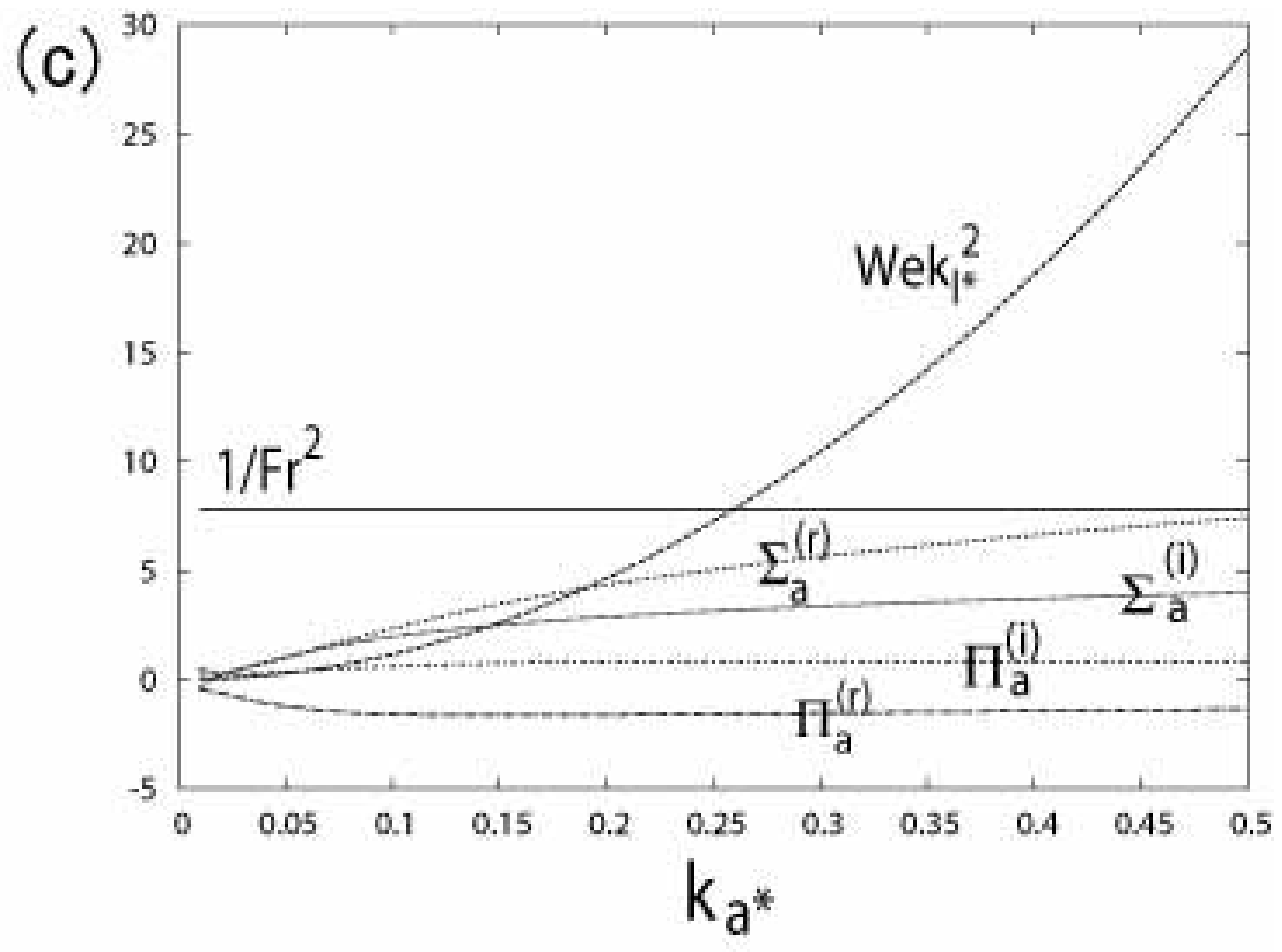}\qquad
\includegraphics[width=7cm,height=7cm,keepaspectratio,clip]{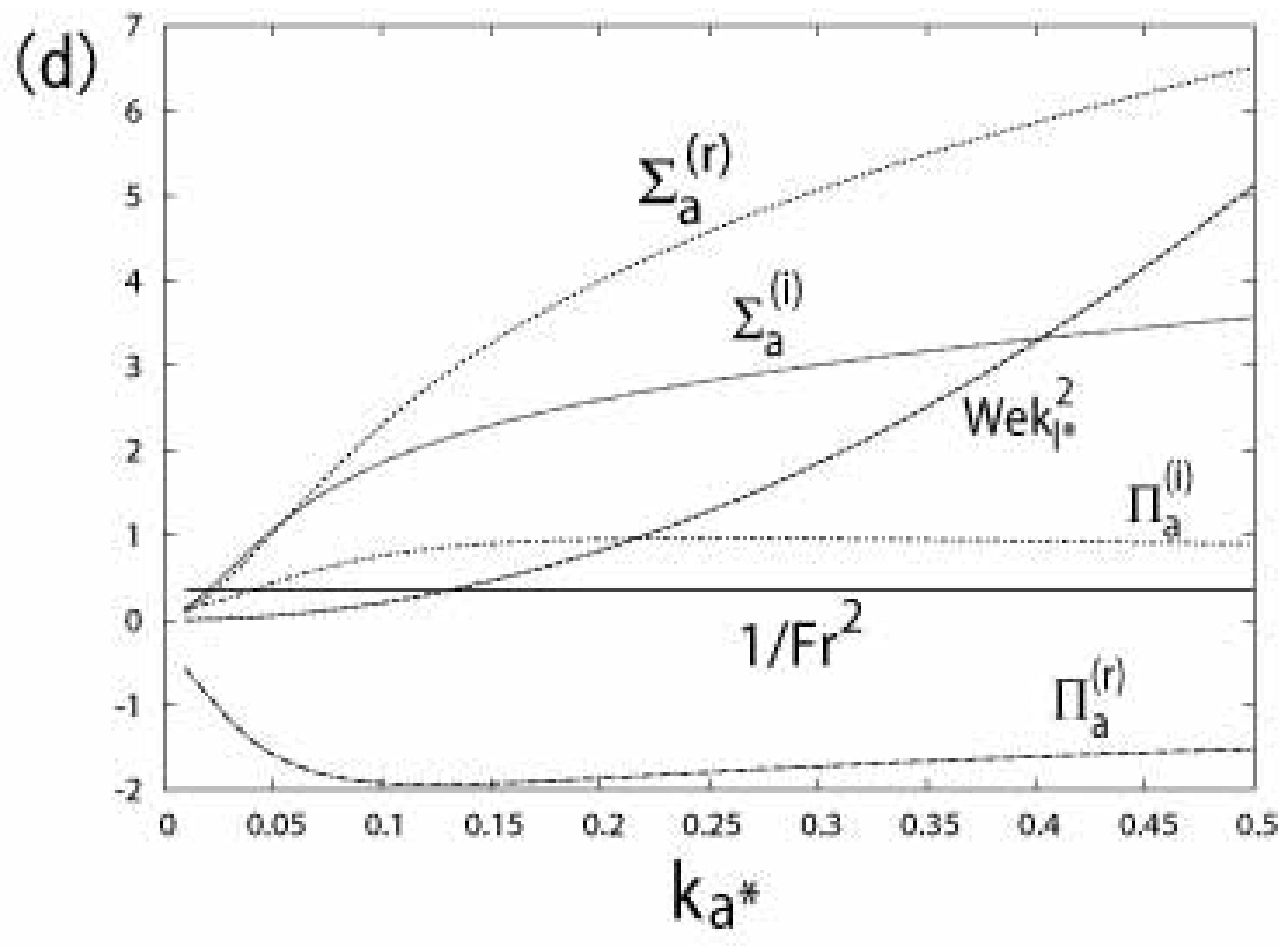}
\end{center}
\caption{Variation of $1/Fr^{2}$, $Wek_{l*}^{2}$, $\Sigma_{a}^{(r)}$, $\Sigma_{a}^{(i)}$, $\Pi_{a}^{(r)}$ and $\Pi_{a}^{(i)}$ against $Q/l_{w}$ for $k_{a*}=0.2$ and $x=0.1$ m, in the case of (a) $u_{\infty}=5$ m/s and (b) $u_{\infty}=20$ m/s. 
The variation of $1/Fr^{2}$, $Wek_{l*}^{2}$, $\Sigma_{a}^{(r)}$, $\Sigma_{a}^{(i)}$, $\Pi_{a}^{(r)}$ and $\Pi_{a}^{(i)}$ against $k_{a*}$ for $Q/l_{w}=1000$ $[{\rm (ml/h)/cm}]$ and $x=0.1$ m, in the case of (c) $u_{\infty}=5$ m/s and (d) $u_{\infty}=20$ m/s. 
Here $k_{a*}=0.2$ corresponds to a wavelength of about 2 cm for $u_{\infty}=5$ m/s and about 1 cm for $u_{\infty}=20$ m/s at $x=0.1$ m.  
}
\label{fig:qoverl-ka-Sigma-Pi}
\end{figure}

Figures \ref{fig:qoverl-ka-Sigma-Pi} (a) and (b) show the variation of the values $1/Fr^{2}$, $Wek_{l*}^{2}$, $\Sigma_{a}^{(r)}$, $\Sigma_{a}^{(i)}$, $\Pi_{a}^{(r)}$ and $\Pi_{a}^{(i)}$ against the water supply rate per width $Q/l_{w}$ in the range of 10 to 1000 [(ml/h)/cm], in the case of $u_{\infty}=5$ m/s and $u_{\infty}=20$ m/s, respectively. Here, $\Sigma_{a}^{(r)}$, $\Pi_{a}^{(r)}$and $\Sigma_{a}^{(i)}$, $\Pi_{a}^{(i)}$ are the real and imaginary parts of Eqs. (\ref{eq:shearstress_a}) and (\ref{eq:normalstress_a}), respectively. 
From Eq. (\ref{eq:h0-ula}), the variation of Eqs. (\ref{eq:shearstress_a}), (\ref{eq:normalstress_a}), and parameters $Fr$ and $We$ with respect to $Q/l_{w}$ and $u_{\infty}$ are as follows: 
$\Sigma_{a} \sim (Q/l_{w})^{1/2}$, 
$\Pi_{a} \sim (Q/l_{w})^{-1/2}$,
$1/Fr^{2} \sim (Q/l_{w})^{-1/2}$,
$We \sim (Q/l_{w})^{-3/2}$ 
for a given $u_{\infty}$, 
and 
$\Sigma_{a} \sim u_{\infty}^{-1/4}$, 
$\Pi_{a} \sim u_{\infty}^{-3/4}$,
$1/Fr^{2} \sim u_{\infty}^{-9/4}$,
$We \sim u_{\infty}^{-3/4}$ 
for a given $Q/l_{w}$. 
It is found that as $Q/l_{w}$ increases, the value of $\Sigma_{a}$ increases, while other parameters decrease. Also, as $u_{\infty}$ increases, the value of $\Sigma_{a}$ decreases much slower than the other parameters.
Hence, Fig. \ref{fig:qoverl-ka-Sigma-Pi} (b) shows that $\Sigma_{a}$ is not negligible compared to the other parameters as $Q/l_{w}$ and $u_{\infty}$ increase.
Therefore, we use the value $Q/l_{w}=1000$ [(ml/h)/cm] throughout this paper, which is also in the same order as that employed in the experiments. \cite{Streitz02} 

On the other hand, Fig. \ref{fig:qoverl-ka-Sigma-Pi} (c) and (d) show the variation of the values $1/Fr^{2}$, $Wek_{l*}^{2}$, $\Sigma_{a}^{(r)}$, $\Sigma_{a}^{(i)}$, $\Pi_{a}^{(r)}$ and $\Pi_{a}^{(i)}$ against the dimensionless wave number $k_{a*}$ for $Q/l_{w}=1000$ [(ml/h)/cm]. When plotting $Wek_{l*}^{2}$ with respect to $k_{a*}$, the relation $k_{l*}=(\bar{h}_{0}/\delta_{0})k_{a*}$ is used. 
In the case of $u_{\infty}=5$ m/s, as shown in Fig. \ref{fig:qoverl-ka-Sigma-Pi} (c), $1/Fr^{2}$ and $Wek_{l*}^{2}$ are dominant terms in Eq. (\ref{eq:bc-normalstress}). As $u_{\infty}$ increases, the values of $1/Fr^{2}$ and $We$ decrease as shown in Table \ref{tab:tableI}. For example, for $u_{\infty}=20$ m/s, Fig. \ref{fig:qoverl-ka-Sigma-Pi} (d) shows that the magnitude of $\Sigma_{a}$ and $\Pi_{a}$ in Eqs. (\ref{eq:bc-shearstress}) and (\ref{eq:bc-normalstress}) are not negligible compared to $1/Fr^{2}$ and $Wek_{l*}^{2}$. 
In Figs. \ref{fig:ka-xi-Thetaxi}, \ref{fig:ka-amp-lambda} and \ref{fig:heat-coefficient} 
we consider two cases: One takes into account the effect of the air stress disturbances $\Sigma_{a}$ and $\Pi_{a}$, and
the other does not. 
We will demonstrate in the following sections that this leads to critically different results for the shape of the water-air interface, the growth conditions of the ice-water interface and the heat transfer coefficient at the water-air interface.  

%%%%%% Fig.2 %%%%%%%%%%%%%%%%%%%%%%%%%%%%%%%%%%%%%%%%%%%%%%%%%%%%%%%%%%%%%%%%%%%%%%%%%%%%%%%%%%%%%%%%%%%%%%%%%%%
%%%%%%%%%%%%%%%%%%%%%%%%%%%%%%%%%%%%%%%%%%%%%%%%%%%%%%%%%%%%%%%%%%%%%%%%%%%%%%%%%%%%%%%%%%%%%%%%%%%%%%%%%%%%%%%%%%

%%%%%%%%%%%%%%%%%%%%%%%%%%%%%%%%%%%%%%%%%%%%%%%%%%%%%%%%%%%%%%%%%%%%%%
\subsection{\label{sec:shape}The shape of the water-air interface}
%%%%%%%%%%%%%%%%%%%%%%%%%%%%%%%%%%%%%%%%%%%%%%%%%%%%%%%%%%%%%%%%%%%%%%

%%%%%%%% Fig.3 %%%%%%%%%%%%%%%%%%%%%%%%%%%%%%%%%%%%%%%%%%%%%%%%%%%%%%%%%%%%%%%%%%%%%%%%%%%%%%%%%%%%%%%%%%%%%%%%%%%%
\begin{figure}[ht]
\begin{center}
\includegraphics[width=8cm,height=8cm,keepaspectratio,clip]{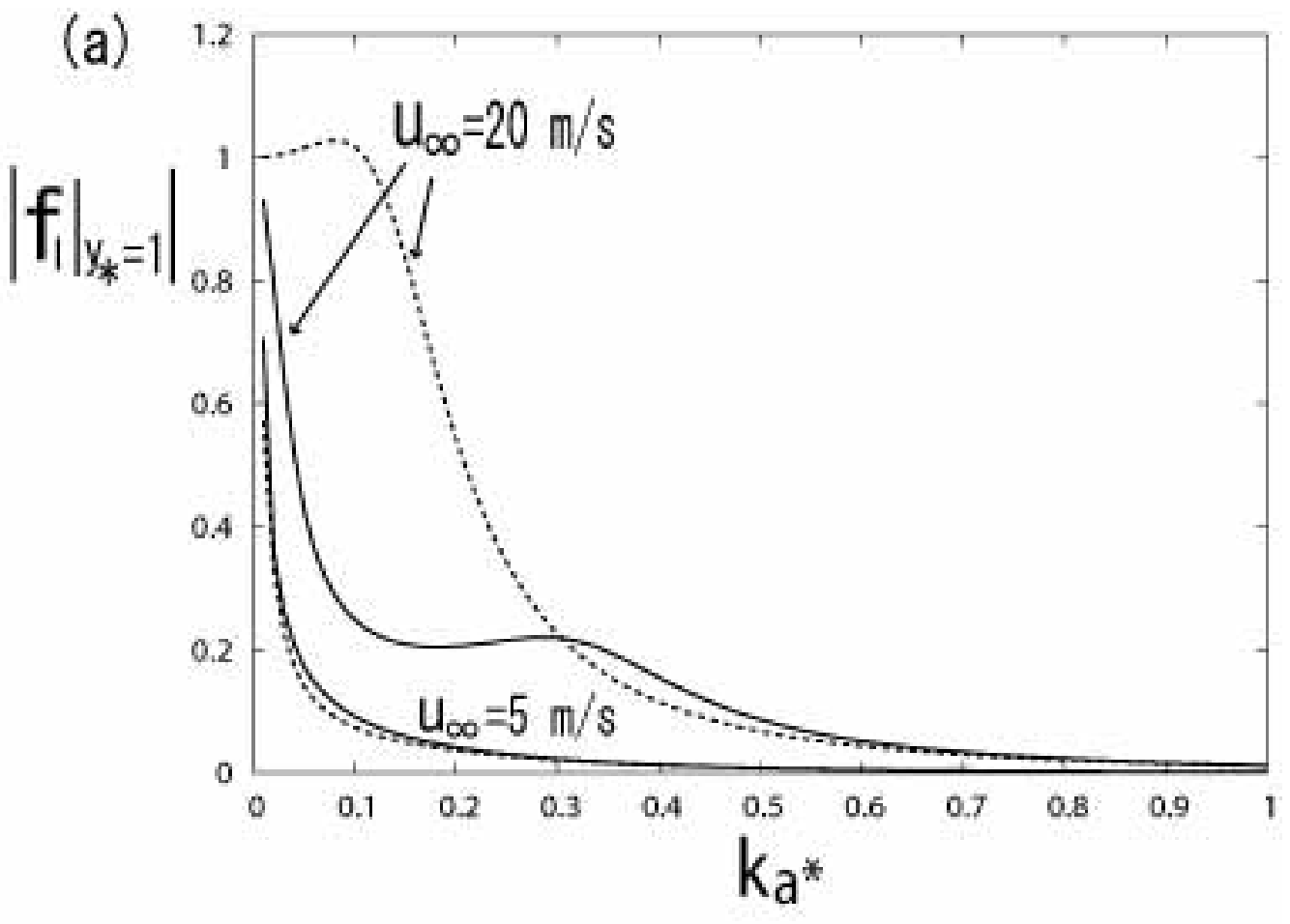}\hspace{5mm}
\includegraphics[width=7cm,height=7cm,keepaspectratio,clip]{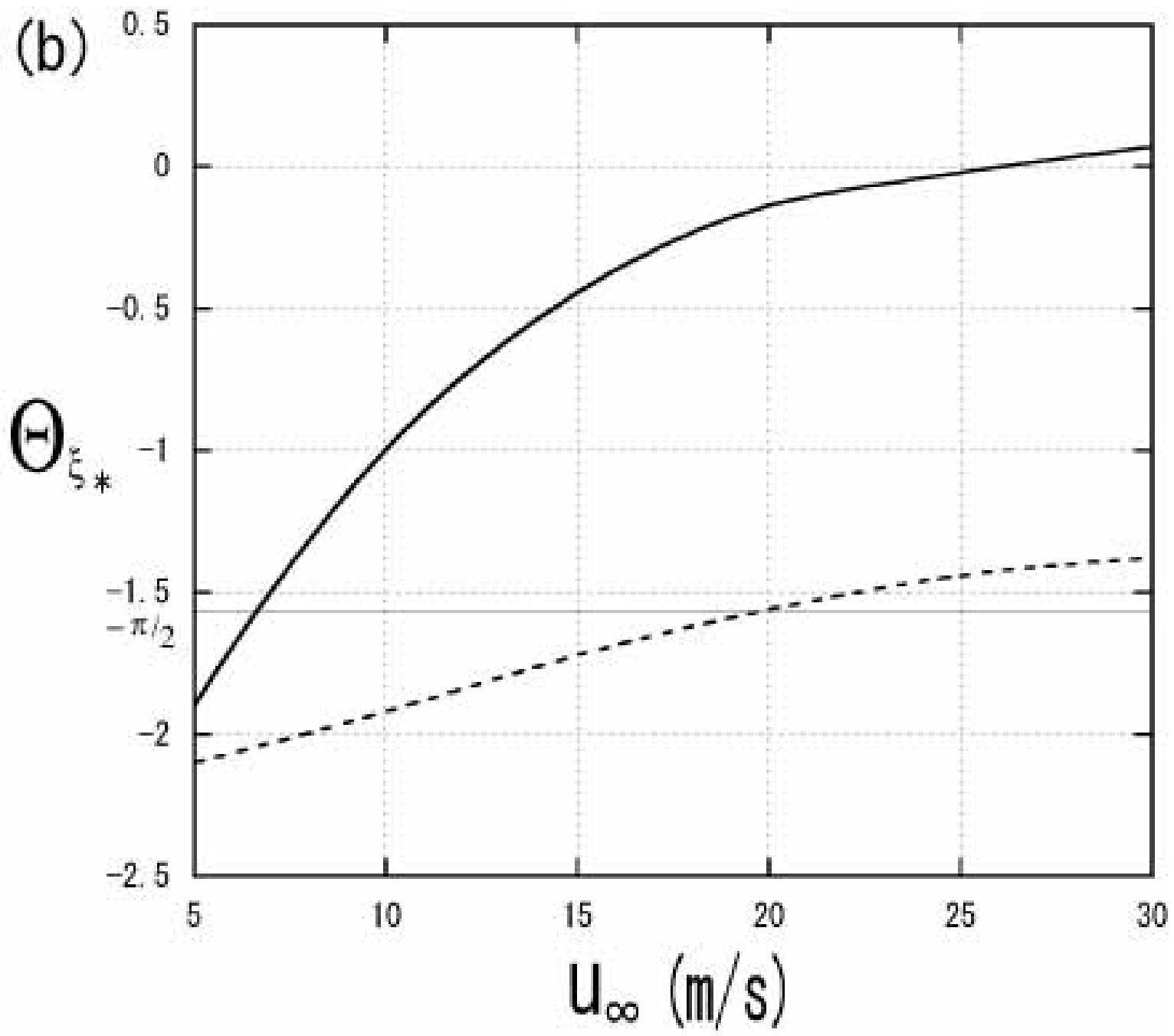}\hspace{5mm}
\end{center}
\caption{
For $Q/l_{w}=1000$ $[{\rm (ml/h)/cm}]$, $u_{\infty}=5$ m/s and $u_{\infty}=20$ m/s, 
(a) represents the variation of amplitude $|f_{l}|_{y_{*}=1}|$ against $k_{a*}$.
Here $k_{a*}=1.0$ corresponds to a wavelength of about 4.5 mm for $u_{\infty}=5$ m/s, and about 2.3 mm for $u_{\infty}=20$ m/s at $x=0.1$ m.
(b) represents the variation of phase difference $\Theta_{\xi_{*}}$ between the ice-water and water-air interfaces against the free stream velocity $u_{\infty}$ for the most unstable mode. 
The solid curves consider the effect of the tangential and normal air shear stress disturbances on the water-air interface. The dashed curves do not consider this effect.
}
\label{fig:ka-xi-Thetaxi}
\end{figure}
%%%%%%%%%%%%%%%%%%%%%%%%%%%%%%%%%%%%%%%%%%%%%%%%%%%%%%%%%%%%%%%%%%%%%%%%%%%%%%%%%%%%%%%%%%%%%%%%%%%%%%%%%%%%%%%%%%%%

It is supposed that a dimensionless small disturbance of the ice-water interface has a sinusoidal form: 
\begin{equation} 
y_{*}=\zeta_{*}=\delta_{b}\Imag[{\rm exp}(\sigma_{*}t_{*}+ik_{l*}x_{*})]
=\delta_{b}(t_{*})\sin[k_{l*}(x_{*}-v_{p*}t_{*})], 
\label{eq:zeta}
\end{equation}
where $\delta_{b}\equiv \zeta_{k}/\bar{h}_{0}$ is an infinitesimal initial amplitude,
$\sigma_{*}=\sigma/(K_{a}\bar{G}_{a}/L\bar{h}_{0})$, 
$t_{*}=(\bar{V}/\bar{h}_{0})t$, $x_{*}=x/\bar{h}_{0}$, 
$\delta_{b}(t_{*})\equiv \delta_{b}{\rm exp}(\sigma^{(r)}_{*}t_{*})$, 
and $\Imag$ denotes the imaginary part of its argument.
Since the water film is very thin, the deformed ice-water interface causes a disturbance of the water-air interface:
\begin{eqnarray} 
y_{*}=\xi_{*}&=&1+\Imag[\delta_{t}{\rm exp}(\sigma_{*}t_{*}+ik_{l*}x_{*})] \nonumber \\
&=&1-\delta_{b}(t_{*})
\left\{
f_{l}^{(r)}|_{y_{*}=1}\sin[k_{l*}(x_{*}-v_{p*}t_{*})]
+f_{l}^{(i)}|_{y_{*}=1}\cos[k_{l*}(x_{*}-v_{p*}t_{*})]
\right\} \nonumber \\
&=&1+\delta_{b}(t_{*})|f_{l}|_{y_{*}=1}|\sin[k_{l*}(x_{*}-v_{p*}t_{*})-\Theta_{\xi_{*}}],
\label{eq:xi}
\end{eqnarray}
where 
$f_{l}^{(r)}$ and $f_{l}^{(i)}$ are the real and imaginary parts of $f_{l}$, respectively, $|f_{l}|_{y_{*}=1}|=[(f_{l}^{(r)}|_{y_{*}=1})^{2}+(f_{l}^{(i)}|_{y_{*}=1})^2]^{1/2}$ is the amplitude and 
$\cos\Theta_{\xi_{*}}=-f_{l}^{(r)}|_{y_{*}=1}/|f_{l}|_{y_{*}=1}|$, $\sin\Theta_{\xi_{*}}=f_{l}^{(i)}|_{y_{*}=1}/|f_{l}|_{y_{*}=1}|$, 
$\Theta_{\xi_{*}}$ represents a phase difference between the water-air and ice-water interfaces. 
When deriving the second equation in Eq. (\ref{eq:xi}), the relation 
$\delta_{t}\equiv \xi_{k}/\bar{h}_{0}=-f_{l}|_{{y_{*}}=1}\delta_{b}$ 
is used. \cite{Ueno03, Ueno04, Ueno07, UFYT10, Ueno10}

Since $f_{l}|_{y_{*}=1}$ depends on the wave number, 
the amplitude and phase change according to the wavelength of the ice-water interface disturbance.  
Figure \ref{fig:ka-xi-Thetaxi} (a) shows that the water-air interface tends to become flat as $k_{a*}$ increases, due to the action of gravity, surface tension and tangential and normal air shear stress disturbances on the water-air interface. 
In the case of $u_{\infty}=5$ m/s, 
since the effect of air shear stress disturbances can be neglected, as shown in Fig. \ref{fig:qoverl-ka-Sigma-Pi} (c), gravity and surface tension are dominant resisting forces for displacement of the water-air interface. 
On the other hand, as $u_{\infty}$ increases, a region of $k_{a*}$ appears, where the action of tangential and normal air shear stress disturbances on the water-air interface is dominant compared to that of gravity and surface tension, as shown in Fig. \ref{fig:qoverl-ka-Sigma-Pi} (d). For example, in the case of $u_{\infty}=20$ m/s in Fig. \ref{fig:ka-xi-Thetaxi} (a), there is a region in the solid curve where the amplitude does not decrease with an increase in $k_{a*}$. 
If we neglect the air shear stress disturbances, the amplitude is overestimated as shown by the dashed curve in Fig. \ref{fig:ka-xi-Thetaxi} (a). As $k_{a*}$ increases, the difference between the solid and dashed curves becomes small because the surface tension $Wek_{l*}^{2}$ is finally most dominant. 

Figure \ref{fig:ka-xi-Thetaxi} (b) shows the variation of the phase difference $\Theta_{\xi_{*}}$ between the ice-water and water-air interfaces against $u_{\infty}$ for the most unstable mode (see {\ref{sec:amplification}), with (solid curve) and without (dashed curve) the effect of air shear stress disturbances. In the case of $u_{\infty}=5$ m/s, an upward phase shift of the water-air interface relative to the ice-water interface is large, as shown in Fig. \ref{fig:heatflux-sla} (a). 
The solid curve shows that the phase difference 
decreases with increasing $u_{\infty}$ and the sign of $\Theta_{\xi_{*}}$ changes from negative to positive at about $u_{\infty}=27$ m/s. 
An example of the configuration of two interfaces at $u_{\infty}=20$ m/s appears in Fig. \ref{fig:heatflux-sla} (b). The decrease of phase difference $\Theta_{\xi_{*}}$ is also due to the effect of air shear stress disturbances on the water-air interface. On the other hand, if we neglect the effect of the air shear stress disturbances, the phase shift is still large even for large $u_{\infty}$, as shown by the dashed curve in Fig. \ref{fig:ka-xi-Thetaxi} (b). 

%%%%%%%%%%%%%%%%%%%%%%%%%%%%%%%%%%%%%%%%%%%%%%%%%%%%%%%%%%%%%%%%%%%%%%%%%%%%%%%%%%%%%%%%%%%%%%%%%
\subsection{\label{sec:amplification}Amplification rate of the ice-water interface disturbance}
%%%%%%%%%%%%%%%%%%%%%%%%%%%%%%%%%%%%%%%%%%%%%%%%%%%%%%%%%%%%%%%%%%%%%%%%%%%%%%%%%%%%%%%%%%%%%%%%%

Figure \ref{fig:ka-amp-lambda} (a) shows the variation of numerically obtained dimensionless amplification rates $\sigma_{*}^{(r)}$ against the dimensionless wave number $k_{a*}$. 
The solid curves represent $\sigma_{*}^{(r)}$, taking into account the effect of tangential and normal air shear stress disturbances on the water-air interface. If we neglect the air shear stress disturbances, the dashed curves are obtained.
In the case of $u_{\infty}=5$ m/s, the difference is negligible. 
However, if the air shear stress disturbances are neglected, the magnitude of $\sigma_{*}^{(r)}$ is overestimated with increasing $u_{\infty}$.
One expects to observe an ice pattern with a wave number at which the amplification rate is the maximum. 
For example, at $u_{\infty}=20$ m/s, $\sigma_{*}^{(r)}$ acquires a maximum value $\sigma^{(r)}_{*\rm max}=17.7$ at $k_{a*}=0.22$. Since the wave number $k$ is normalized by $\delta_{0}$, the corresponding wavelength of the ice pattern is 1.03 cm from $\lambda=2\pi\delta_{0}/k_{a*}$. Here, the value of $\delta_{0}=(2\nu_{a}x/u_{\infty})^{1/2}=361$ $\mu$m estimated from $x=0.1$ m and $u_{\infty}=20$ m/s is used. 
The magnitude of $v_{p*}$ is defined from the wave number at which $\sigma_{*}^{(r)}$ acquires a maximum value. 
At $k_{a*}=0.22$, we obtain $v_{p*}=-96.6$, and hence the displacement of the ice-water interface after the time $t_{*}=1/\sigma_{* \rm max}^{(r)}$ is $\Delta x_{*}=v_{p*}/\sigma_{* \rm max}^{(r)}=-5.4$. The ice pattern will move in the direction opposite to the water flow (see Fig. \ref{fig:heatflux-sla}). The variation of $\sigma_{* \rm max}^{(r)}$, $\lambda$,  $v_{p*}$ and $\Delta x_{*}$ against $u_{\infty}$ is shown in Table \ref{tab:tableII}.

%%%%%%% Fig. 4 %%%%%%%%%%%%%%%%%%%%%%%%%%%%%%%%%%%%%%%%%%%%%%%%%%%%%%%%%%%%%%%%%%%%%%%%%%%%%%%%%%%%%%%%%%%%%%%%%%%%%
\begin{figure}[ht]
\begin{center}
\includegraphics[width=8cm,height=8cm,keepaspectratio,clip]{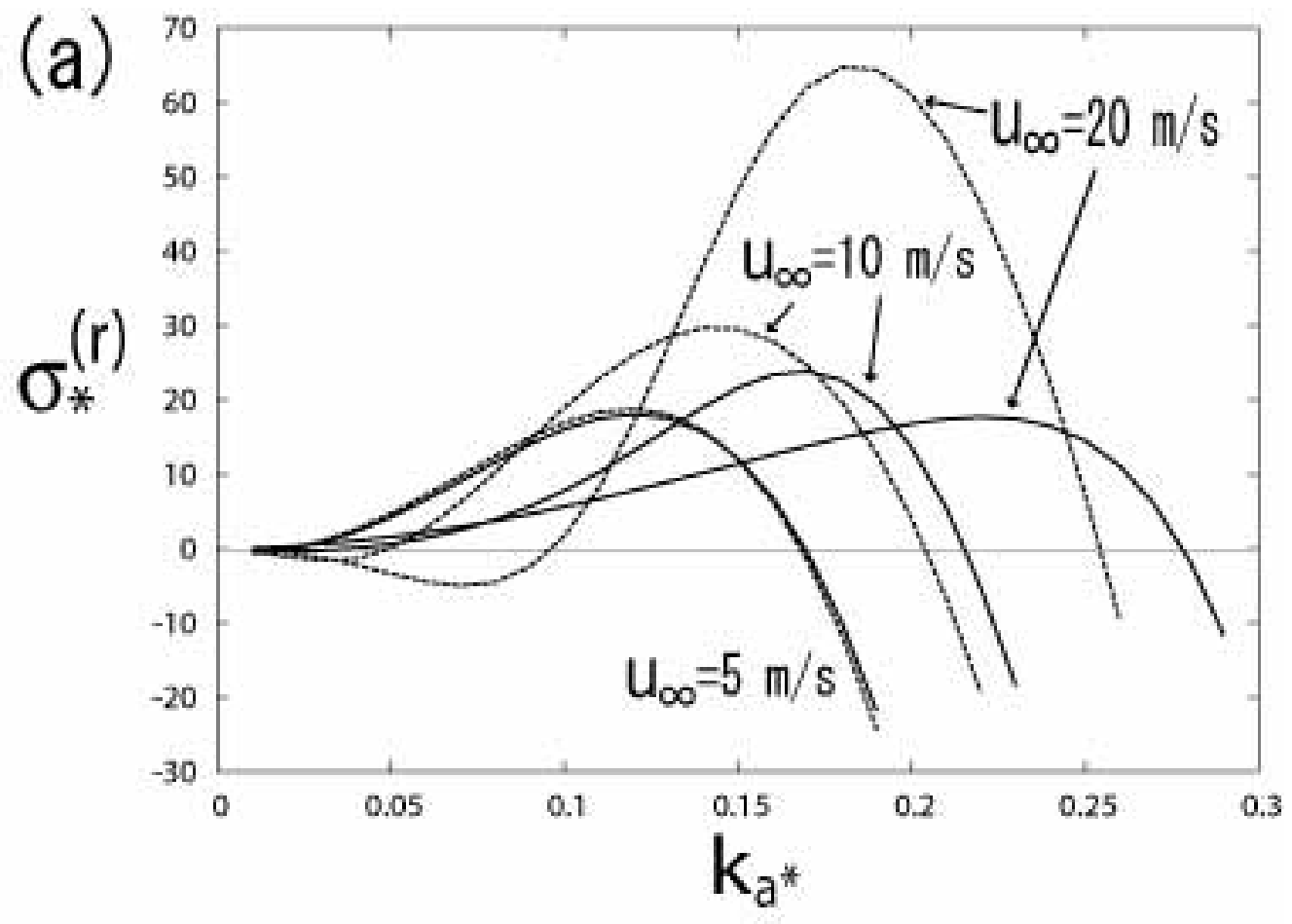}\hspace{5mm}
\includegraphics[width=7cm,height=7cm,keepaspectratio,clip]{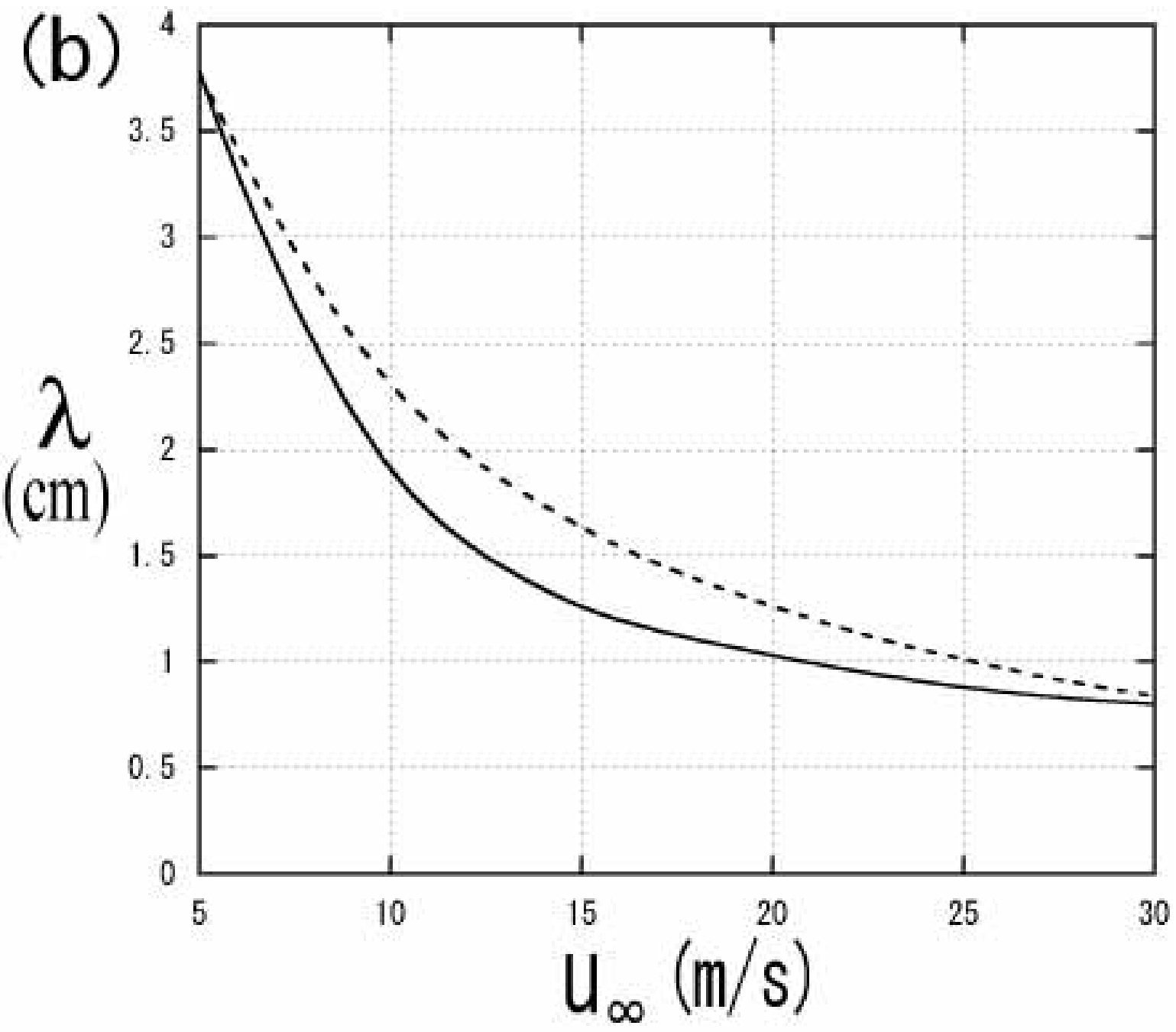}
\end{center}
\caption{For $Q/l_{w}=1000$ $[{\rm (ml/h)/cm}]$  and $x=0.1$ m,
(a) dimensionless amplification rate $\sigma_{*}^{(r)}=\sigma^{(r)}/(K_{a}\bar{G}_{a}/L\bar{h}_{0})$ versus dimensionless wave number $k_{a*}=k\delta_{0}$,
(b) variation of wavelength of ripples against the free stream velocity $u_{\infty}$.
The solid curves consider the effect of the tangential and normal air shear stress disturbances on the water-air interface, and the dashed curves do not consider this effect.}
\label{fig:ka-amp-lambda}
\end{figure}
%%%%%%%%%%%%%%%%%%%%%%%%%%%%%%%%%%%%%%%%%%%%%%%%%%%%%%%%%%%%%%%%%%%%%%%%%%%%%%%%%%%%%%%%%%%%%%%%%%%%%%%%%%%%%%%%%%%%

It is found from Fig. \ref{fig:ka-amp-lambda} (b) and Table \ref{tab:tableII} that the wavelength shortens with increasing $u_{\infty}$. 
Wavelike ice patterns with various roughness spacings and heights were experimentally observed by changing the wind speed and slope of an inclined plane. \cite{Streitz02} For a wind speed of 16 km/h=4.4 m/s, the roughness spacing is increased as the plane slope is decreased (the roughness spacing for smooth-ice base is about 3 cm at about $3^{\circ}$, see Fig. 10 in Ref. \onlinecite{Streitz02}) and for the plane slope of $8^{\circ}$, the roughness spacing is decreased as the wind speed is increased (see Fig. 11 in Ref. \onlinecite{Streitz02}). The latter result is consistent with the theoretical prediction, except that the results herein are only those obtained with a slope of $0^{\circ}$.

Since the values $\sigma_{*}^{(r)}$ in Table \ref{tab:tableII} are much larger than those
predicted by the morphological instability triggered by thermal diffusion at the water-air interface in previous papers, \cite{Ueno03, Ueno04, Ueno07, UFYT10} the ice-water interface instability herein is enhanced by the flow in the water film. 
On the other hand, as shown in Figs. \ref{fig:qoverl-ka-Sigma-Pi} (c) and (d), the surface tension $Wek_{l*}^{2}$ is most dominant to suppress the water-air disturbance with increasing $k_{a*}$ and stabilizes the corresponding ice-water interface disturbance. 
The solid and dashed curves in Fig. \ref{fig:ka-amp-lambda} (a) show that as $u_{\infty}$ increases, the wave number at which $\sigma_{*}^{(r)}$ vanishes, that is, the neutral stability point is shifted to higher wave number.
This is because the value of the Weber number $We$ decreases with an increase in $u_{\infty}$, as indicated in Table \ref{tab:tableI}, hence the stabilization due to the surface tension $Wek_{l*}^{2}$ becomes more effective for higher wave numbers. 
As shown in Figs. \ref{fig:qoverl-ka-Sigma-Pi} (c) and (d), since $\Pi_{a}^{(r)}$ has negative values with respect to $k_{a*}$, 
the value of $Wek_{l*}^{2}+\Pi_{a}^{(r)}$ in Eq. (\ref{eq:bc-normalstress}) decreases as $u_{\infty}$ increases.
This means that the stabilization due to the surface tension $Wek_{l*}^{2}$ is weakened by normal air shear stress disturbance $\Pi_{a}^{(r)}$. Hence, the wave number of the neutral stability point in the solid curves is shifted to the higher wave number compared to that in the dashed curves with an increase in $u_{\infty}$. 
That is why the wavelength evaluated from the most unstable mode becomes shorter as $u_{\infty}$ increases. 
It should be noted that the magnitude of $\sigma_{*}^{(r)}$ is decreased by the effect of the tangential and normal air shear stress disturbances. However, the effect of air shear stress disturbances on the wavelength does not make a significant difference, as shown in Fig. \ref{fig:ka-amp-lambda} (b). 

On a static structure, typical wind speed is in the order of 20 m/s. \cite{Myers02_1} The wind speed in the aufeis formation experiments by Streitz and Ettema \cite{Streitz02} was varied up to 48 km/h=13.3 m/s. On the other hand, in aircraft icing a typical value is around 100 m/s. \cite{Myers02_1}
By applying our analysis to the case $u_{\infty}=100$ m/s, assuming that the air stream flow remains laminar as shown in Fig. \ref{fig:ice-water-air}, the results shown in 
Tables \ref{tab:tableI} and \ref{tab:tableII} are obtained.
On the other hand, in the limit $u_{\infty}\rightarrow 0$, there is no driving force to move the water film. In this case, $\delta_{0}=(2\nu_{a}x/u_{\infty})^{1/2}$ and $\bar{h}_{0}$ in Eq. (\ref{eq:h0-ula}) have an infinite value and hence the corresponding wavelength is not defined. The same issue arises for gravity driven water flows found in previous papers, \cite{Ogawa02, Ueno03, Ueno04, Ueno07, UFYT10, Ueno10} in which the thickness of water film is determined from $\bar{h}_{0}=[3\nu_{l}/(g\sin\theta)Q/l_{w}]^{1/3}$, where $\theta$ is the inclination angle with respect to the horizontal. In the limit $\theta \rightarrow 0$, there is no driving force to move the water film. Then the wavelength at $\theta=0$ is not defined (see Fig. 8 (a) in Ref. \onlinecite{UFYT10}).  

%%%%%% Table.2 %%%%%%%%%%%%%%%%%%%%%%%%%%%%%%%%%%%%%%%%%%%%%%%%%%%%%%%%%%%%%%%%%%%%%%%%%%%%%%%%%%%%%%%%%%%
\begin{table}[ht]
\caption{\label{tab:tableII}  
Variation of 
temperature at the water-air interface, $T_{la}$, 
undisturbed ice growth rate, $\bar{V}$,
maximum value of dimensionless amplification rate $\sigma_{*\rm max}^{(r)}$ at a dimensionless wave number $k_{a*}$, 
the corresponding wavelength, $\lambda$,
dimensionless phase velocity, $v_{p*}$,
dimensionless displacement of the ice-water interface, $\Delta x_{*}=v_{p*}/\sigma_{*\rm max}^{(r)}$ 
after the dimensionless time $t_{*}=1/\sigma_{*\rm max}^{(r)}$, 
against the free stream velocity, $u_{\infty}$,
for $x=0.1$ m, $Q/l_{w}=1000$ $[{\rm (ml/h)/cm}]$ and $T_{\infty}=-10$ $^{\circ}$C.}
\begin{ruledtabular}
\begin{tabular}{cccccccc}
$u_{\infty}$ (m/s) & $T_{la}$ ($^{\circ}$C) & $\bar{V}$ ($\times 10^{-6}$ m/s) 
& $k_{a*}$ &$\sigma_{*\rm max}^{(r)}$ & $\lambda$ (cm) & $v_{p*}$ & $\Delta x_{*}$\\ 
  5 & -0.32 & 0.40 & 0.12 & 18.1 & 3.78 &  -70.3 & -3.9 \\ 
 10 & -0.27 & 0.57 & 0.17 & 23.9 & 1.88 &  -97.9 & -4.1 \\
 15 & -0.25 & 0.70 & 0.21 & 21.0 & 1.25 & -105.1 & -5.0 \\
 20 & -0.23 & 0.81 & 0.22 & 17.7 & 1.03 &  -96.6 & -5.4 \\ 
 25 & -0.22 & 0.91 & 0.23 & 16.0 & 0.88 &  -91.8 & -5.7 \\
 30 & -0.21 & 1.00 & 0.23 & 15.2 & 0.80 &  -80.1 & -5.3 \\ 
100 & -0.15 & 1.83 & 0.29 & 13.8 & 0.35 &  -68.7 & -5.0 \\  
\end{tabular}
\end{ruledtabular}
\end{table}
%%%%%%%%%%%%%%%%%%%%%%%%%%%%%%%%%%%%%%%%%%%%%%%%%%%%%%%%%%%%%%%%%%%%%%%%%%%%%%%%%%%%%%%%%%%%%%%%%%%%%%%%%

%%%%%%%%%%%%%%%%%%%%%%%%%%%%%%%%%%%%%%%%%%%%%%%%%%%%%%%%%%%%%%%%%%%%%%%%%%%%%%%%%%%%%%%%%%
\subsection{\label{sec:Heat}Heat transfer at disturbed ice-water and water-air interfaces}
%%%%%%%%%%%%%%%%%%%%%%%%%%%%%%%%%%%%%%%%%%%%%%%%%%%%%%%%%%%%%%%%%%%%%%%%%%%%%%%%%%%%%%%%%%

Using the assumed forms of $T'_{l}$ and $T'_{s}$ in Eq. (\ref{eq:pertset}) and considering their imaginary parts, the temperature in the water layer and ice can be expressed in dimensionless form as follows:
\begin{eqnarray}
T_{l*}(y_{*})\equiv \frac{T_{l}(y_{*})-T_{sl}}{T_{sl}-T_{la}}
&=& -y_{*}
+\delta_{b}(t_{*})\left\{H_{l}^{(r)}(y_{*})\sin[k_{l*}(x_{*}-v_{p*}t_{*})] \right.
\nonumber \\
&& \left.               +H_{l}^{(i)}(y_{*})\cos[k_{l*}(x_{*}-v_{p*}t_{*})]\right\},
\label{eq:Tl}
\end{eqnarray}
\begin{eqnarray}
T_{s*}(y_{*}) \equiv \frac{T_{s}(y_{*})-T_{sl}}{T_{sl}-T_{la}}
&=& \delta_{b}(t_{*}){\rm exp}(k_{l*}y_{*})
\left\{(H_{l}^{(r)}|_{y_{*}=0}-1)\sin[k_{l*}(x_{*}-v_{p*}t_{*})] \right.
\nonumber \\
&& \left. +H_{l}^{(i)}|_{y_{*}=0}\cos[k_{l*}(x_{*}-v_{p*}t_{*})]\right\},
\label{eq:Ts}
\end{eqnarray}
where we used $\bar{T}_{s}=T_{sl}$ and the solution $H_{s}(y_{*})=(H_{l}|_{y_{*}=0}-1){\rm exp}(k_{l*}y_{*})$. \cite{Ueno03, Ueno04, Ueno07, UFYT10, Ueno10}

A microscopic energy balance have to be considered to explain fine details of ice morphology.
We define the disturbed parts of heat flux from the ice-water interface to the water, from the ice to the ice-water interface, and from the water-air interface to the air, as  
$q'_{l}\equiv \Imag[-K_{l}\partial T'_{l}/\partial y|_{y=\zeta}]$, 
$q'_{s}\equiv \Imag[-K_{s}\partial T'_{s}/\partial y|_{y=\zeta}]$ and 
$q'_{a}\equiv \Imag[-K_{a}\partial T'_{a}/\partial y|_{y=\xi}]$, respectively.
These can be expressed in dimensionless form as follows: 
\begin{eqnarray}
q'_{l*}\equiv
\frac{q'_{l}}{K_{l}\bar{G}_{l}}
&=&-\delta_{b}(t_{*})\left\{\frac{dH_{l}^{(r)}}{dy_{*}}\Big|_{y_{*}=0}\sin[k_{l*}(x_{*}-v_{p*}t_{*})]  \right. \nonumber \\
&& \left.                  +\frac{dH_{l}^{(i)}}{dy_{*}}\Big|_{y_{*}=0}\cos[k_{l*}(x_{*}-v_{p*}t_{*})]\right\},
\label{eq:ql}
\end{eqnarray}
\begin{eqnarray}
q'_{s*}\equiv
\frac{q'_{s}}{K_{l}\bar{G}_{l}}
&=&-\delta_{b}(t_{*})K^{s}_{l}k_{l*}\left\{(H_{l}^{(r)}|_{y_{*}=0}-1)\sin[k_{l*}(x_{*}-v_{p*}t_{*})] \right. \nonumber \\
&& \left.                                  +H_{l}^{(i)}|_{y_{*}=0}\cos[k_{l*}(x_{*}-v_{p*}t_{*})]\right\},                                          
\label{eq:qs}
\end{eqnarray}
\begin{eqnarray}
q'_{a*}\equiv 
\frac{q'_{a}}{K_{l}\bar{G}_{l}} 
&=&-\delta_{b}(t_{*})\left\{\left(G'^{(r)}_{a}f_{l}^{(r)}|_{y_{*}=1}-G'^{(i)}_{a}f_{l}^{(i)}|_{y_{*}=1}\right)
                           \sin[k_{l*}(x_{*}-v_{p*}t_{*})] \right. \nonumber \\
&& \left.    +\left(G'^{(r)}_{a}f_{l}^{(i)}|_{y_{*}=1}+G'^{(i)}_{a}f_{l}^{(r)}|_{y_{*}=1}\right)
                           \cos[k_{l*}(x_{*}-v_{p*}t_{*})]\right\},
\label{eq:qa}
\end{eqnarray}
where
$G'^{(r)}_{a}\equiv (\bar{h}_{0}/\delta_{0})(-dH_{a}^{(r)}/d\eta)|_{\eta=0}$ and
$G'^{(i)}_{a}\equiv (\bar{h}_{0}/\delta_{0})(-dH_{a}^{(i)}/d\eta)|_{\eta=0}$
represent the real and imaginary parts of the disturbed part of the air temperature gradient 
$G'_{a}\equiv (\bar{h}_{0}/\delta_{0})(-dH_{a}/d\eta)|_{\eta=0}$ at the water-air interface.
From Eq. (\ref{eq:heatflux-zeta}), the disturbed part of the Stephan condition in dimensionless form
can be written as $\partial \zeta_{*}/\partial t_{*}=q'_{l*}-q'_{s*}$. Substituting Eqs. (\ref{eq:zeta}), (\ref{eq:ql}) and (\ref{eq:qs}) into this condition, Eqs. (\ref{eq:amp}) and (\ref{eq:phasevel}) are obtained.

%%%%%%% Fig. 5 %%%%%%%%%%%%%%%%%%%%%%%%%%%%%%%%%%%%%%%%%%%%%%%%%%%%%%%%%%%%%%%%%%%%%%%%%%%%%%%%%%%%%%%%%%%%%%%%%%%%%%
\begin{figure}[ht]
\begin{center} 
\includegraphics[width=8cm,height=8cm,keepaspectratio,clip]{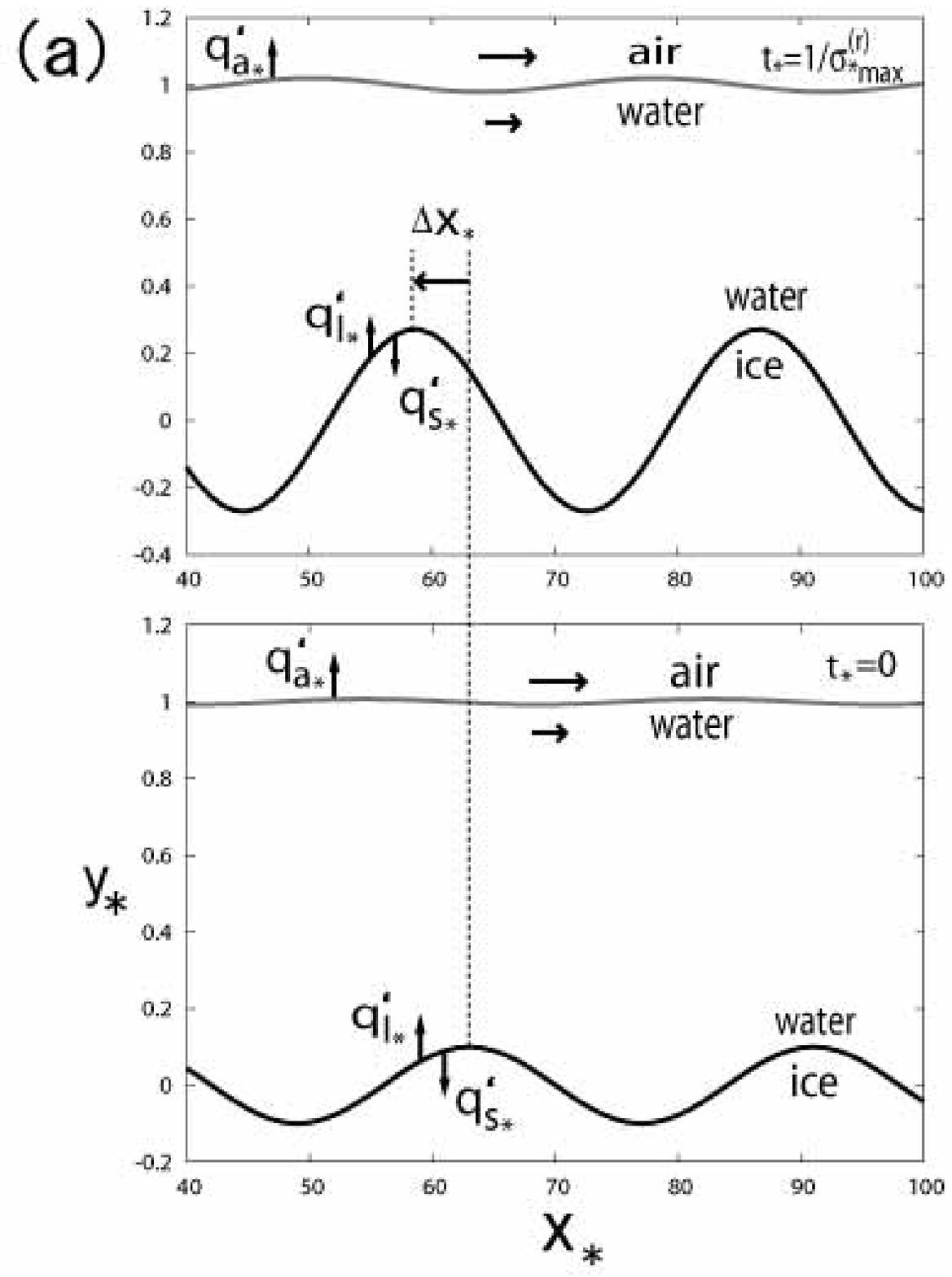}\hspace{5mm} 
\includegraphics[width=8cm,height=8cm,keepaspectratio,clip]{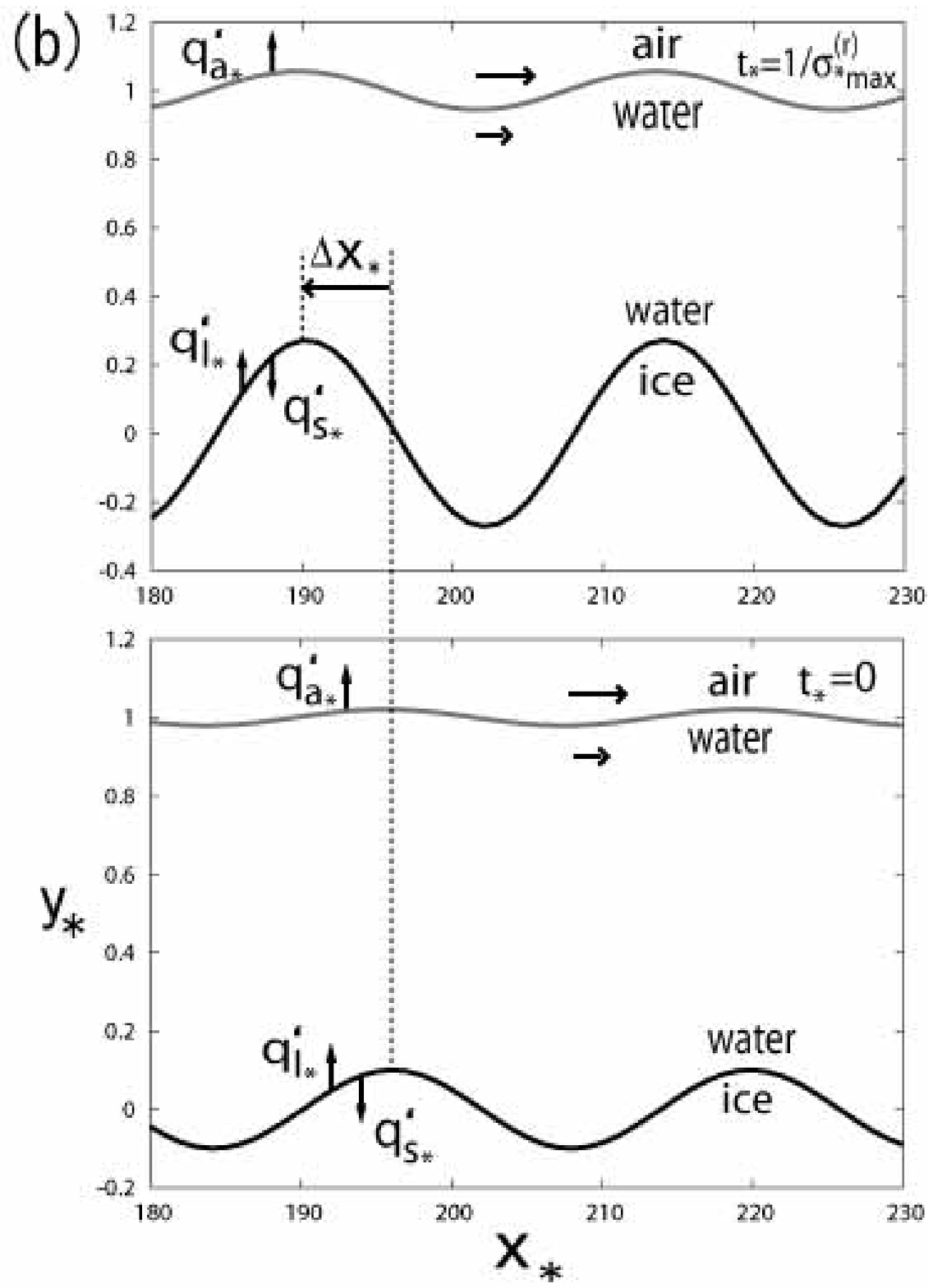} 
\end{center}
\caption{For $Q/l_{w}=1000$ [(ml/h)/cm], 
(a) and (b) are illustrations of the time evolution of an initial disturbance of the ice-water interface from $t_{*}=0$ to $t_{*}=1/\sigma^{(r)}_{*\rm max}$. The arrows indicate the position of maximum point of disturbed heat flux $q'_{l*}$, $q'_{s*}$ at the ice-water interface and that of $q'_{a*}$ at the water-air interface.
(a) represents the disturbance of $k_{a*}=0.12$ in the case of $u_{\infty}=5$ m/s. 
(b) represents the disturbance of $k_{a*}=0.22$ in the case of $u_{\infty}=20$ m/s. 
$\Delta x_{*}$ is the displacement of the ice-water interface after the time $t_{*}=1/\sigma^{(r)}_{*\rm max}$.
Vertical height is not to scale.}
\label{fig:heatflux-sla}
\end{figure}
%%%%%%%%%%%%%%%%%%%%%%%%%%%%%%%%%%%%%%%%%%%%%%%%%%%%%%%%%%%%%%%%%%%%%%%%%%%%%%%%%%%%%%%%%%%%%%%%%%%%%%%%%%%%%%%%%%%%%%

Figures \ref{fig:heatflux-sla} (a) and (b) illustrate the time evolution of the ice-water interface disturbance with an initial amplitude of $\delta_{b}=0.1$ in the case of $u_{\infty}=5$ m/s and $u_{\infty}=20$ m/s, respectively. 
The respective wave numbers of disturbance are $k_{a*}=0.12$ in Fig. \ref{fig:heatflux-sla} (a) and $k_{a*}=0.22$ in Fig. \ref{fig:heatflux-sla} (b). These are the fastest growing modes, at which $\sigma^{(r)}_{*}$ acquires a maximum value, as shown by the solid curves in Fig. \ref{fig:ka-amp-lambda} (a). 
The phase shift of the water-air interface relative to the ice-water interface in Fig. \ref{fig:heatflux-sla} (b) is negligibly small compared to that in Fig.  \ref{fig:heatflux-sla} (a), as shown by the solid curve in Fig. \ref{fig:ka-xi-Thetaxi} (b).
Due to the left-to right air and water flows indicated by arrows in Fig. \ref{fig:heatflux-sla}, the isotherms in the air and water boundary layers are no longer symmetrical around each protruded part of the water-air and ice-water interfaces. Since the isotherms become closer on the upstream side of each protruded part of the interfaces, $q'_{a*}$, $q'_{l*}$ and $q'_{s*}$ are largest on the upstream side of each protruded part, as indicated by the vertical arrows in Fig. \ref{fig:heatflux-sla}. Hence, the ice growth rate on the upstream side of each protruded part is faster than that on the downstream side, and this results in the translation of the ice-water interface in the direction opposite to the water flow. As mentioned in \ref{sec:amplification}, the displacements after the time $t_{*}=1/\sigma^{(r)}_{*\rm max}$ are $\Delta x_{*}=-3.9$ and $\Delta x_{*}=-5.4$ in Figs. \ref{fig:heatflux-sla} (a) and (b), respectively. 

We separate the local heat transfer coefficient at the water-air interface, $h_{x}$, 
into the undisturbed part 
$\bar{h}_{x}=-K_{a}\partial\bar{T}_{a}/\partial y|_{y=\bar{h}_{0}}/(T_{la}-T_{\infty})$ and 
the disturbed part 
$h'_{x}=\Imag[-K_{a}\partial T'_{a}/\partial y|_{y=\bar{h}_{0}}/(T_{la}-T_{\infty})]$.
The former can be written as 
$\bar{h}_{x}=K_{a}/\delta_{0}(-d\bar{T}_{a*}/d\eta|_{\eta=0})$.
Using the value of $\bar{G}_{a*}=-d\bar{T}_{a*}/d\eta|_{\eta=0}=0.413$ obtained in \ref{sec:LSA}, 
the value of local Nusselt number scaled by $\sqrt{Re_{ax}}$ is 
$\bar{Nu}_{x}/\sqrt{Re_{ax}}
=(\bar{h}_{x}x/K_{a})/\sqrt{u_{\infty}x/\nu_{a}}
=-(1/\sqrt{2})d\bar{T}_{a*}/d\eta|_{\eta=0}=0.292$,
from which we obtain $\bar{h}_{x}=0.292K_{a}\sqrt{u_{\infty}/(\nu_{a}x})$. 
A similar expression for the laminar convective heat transfer coefficient,
$\bar{h}_{s}=0.296(K_{a}/\sqrt{\nu_{a}})\sqrt{V_{e}^{2.87}/(\int_{0}^{s}V_{e}^{1.87}ds)}$,
is used in aircraft icing models, \cite{Gent00, Myers04} 
where $s$ is the surface distance from the stagnation point 
and $V_{e}$ is the velocity at edge of air boundary layer.
When $V_{e}=u_{\infty}$ (constant) and replacing $s$ with $x$, $\bar{h}_{s}$ yields the same expression as $\bar{h}_{x}$ except for the very slight difference of numerical factor.
Using $\bar{h}_{x}$, the undisturbed ice growth rate $\bar{V}$ in Eq. (\ref{eq:V}) can be expressed as $\bar{V}=-\bar{h}_{x}T_{\infty}/L$. 
On the other hand, the disturbed part of the heat transfer coefficient normalized by the undisturbed one can be written as  
\begin{eqnarray} 
h'_{x}/\bar{h}_{x}
&=&-\Imag\left[G'_{a}f_{l}|_{y_{*}=1}\delta_{b}{\rm exp}(\sigma_{*}t_{*}+ik_{l*}x_{*})\right] \nonumber \\
&=&\delta_{b}(t_{*})\left[(h'_{x}/\bar{h}_{x})^{(r)}\sin[k_{l*}(x_{*}-v_{p*}t_{*})\right]
+(h'_{x}/\bar{h}_{x})^{(i)}\cos[k_{l*}(x_{*}-v_{p*}t_{*})]]\nonumber \\
&=&\delta_{b}(t_{*})|h'_{x}/\bar{h}_{x}|\sin[k_{l*}(x_{*}-v_{p*}t_{*})-\Theta_{q'_{a*}}].
\label{eq:h'x}
\end{eqnarray}
Equation (\ref{eq:h'x}) becomes the same form as Eq. (\ref{eq:qa}) by putting
$(h'_{x}/\bar{h}_{x})^{(r)}=-(G'^{(r)}_{a}f_{l}^{(r)}|_{y_{*}=1}-G'^{(i)}_{a}f_{l}^{(i)}|_{y_{*}=1})$ and
$(h'_{x}/\bar{h}_{x})^{(i)}=-(G'^{(r)}_{a}f_{l}^{(i)}|_{y_{*}=1}+G'^{(i)}_{a}f_{l}^{(r)}|_{y_{*}=1})$. 
Here
\begin{equation} 
|h'_{x}/\bar{h}_{x}|=[\{(h'_{x}/\bar{h}_{x})^{(r)}\}^{2}+\{(h'_{x}/\bar{h}_{x})^{(i)}\}^{2}]^{1/2}
\label{eq:amplitude-h'x}
\end{equation}  
is the amplitude, and 
$\cos\Theta_{q'_{a*}}=(h'_{x}/\bar{h}_{x})^{(r)}/|h'_{x}/\bar{h}_{x}|$ and 
$\sin\Theta_{q'_{a*}}=-(h'_{x}/\bar{h}_{x})^{(i)}/|h'_{x}/\bar{h}_{x}|$, 
$\Theta_{q'_{a*}}$ is a phase difference between $q'_{a*}=h'_{x}/\bar{h}_{x}$ and the ice-water interface. 

%%%%%%% Fig. 6 %%%%%%%%%%%%%%%%%%%%%%%%%%%%%%%%%%%%%%%%%%%%%%%%%%%%%%%%%%%%%%%%%%%%%%%%%%%%%%%%%%%%%%%%%%%%%%%%%%%%
\begin{figure}[ht]
\begin{center}
\includegraphics[width=7cm,height=7cm,keepaspectratio,clip]{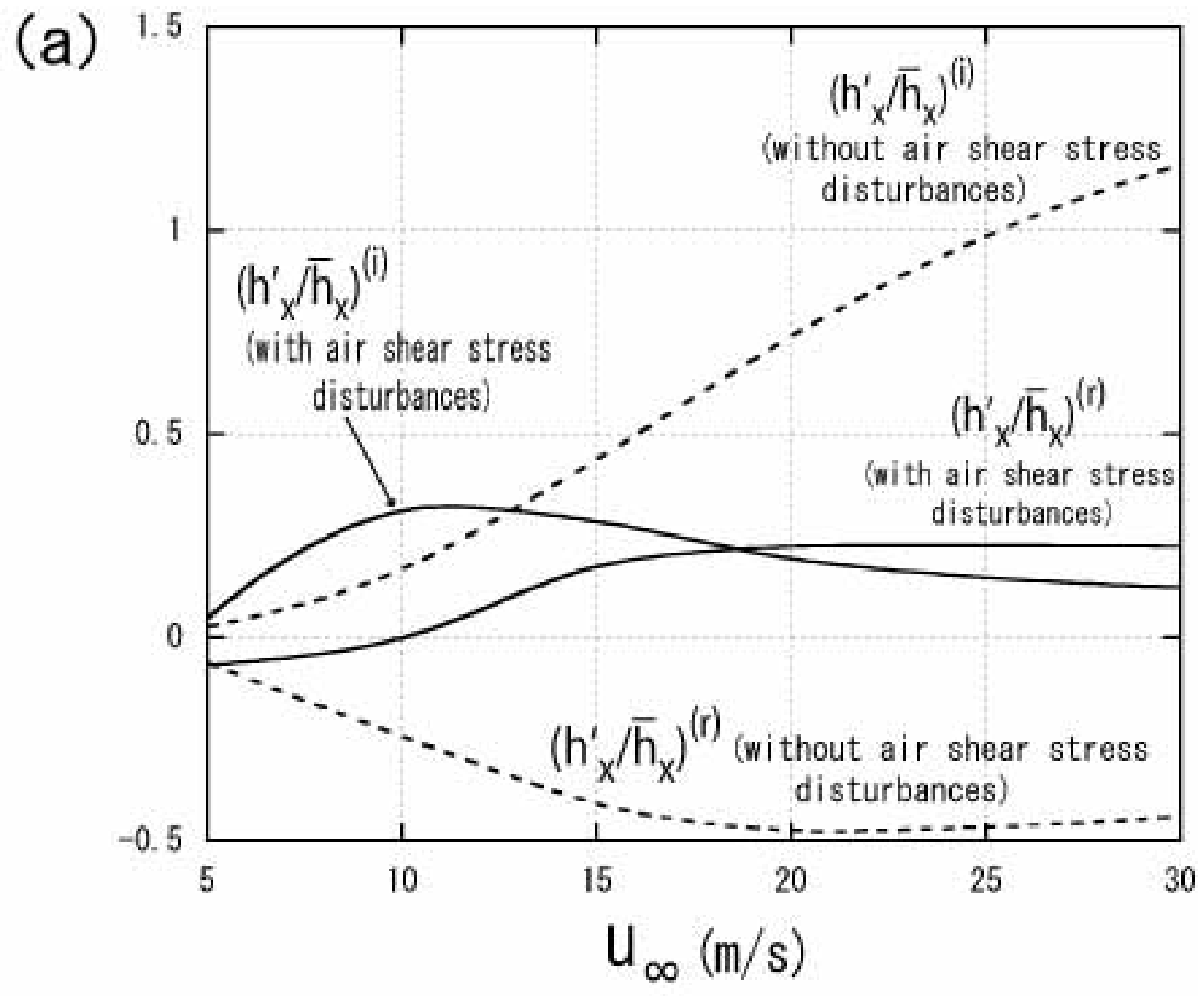}\hspace{5mm}
\includegraphics[width=7cm,height=7cm,keepaspectratio,clip]{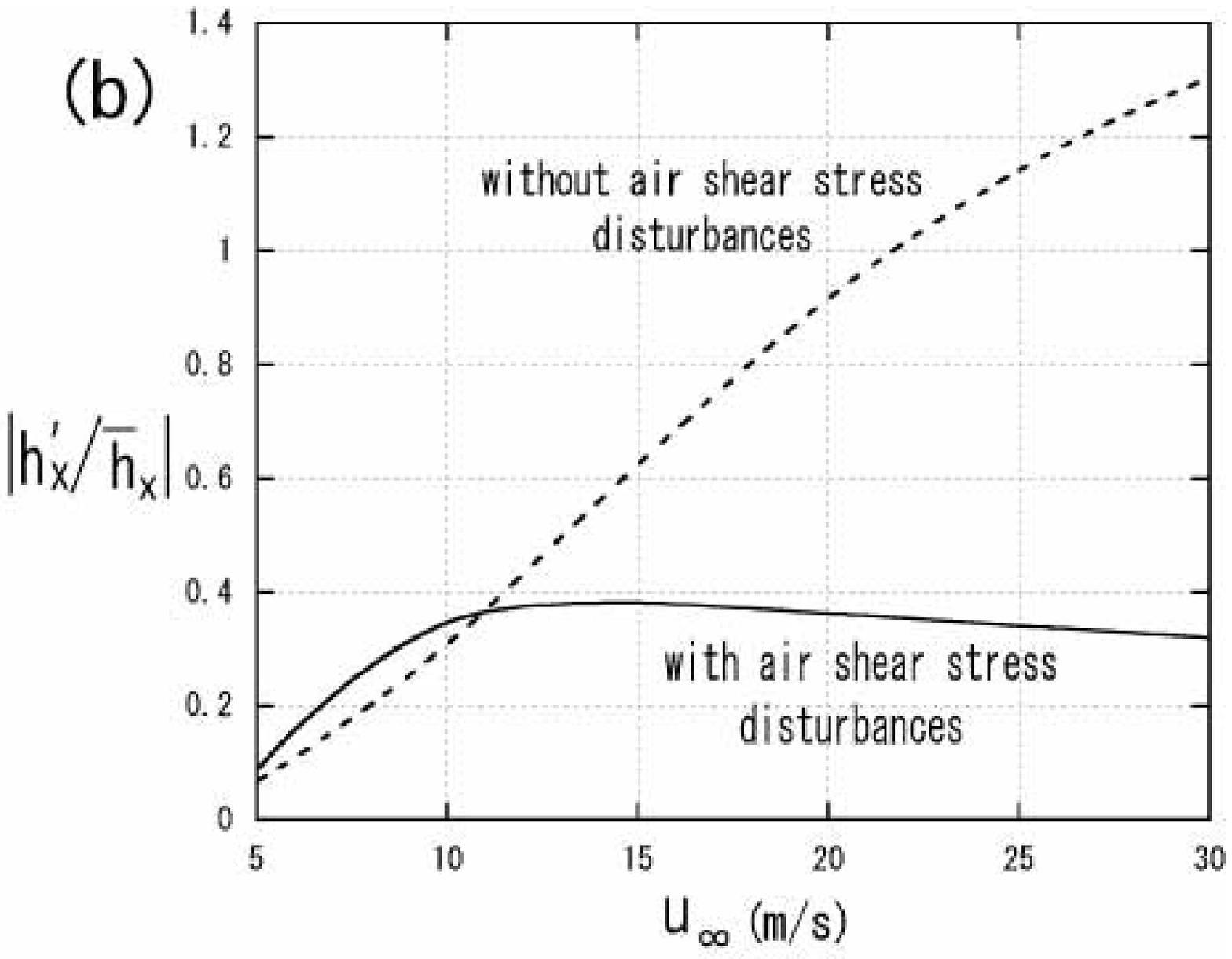}\\[5mm] 
\includegraphics[width=8cm,height=8cm,keepaspectratio,clip]{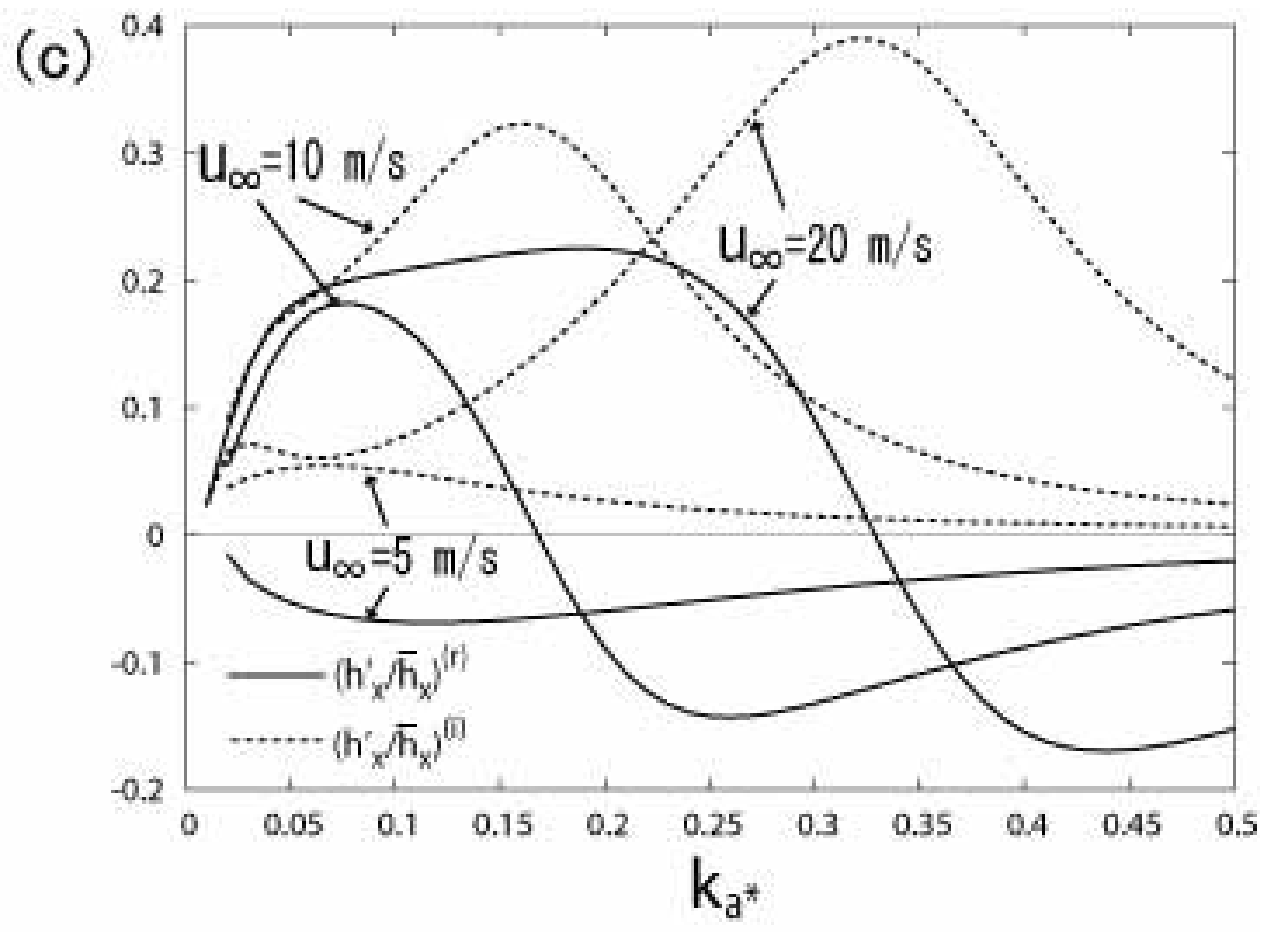} 
\end{center}
\caption{For $Q/l_{w}=1000$ [(ml/h)/cm] and $x=0.1$ m, 
(a) represents the variation of the disturbed part of the heat transfer coefficient normalized by the undisturbed heat transfer coefficient against the free stream velocity $u_{\infty}$: $(h'_{x}/\bar{h}_{x})^{(r)}$ is real part and $(h'_{x}/\bar{h}_{x})^{(i)}$ is imaginary part.
(b) represents the variation of $|h'_{x}/\bar{h}_{x}|$ against $u_{\infty}$. The solid curves consider the effect of the tangential and normal air shear stress disturbances on the water-air interface, and the dashed curves do not consider this effect.
(c) represents the variation of $(h'_{x}/\bar{h}_{x})^{(r)}$ (solid curves) and $(h'_{x}/\bar{h}_{x})^{(i)}$ (dashed curves) against dimensionless wave number $k_{a*}$ for free stream velocities $u_{\infty}$=5, 10, 20 m/s.
}
\label{fig:heat-coefficient}
\end{figure}
%%%%%%%%%%%%%%%%%%%%%%%%%%%%%%%%%%%%%%%%%%%%%%%%%%%%%%%%%%%%%%%%%%%%%%%%%%%%%%%%%%%%%%%%%%%%%%%%%%%%%%%%%%%%%%%%%%%%

The solid curves in Fig. \ref{fig:heat-coefficient} (a) show the variations of $(h'_{x}/\bar{h}_{x})^{(r)}$ and $(h'_{x}/\bar{h}_{x})^{(i)}$ against $u_{\infty}$, taking into account the air shear stress disturbances. These values are estimated for $Q/l_{w}=1000$ [(ml/h)/cm] and $x=0.1$ m, and for $k_{a*}$ at which $\sigma_{*}^{(r)}$ acquires a maximum value. It should be noted that $q'_{a*}=h'_{x}/\bar{h}_{x}$ includes $G'_{a}$ and $f_{l}|_{y_{*}=1}$. $\bar{h}_{x}$ depends on only two parameters, free stream velocity $u_{\infty}$ and position $x$. On the other hand,  $h'_{x}$ depends on many parameters. 
$G'_{a}=(\bar{h}_{0}/\delta_{0})(-dH_{a}/d\eta)|_{\eta=0}$ is determined from the disturbed airflow and temperature fields, and $f_{l}|_{y_{*}=1}$ determines the magnitude of amplitude and phase of the water-air interface
by using the relation $\xi_{k}=-f_{l}|_{y_{*}=1}\zeta_{k}$. The shape of the water-air interface changes by the action of gravity, surface tension and air shear stress disturbances. 
As shown in Fig. \ref{fig:ka-xi-Thetaxi}, if we neglect the effect of the air shear stress disturbances, the amplitude and phase of the water-air interface relative to the ice-water interface are not correctly evaluated as $u_{\infty}$ increases. This results in an overestimated value of $|h'_{x}/\bar{h}_{x}|$
compared to that taking into account the effect of air shear stress disturbances with increasing $u_{\infty}$, as shown by the dashed and solid curves in Fig. \ref{fig:heat-coefficient} (b).

The solid curves in Fig. \ref{fig:heat-coefficient} (a) shows that $(h'_{x}/\bar{h}_{x})^{(i)}$ is positive for any $u_{\infty}$, while $(h'_{x}/\bar{h}_{x})^{(r)}$ is negative when $u_{\infty}<10$ m/s and is positive when $u_{\infty}>10$ m/s. 
From $\cos\Theta_{q'_{a*}}=(h'_{x}/\bar{h}_{x})^{(r)}/|h'_{x}/\bar{h}_{x}|$ and 
$\sin\Theta_{q'_{a*}}=-(h'_{x}/\bar{h}_{x})^{(i)}/|h'_{x}/\bar{h}_{x}|$, the corresponding phase difference between $q'_{a*}=h'_{x}/\bar{h}_{x}$ and the ice-water interface is 
$-\pi<\Theta_{q'_{a*}}<-\pi/2$ and $-\pi/2<\Theta_{q'_{a*}}<0$, respectively.
On the other hand, if we neglect the air shear stress disturbances, as shown by the dashed curves in Fig. \ref{fig:heat-coefficient} (a), $(h'_{x}/\bar{h}_{x})^{(r)}$ is negative and $(h'_{x}/\bar{h}_{x})^{(i)}$ is positive for any $u_{\infty}$. 
Then, the phase difference between $q'_{a*}=h'_{x}/\bar{h}_{x}$ and the ice-water interface is always $-\pi<\Theta_{q'_{a*}}<-\pi/2$
with respect to any $u_{\infty}$. This means that the position of maximum point of heat flux $q'_{a*}$ or that of the heat transfer coefficient $h'_{x}/\bar{h}_{x}$ depend significantly on the air shear stress disturbances.

Figure \ref{fig:heat-coefficient} (c) shows the variation of $(h'_{x}/\bar{h}_{x})^{(r)}$ (solid curves) and $(h'_{x}/\bar{h}_{x})^{(i)}$ (dashed curves) against $k_{a*}$ for the free stream velocities of $u_{\infty}$=5, 10, 20 m/s.
In the case of $u_{\infty}=5$ m/s, $(h'_{x}/\bar{h}_{x})^{(r)}$ is negative and $(h'_{x}/\bar{h}_{x})^{(i)}$ is positive for any $k_{a*}$. From $\cos\Theta_{q'_{a*}}=(h'_{x}/\bar{h}_{x})^{(r)}/|h'_{x}/\bar{h}_{x}|$ and 
$\sin\Theta_{q'_{a*}}=-(h'_{x}/\bar{h}_{x})^{(i)}/|h'_{x}/\bar{h}_{x}|$, the phase difference between $q'_{a*}=h'_{x}/\bar{h}_{x}$ and the ice-water interface is 
$-\pi<\Theta_{q'_{a*}}<-\pi/2$ 
as shown in Fig. \ref{fig:heatflux-sla} (a).   
On the other hand, in the case of $u_{\infty}$= 10 and 20 m/s, $(h'_{x}/\bar{h}_{x})^{(i)}$ is positive for any $k_{a*}$, but $(h'_{x}/\bar{h}_{x})^{(r)}$ changes from positive to negative at $k_{a*}=0.17$ and $k_{a*}=0.33$, respectively.
Then, in the case of $u_{\infty}=10$ m/s, the phase difference is 
$-\pi/2<\Theta_{q'_{a*}}<0$ for $k_{a*}<0.17$ and 
$-\pi<\Theta_{q'_{a*}}<-\pi/2$ for $k_{a*}>0.17$. 
Likewise, in the case of $u_{\infty}=20$ m/s, the phase difference is  
$-\pi/2<\Theta_{q'_{a*}}<0$ for $k_{a*}<0.33$ and 
$-\pi<\Theta_{q'_{a*}}<-\pi/2$ for $k_{a*}>0.33$, 
as shown in Fig. \ref{fig:heatflux-sla} (b). This means that the position of maximum point of $q'_{a*}=h'_{x}/\bar{h}_{x}$ on the water-air interface changes according to the wavelength of the ice-water interface disturbance and the free stream velocity. 
Furthermore, as shown in Fig. \ref{fig:heatflux-sla}, the position of the maximum point of $q'_{a*}=h'_{x}/\bar{h}_{x}$ moves in the direction opposite to the water flow with time. 

%%%%%%%%%%%%%%%%%%%%%%%%%%%%%%%%%%
\section{Summary and Discussion}
%%%%%%%%%%%%%%%%%%%%%%%%%%%%%%%%%%

We have proposed a theoretical model for ice growth under a supercooled water film driven by wind drag. The thickness and surface velocity of the water layer are variable by changing air stream velocity and water supply rate.  
For a given water supply rate, we investigated the morphological instability of the ice-water interface for various air stream velocities using a linear stability analysis, taking into account the effect of gravity, surface tension and the tangential and normal air shear stress disturbances due to the airflow on the shape of the water-air interface. 

Even for the simple model developed here, the form of heat transfer coefficient at the disturbed water-air interface is too complicated, which depends on the disturbed air flow and temperature fields, the shape of the disturbed water-air interface, as well as the shape of the ice-water interface. By considering the interaction between the air and water flows, we have found that the heat transfer coefficient at the water-air interface is significantly affected by the air shear stress disturbances, which suppresses the dimensionless amplification rate of the ice-water interface disturbance as the air stream velocity increases. However, the air shear stress disturbances do not significantly change the wavelength of an ice pattern occurring as a result of morphological instability of the ice-water interface. The model herein predicts that a centimeter scale ice pattern will appear, and its wavelength will decrease with increasing air stream velocity. Moreover, the ice pattern will translate towards the water source with time. 
At higher airspeed, the theoretical predictions obtained here might be relevant to the experiments for surface roughness characteristics associated with leading edge ice accretion on a NACA 0012 airfoil at a 0-deg angle of attack. \cite{Shin96} In that experiment, the height and spacing of roughness elements were measured with various icing parameters in glaze icing conditions. It was observed that the roughness spacing is about 1 mm, and that smooth-to-rough zones move upstream towards the stagnation region with time. 

Here, first we mention some differences between previous wet icing models \cite{Bourgault00, Myers02_1, Myers02_2, Myers04} and the current model:
(1) The undisturbed part of water film velocity profile derived herein, $\bar{u}_{l*}=y_{*}$, is the same as that used in the shallow-water icing model. \cite{Bourgault00} However, the disturbed part of water flow due to the disturbance of the ice-water interface is taken into account herein. 
(2) The current undisturbed part of temperature in the water film has a linear profile, as used in the models. \cite{Myers02_1, Myers02_2, Myers04} However, in this model, 
the disturbed part $T'_{l}$ due to the disturbance of the ice-water interface is considered, as shown in Eq. (\ref{eq:Tl}). 
Since the Peclet number herein is large, the disturbed part of temperature distribution in the water film is affected by the advection due to $\bar{u}_{l*}$ and $v'_{l}$, as indicated in the terms with $\Pec_{l}$ in Eq. (\ref{eq:geq-Hl}). 
(3) In previous icing models, detailed calculations concerning the effect of the interaction between the air and water flows on the temperature distribution in the water film were not carried out. The air shear stress disturbances influence the disturbed part of stream function of the water flow, $f_{l}$. As a result, 
the disturbed part of temperature distribution in the water film, $H_{l}$, is affected by both air and water flows. 

If we neglect the disturbed part of temperature distribution in the water film flow 
and focus on only the influence of the temperature distribution in the air on the growth condition of the ice-water interface disturbance, Eq. (\ref{eq:heatflux-zeta}) may be replaced by 
$L(\bar{V}+\partial \zeta/\partial t)
=-K_{a}\partial T_{a}/\partial y|_{y=\xi}$. 
Linearizing this equation at $y=\bar{h}_{0}$, the zeroth order yields
$\bar{V}=-K_{a}T_{\infty}/(L\delta_{0}/\bar{G}_{a*})$,
which is identical to Eq. (\ref{eq:V}). 
It is found that since the undisturbed part of ice growth rate, $\bar{V}$, does not include any parameter associated with the water film, $\bar{V}$ is determined without considering the details of the water film, 
and the heat transfer through the air boundary layer is the deciding factor in $\bar{V}$. 
From the first order in $\xi_{k}$, $\partial \zeta_{*}/\partial t_{*}=q'_{a*}=h'_{x}/\bar{h}_{x}$ yields 
$\sigma_{*}^{(r)}=(h'_{x}/\bar{h}_{x})^{(r)}$ and 
$v_{p*}=-\sigma_{*}^{(i)}/k_{l*}=-(h'_{x}/\bar{h}_{x})^{(i)}/k_{l*}$.
When we neglect the details of the water film, $(h'_{x}/\bar{h}_{x})^{(r)}$ (solid curves) in Fig. \ref{fig:heat-coefficient} (c) directly represents the amplification rate. 
However, there is a significant difference between $\sigma_{*}^{(r)}$ in Fig. \ref{fig:ka-amp-lambda} (a) and that in Fig. \ref{fig:heat-coefficient} (c). 
$\sigma_{*}^{(r)}$ in Fig. \ref{fig:ka-amp-lambda} (a) takes into account the effect of the disturbed part of  temperature distribution in the water layer as well as that in the air boundary layer on the ice growth conditions. 
On the other hand, $\sigma_{*}^{(r)}=(h'_{x}/\bar{h}_{x})^{(r)}$ in Fig. \ref{fig:heat-coefficient} (c) is obtained without considering the details of the disturbed temperature distribution in the water film. 
This suggests that the wavelength of ice pattern and its translation velocities shown in Table \ref{tab:tableII} cannot be evaluated correctly if we neglect the details of the water film. 
It should be emphasized that the same issue arose in a previous paper; \cite{Ueno10} 
If we neglected the influence of the disturbed temperature distribution in the water film flow on the growth condition for the ice-water interface disturbance, the amplification rate $\sigma_{*}^{(r)}$ had positive values for all wave numbers and hence a characteristic wavelength of icicle ripples was not obtained. 
A centimeter scale of ripples in wavelength was obtained from only $\sigma_{*}^{(r)}$ being taken into account the disturbed temperature distribution in the water film. \cite{Ueno03, Ueno04, Ueno07, UFYT10, Ueno10}

%************************************************************************************************************
Second, we mention the temperature at the ice-water interface, $T_{i}$, in Eq. (\ref{eq:Tsl}). 
Within linear stability analysis, $T_{i}=T_{sl}+\Delta T_{sl}$,
where $T_{sl}$ is the temperature at an undisturbed ice-water interface
and $\Delta T_{sl}$ is a deviation from it when the ice-water interface is disturbed. 
Its dimensionless form $\Delta T_{sl*}\equiv\Imag[\Delta T_{sl}/(T_{sl}-T_{la})]$ can be expressed as follows by evaluating Eq. (\ref{eq:Tl}) at the disturbed ice-water interface $y_{*}=\zeta_{*}$:
\begin{equation}
\Delta T_{sl*}
= \delta_{b}(t_{*})\left\{(H_{l}^{(r)}|_{y_{*}=0}-1)\sin[k_{l*}(x_{*}-v_{p*}t_{*})] 
+H_{l}^{(i)}|_{y_{*}=0}\cos[k_{l*}(x_{*}-v_{p*}t_{*})]\right\}.
\label{eq:DeltaTsl}
\end{equation} 
Figure \ref{fig:isotherms} (a) represents the isotherms in the water film. The real and imaginary parts of the disturbed part of temperature distribution in the water film, $H_{l}^{(r)}$ and $H_{l}^{(i)}$ in Eq. (\ref{eq:Tl}), are determined by solving Eq. (\ref{eq:geq-Hl}) subject to the boundary conditions (\ref{eq:Tla-xi}) and (\ref{eq:heatflux-xi-h0}), which were derived from the continuity of temperature and heat flux at the water-air interface, Eqs. (\ref{eq:bc-Tla}) and (\ref{eq:heatflux-xi}), respectively.
Since the temperature distribution in the water film is affected by both air and water flows, $\Delta T_{sl*}$ varies. 
When the ice-water interface is flat, $\Delta T_{sl*}$ must be zero. Indeed, $H_{l}^{(r)}|_{y_{*}=0} \rightarrow 1$ and $H_{l}^{(i)}|_{y_{*}=0} \rightarrow 0$ in the limit $k_{l*} \rightarrow 0$ were numerically confirmed.

%%%%%%% Fig. 7 %%%%%%%%%%%%%%%%%%%%%%%%%%%%%%%%%%%%%%%%%%%%%%%%%%%%%%%%%%%%%%%%%%%%%%%%%%%%%%%%%%%%%%%%%%%%%%%%%%%%
\begin{figure}[ht]
\begin{center}
\includegraphics[width=7cm,height=7cm,keepaspectratio,clip]{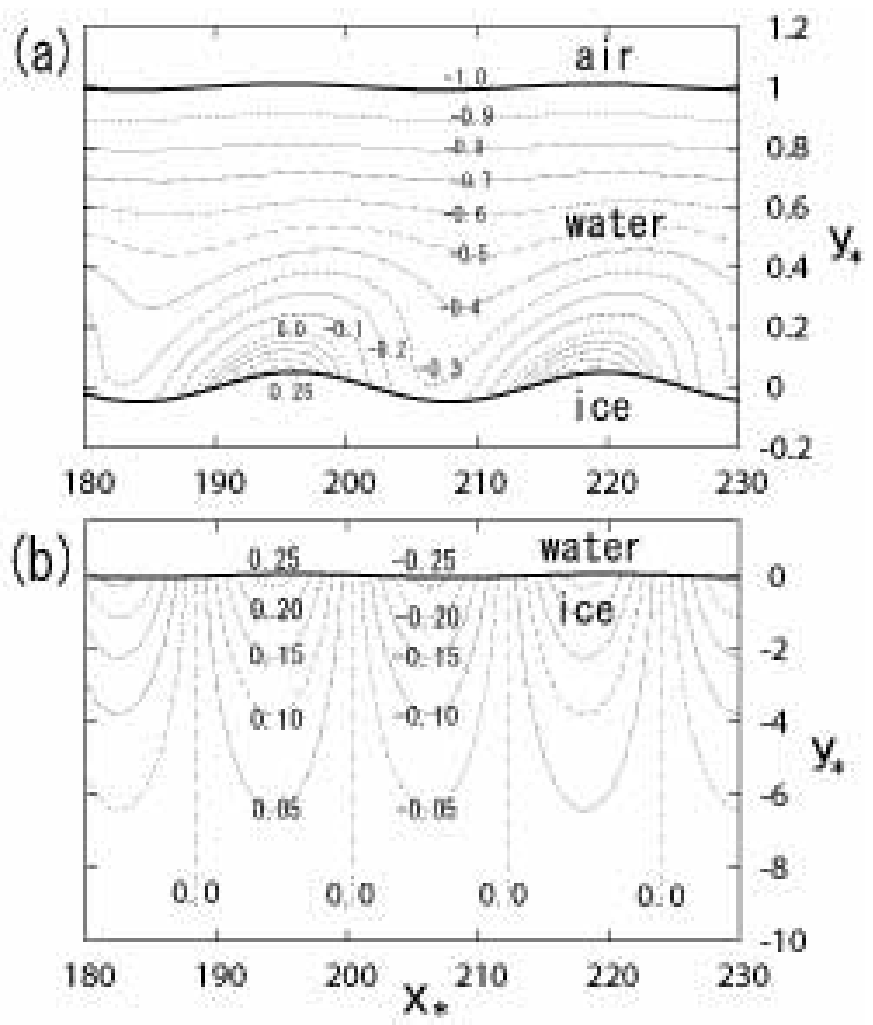}\hspace{5mm}
\includegraphics[width=7cm,height=7cm,keepaspectratio,clip]{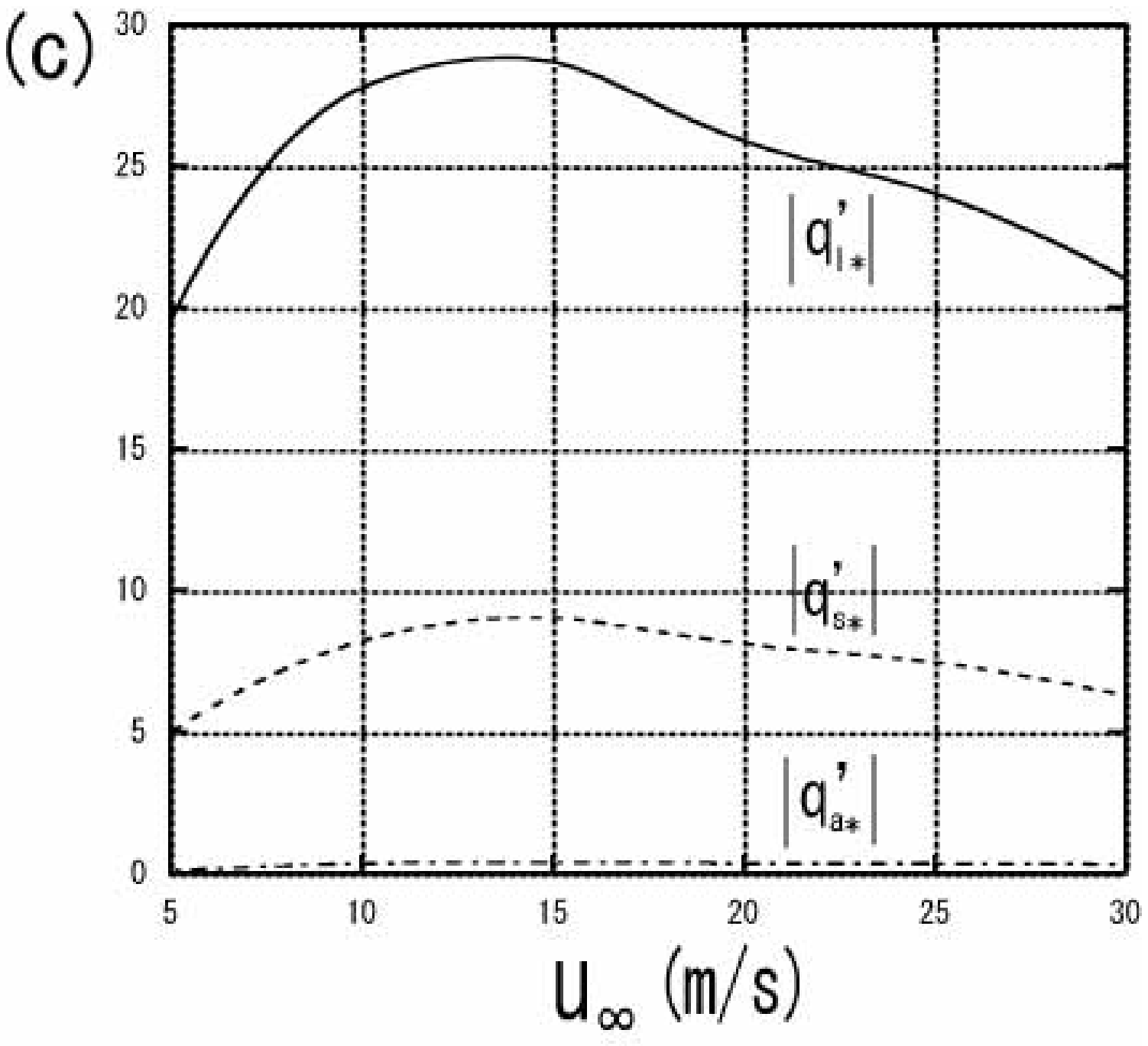}\hspace{5mm}
\includegraphics[width=7cm,height=7cm,keepaspectratio,clip]{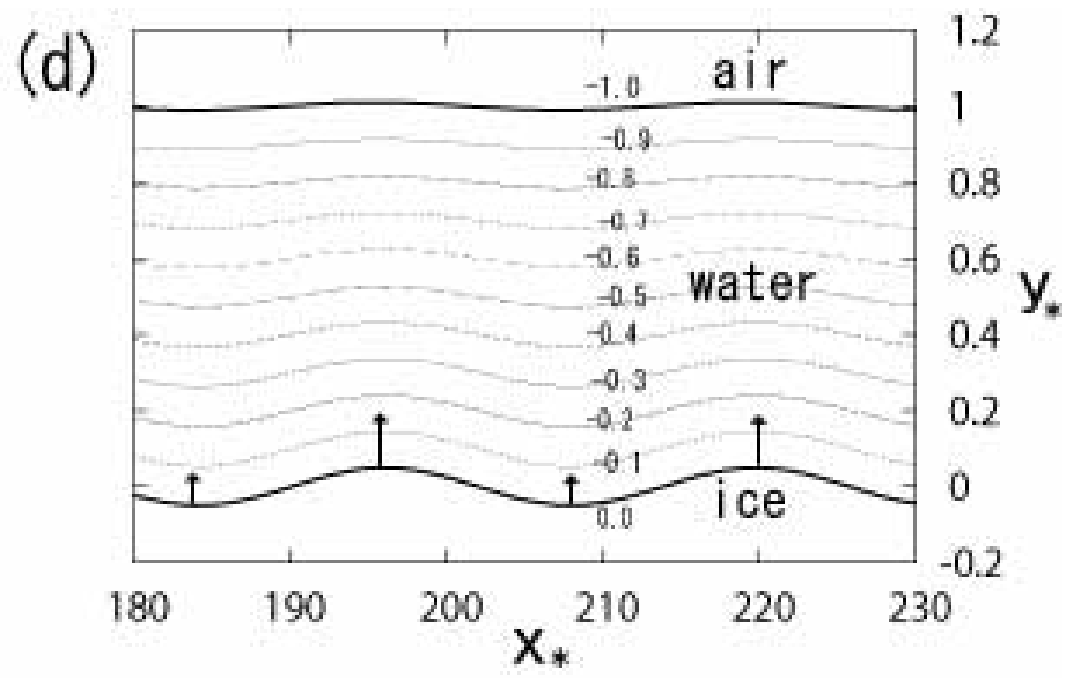} 
\includegraphics[width=7cm,height=7cm,keepaspectratio,clip]{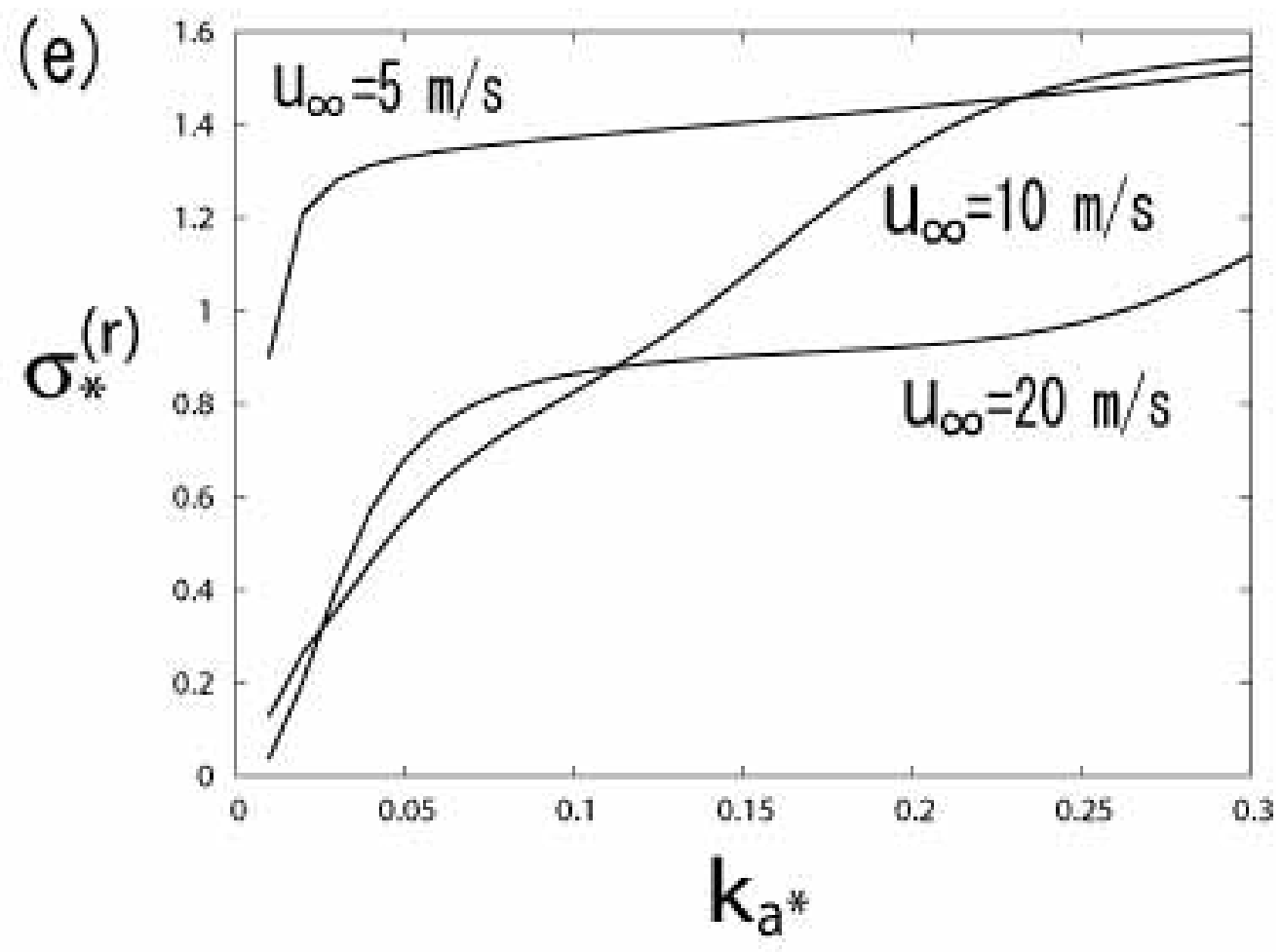} 
\end{center}
\caption{For $Q/l_{w}=1000$ [(ml/h)/cm], $u_{\infty}=20$ m/s and $\delta_{b}(t_{*})=0.05$, 
(a) represents the isotherms in the water film and (b) the isotherms in the ice, for the boundary condition 
$T_{l}|_{y=\xi}=T_{a}|_{y=\xi}=T_{la}$ (constant). 
(c) represents the variation of disturbed heat flux $|q'_{l*}|$, $|q'_{s*}|$ and $|q'_{a*}|$ against the free stream velocity $u_{\infty}$.
For the boundary condition $T_{l}|_{y=\zeta}=T_{s}|_{y=\zeta}=T_{sl}$ (constant),
(d) represents the isotherms in the water film and 
(e) dimensionless amplification rate $\sigma_{*}^{(r)}$ versus dimensionless wave number $k_{a*}$.
The numbers in the isotherms are the values of dimensionless temperatures in water (\ref{eq:Tl}) and in ice (\ref{eq:Ts}).
}
\label{fig:isotherms}
\end{figure}
%%%%%%%%%%%%%%%%%%%%%%%%%%%%%%%%%%%%%%%%%%%%%%%%%%%%%%%%%%%%%%%%%%%%%%%%%%%%%%%%%%%%%%%%%%%%%%%%%%%%%%%%%%%%%%%%%%%%
                        
From Eqs. (\ref{eq:ql}), (\ref{eq:qs}) and (\ref{eq:qa}), we define the magnitude of $q'_{l*}$, $q'_{s*}$ and $q'_{a*}$ as follows:
\begin{eqnarray}
|q'_{l*}|&\equiv & \{(dH_{l}^{(r)}/dy_{*}|_{y_{*}=0})^{2}+(dH_{l}^{(i)}/dy_{*}|_{y_{*}=0})^{2}\}^{1/2}, \\
|q'_{s*}|&\equiv & K^{s}_{l}k_{l*}\{(H_{l}^{(r)}|_{y_{*}=0}-1)^{2}+(H_{l}^{(i)}|_{y_{*}=0})^{2}\}^{1/2}, \\
|q'_{a*}|&\equiv &\{(G'^{(r)}_{a}f_{l}^{(r)}|_{y_{*}=1}-G'^{(i)}_{a}f_{l}^{(i)}|_{y_{*}=1})^{2}
                 +(G'^{(r)}_{a}f_{l}^{(i)}|_{y_{*}=1}+G'^{(i)}_{a}f_{l}^{(r)}|_{y_{*}=1})^{2}\}^{1/2}.
\label{eq:amplitude-ql-qa-qs}
\end{eqnarray} 
Since the water flow is affected by the air flow through the air shear stress disturbances, $|q'_{l*}|$ and $|q'_{s*}|$ as well as $|q'_{a*}|$ depend on the free stream velocity $u_{\infty}$. 
Figure \ref{fig:isotherms} (c) represents the variation of $|q'_{l*}|$, $|q'_{s*}|$ and $|q'_{a*}|$ against $u_{\infty}$. These values were estimated for $Q/l_{w}=1000$ [(ml/h)/cm] and $x=0.1$ m, and for $k_{a*}$ at which $\sigma_{*}^{(r)}$ acquires a maximum value for a given $u_{\infty}$. 
From Eq. (\ref{eq:heatflux-zeta}), the disturbed part of the Stephan condition is $\partial \zeta_{*}/\partial t_{*}=q'_{l*}-q'_{s*}$. Therefore, the net heat flux $q'_{l*}-q'_{s*}$ determines the amplification rate of the ice-water interface disturbance.
Since $|q'_{a*}|=|h'_{x}/\bar{h}_{x}|$ from Eq. (\ref{eq:amplitude-h'x}), the dashed-dotted curve in Fig. \ref{fig:isotherms} (c) is the same as the solid curve in Fig. \ref{fig:heat-coefficient} (b).  
When both ice-water and water-air interfaces are flat and the undisturbed temperature in the water film is the linear profile $\bar{T}_{l*}=y_{*}$, the latent heat released at the ice-water interface, $L\bar{V}$, the undisturbed part of heat flux at the ice-water interface, $K_{l}\bar{G}_{l}$ and that at the water-air interface, $K_{a}\bar{G}_{a}$,
must be equal. 
Hence all of the latent heat is released away from the water-air interface through the water film.
However, in the case of the disturbed interfaces, the disturbed part of heat flux at the ice-water
interface, $|q'_{l*}|$, is not necessarily equal to that at the water-air interface, $|q'_{a*}|$. Indeed, as shown in Fig. \ref{fig:isotherms} (c), $|q'_{a*}|$ is much smaller than $|q'_{l*}|$. 
The release of latent heat at the water-air interface, $|q'_{a*}|$, is limited not only by the temperature gradient at the water-air interface but also by the shape of the water-air interface. Hence all of the latent heat released at the disturbed ice-water interface cannot be removed at the water-air interface and most of it is carried by the flow in the water film.  

The effect of morphological instabilities is to increase the surface area of the phase boundary and hence to enhance the release of latent heat. The latent heat that is not completely removed by air and water flows may lead to a local temperature rise in the supercooled water and ice locally. Also, the flow in the water film can carries a supercooled water in the interior towards the ice-water interface. Hence, in Fig. \ref{fig:isotherms} (a), not only the isotherms in the water film are deformed by the advection terms in Eq. (\ref{eq:geq-Hl}) but also alternating patterns of warming and cooling appear in the neighbourhood of the ice-water interface. 
The characteristic time of the deformation associated with the shear rate is $1/(u_{la}/\bar{h}_{0})$.
On the other hand, the thermal diffusion time associated with a wave number $k_{l}$ is $1/(\kappa_{l}k_{l}^{2})$.  
For a disturbed ice-water interface with a 1 cm wavelength shown in Figs. \ref{fig:isotherms} (a) and (d), 
the condition $1/(u_{la}/\bar{h}_{0}) \ll 1/(\kappa_{l}k_{l}^{2})$ is satisfied.
Hence, the temperature distribution in the neighbourhood of the disturbed ice-water interface is deformed by the advection in the water film. 
In order to avoid a temperature discontinuity, $\Delta T_{sl*}$, between water and ice, the disturbed heat flux $q'_{s*}$ due to thermal diffusion occurs in the ice. 
As a result, Fig. \ref{fig:isotherms} (b) also shows alternating patterns of warming and cooling in the ice.
From the comparison of Eqs. (\ref{eq:DeltaTsl}) and (\ref{eq:qs}) as well as that of Figs. \ref{fig:isotherms} (b) and \ref{fig:heatflux-sla} (b), if $\Delta T_{sl*}>0$ (warming), then $q'_{s*}<0$, hence the direction of $q'_{s*}$ is from the ice-water interface to the ice. On the other hand, if $\Delta T_{sl*}<0$ (cooling), then $q'_{s*}>0$, hence the direction of $q'_{s*}$ is from the ice to the ice-water interface.
However, it should be noted that this disturbed part of heat flux in the ice exists only in the vicinity of the ice-water interface, as Eq. (\ref{eq:Ts}) shows that the disturbed temperature in the ice is exponentially attenuated, and the ice temperature approaches $T_{s}=T_{sl}$ ($T_{sl}=$0 $^{\circ}$C for pure water) far from the ice-water interface, as shown in Fig. \ref{fig:isotherms} (b).   
  
On the other hand, the isotherms in Fig. \ref{fig:isotherms} (d) are determined from different solutions $H_{l}^{(r)}$ and $H_{l}^{(i)}$, which are obtained by solving Eq. (\ref{eq:geq-Hl}) subject to the boundary conditions $H_{l}|_{y_{*}=0}=1$ and Eq. (\ref{eq:heatflux-xi-h0}). $H_{l}|_{y_{*}=0}=1$ is derived from the condition $T_{l}|_{y=\zeta}=T_{s}|_{y=\zeta}=T_{sl}$ (constant). In this case, the temperature at the water-air interface is an unknown to be determined, which is deviated from $T_{la}$ in Eq. (\ref{eq:bc-Tla})
and hence the first boundary condition in Eq. (\ref{eq:bc-Ha}) is replaced by
\begin{equation}
H_{a}|_{\eta=0}=1-\frac{K_{a}}{K_{l}}-\frac{K_{a}}{K_{l}}H_{l}|_{y_{*}=1}/f_{l}|_{y_{*}=1}.
\label{eq:DeltaTla}
\end{equation}
However, since $K_{a}/K_{l} \ll 1$, $H_{a}|_{\eta=0} \approx 1$ is a good approximation, which is equivalent to the condition $T_{l}|_{y=\xi}=T_{a}|_{y=\xi}\approx T_{la}$ (constant).
As shown in Fig. \ref{fig:isotherms} (d), the isotherms are slightly deformed from the undisturbed temperature distribution $\bar{T}_{l*}=y_{*}$ and the variation of temperature in the neighbourhood of the ice-water and water-air interfaces are strongly affected by the boundary temperatures, $T_{l*}|_{y_{*}=\zeta_{*}}=T_{s*}|_{y_{*}=\zeta_{*}}=0.0$ (constant) and $T_{l*}|_{y_{*}=\xi_{*}}=T_{a*}|_{y_{*}=\xi_{*}}\approx -1.0$ (constant). In particular, the isotherms in the neighbourhood of the ice-water interface in Fig. \ref{fig:isotherms} (d) are significantly different from those in Fig. \ref{fig:isotherms} (a). Although there exists a shear flow in the water film, the temperature distribution is almost symmetric around any protruded part of the ice-water interface. The long arrows at the ice-water interface in Fig. \ref{fig:isotherms} (d) show that the temperature gradient is largest at each protruded part of the ice-water interface. This promotes the ice growth at the protruded part than at the depressed part, always resulting in an unstable growth of the ice-water interface. 
Consequently, the amplification rate $\sigma^{(r)}_{*}$ obtained from the boundary condition $T_{l}|_{y=\zeta}=T_{s}|_{y=\zeta}=T_{sl}$ (constant) has positive values for all wave numbers, as shown in Fig. \ref{fig:isotherms} (e).

From $1/(u_{la}/\bar{h}_{0})=1/(\kappa_{l}k_{l}^{2})$, the wave number at which the two time scale equals is given by $k_{l*}=\sqrt{\Pec_{l}}\sim 20$ for $Q/l_{w}=1000$ [(ml/h)/cm]. The corresponding wavelength is 150 $\mu$m from $k_{l*}=k\bar{h}_{0}=20$ for $u_{\infty}=20$ m/s.
The effect of the water flow on the isotherms in such a microscopic length scale is negligible. Instead, taking into account the Gibbs-Thomson effect, the temperature at the ice-water interface is expressed as 
$T_{i}=T_{sl}+(T_{sl}\Gamma/L)\partial^{2} \zeta/\partial x^{2}$, from which $H_{l}|_{y_{*}=0}=1-(d_{0}/\bar{h}_{0})(l_{\rm th}/\bar{h}_{0})k_{l*}^{2}$ is derived.
Here $l_{\rm th}=\kappa_{l}/\bar{V}$ and $d_{0}=T_{sl}\Gamma C_{pl}/L^{2}$ are a macroscopic and microscopic characteristic length, respectively, $\Gamma$ is the ice-water interface tension and $C_{pl}$ is the specific heat at constant pressure of the water. Neglecting the advection terms in Eq. (\ref{eq:geq-Hl}) and solving it subject to the boundary conditions $H_{l}|_{y_{*}=0}=1-(d_{0}/\bar{h}_{0})(l_{\rm th}/\bar{h}_{0})k_{l*}^{2}$ and Eq. (\ref{eq:heatflux-xi-h0}) with  $f_{l*}|_{y_{*}=1} \approx 0$, from Eq. (\ref{eq:amp}) we obtain the amplification rate,
$\sigma^{(r)}_{*}=k_{l*}\{1-(d_{0}/\bar{h}_{0})(l_{\rm th}/\bar{h}_{0})k_{l*}^{2}(1+K^{s}_{l})\}$,  
which is just the result of the Mullins-Sekerka instability. \cite{Langer80} 
Here the condition $f_{l*}|_{y_{*}=1} \approx 0$ means that the water-air interface is nearly flat because the surface tension is so dominant in the microscopic length scale of dendrite that the effect of disturbance concerning the dendrite spacing on the shape of the water-air interface is negligible. 
$\sigma^{(r)}_{*}$ acquires a maximum value at 
$k_{l*}=[\bar{h}_{0}^{2}/\{3l_{\rm th}d_{0}(1+K^{s}_{l})\}]^{1/2} \sim 10$
for $\bar{h}_{0} \sim 5 \times 10^{-4}$ m, $l_{\rm th} \sim 10^{-1}$ m, $d_{0} \sim 10^{-9}$ m and $K^{s}_{l}=3.92$.
The dependence of the microscopic wavelength predicted from the Mullins-Sekerka instability on the free stream velocity $u_{\infty}$ is 
$\lambda_{\rm micro}=2\pi\{3l_{\rm th}d_{0}(1+K^{s}_{l})\}^{1/2} \propto u_{\infty}^{-1/4}$ 
because $l_{\rm th} \sim \delta_{0} \sim u_{\infty}^{-1/2}$ in the model herein. This result is contrast to the dependence of the macroscopic wavelength $\lambda_{\rm macro}$ on $u_{\infty}$. Figure \ref{fig:ka-amp-lambda} (b) shows that $\lambda_{\rm macro} \propto u_{\infty}^{-0.88}$. 
It should be noted that the mechanism of the macro-scale morphological instability under a supercooled liquid film herein is quite different from the dendritic growth. $\lambda_{\rm macro}$ depends on the water film thickness $\bar{h}_{0}$ as indicated in Tables \ref{tab:tableI} and \ref{tab:tableII}, while $\lambda_{\rm micro}$ does not depend on it.

The use of the boundary condition $T_{l}|_{y=\zeta}=T_{s}|_{y=\zeta}=T_{sl}$ (constant) caused a serious problem in a model for the icicle ripple formation. \cite{UFYT10} This condition was used in the model proposed by Ogawa and Furukawa \cite{Ogawa02} when determining $H_{l}$. However, our numerical analysis did not reproduce their amplification rate shown in Fig. 4 of Ref. \onlinecite{Ogawa02}. Instead, the numerically obtained amplification rate had positive values for all wave numbers and there was no well-defined maximum amplification rate (for details, see Figure 5 (c) in Ref. \onlinecite{UFYT10}).
The same issue has already arisen in aircraft icing problems. Tsao and Rothmayer developed a physical model to describe the aero-hydro-thermo-dynamic interaction of a cold air boundary layer with glaze ice sheets and water films. \cite{Tsao98} However, their stability analysis showed that the ice-water interface disturbance became unstable for all modes (see Figure 7 in Ref. \onlinecite{Tsao98}), which indicated that there is no dominant amplification rate to select a preferred wavelength. 
The assumption that the disturbed ice-water interface is at the equilibrium freezing temperature was used in their model too.
To overcome this issue, the Gibbs-Thomson effect was introduced to stabilize the smallest scale icing disturbances. \cite{Tsao00} However, the length scale predicted by their theory was much smaller than the ice roughness spacing of the order of millimeters observed in the experiments by Shin. \cite{Shin96}  
On the other hand, the condition $T_{l}|_{y=\zeta}=T_{s}|_{y=\zeta}=T_{sl}$ (constant) was not used in the model for the icicle ripple formation proposed by Ueno \cite{Ueno03, Ueno04, Ueno07, UFYT10, Ueno10} when determining $H_{l}$. That model was able to predict a centimeter-scale wavelength of icicle ripples and upward ripple migration due to an asymmetry in the temperature distribution between the upstream and downstream side of any protruded part of the ice-water interface, which were confirmed by the experiments. \cite{UFYT10, Chen10} 

%************************************************************************************************************************

%%%%%%% Fig. 8 %%%%%%%%%%%%%%%%%%%%%%%%%%%%%%%%%%%%%%%%%%%%%%%%%%%%%%%%%%%%%%%%%%%%%%%%%%%%%%%%%%%%%%%%%%%%%%%%%%%%
\begin{figure}[ht]
\begin{center}
\includegraphics[width=10cm,height=10cm,keepaspectratio,clip]{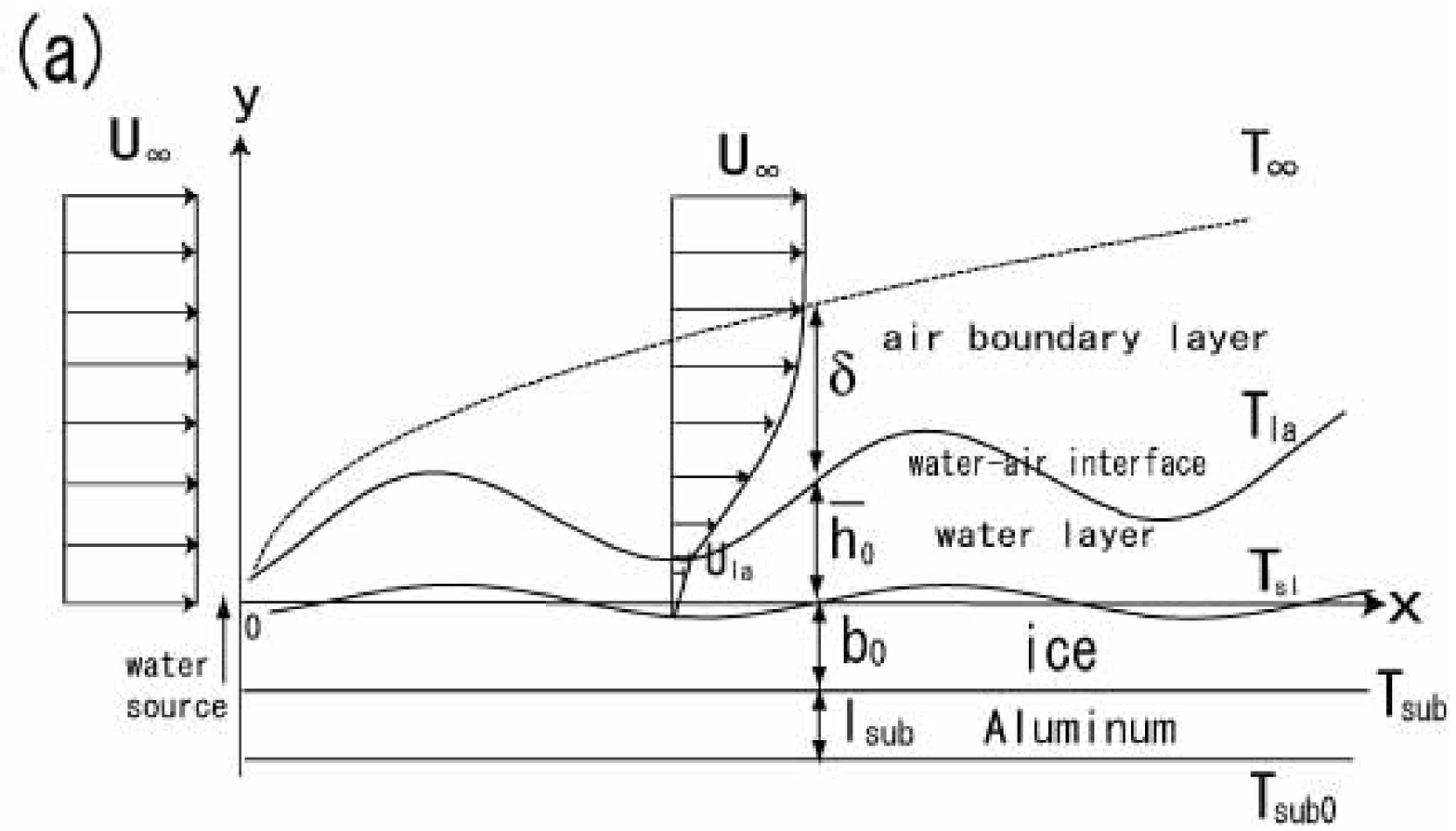}\\[5mm]
\includegraphics[width=8cm,height=8cm,keepaspectratio,clip]{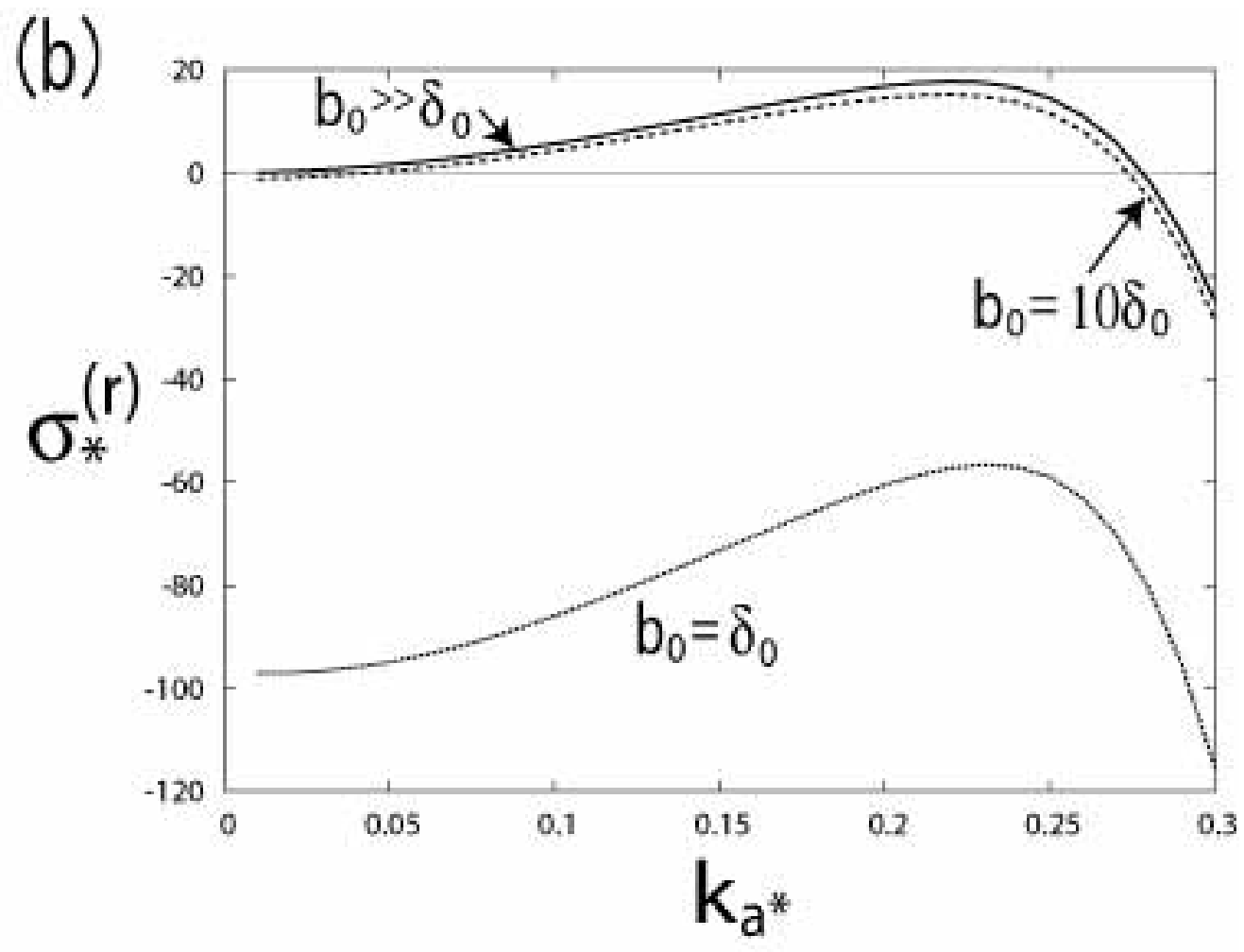} 
\end{center}
\caption{(a) Schematic of an ice growth on an aluminum plate under air and water flows.
(b) For $Q/l_{w}=1000$ [(ml/h)/cm] and $u_{\infty}=20$ m/s, dimensionless amplification rate $\sigma^{(r)}_{*}$ versus dimensionless wave number $k_{a*}$ for various ice thickness $b_{0}$. The thickness of air boundary layer in (a) is defined by $\delta=\delta_{0}/\bar{G}_{a*}$, where $\bar{G}_{a*}=0.413$ and $\delta_{0}=361$ $\mu$m for $u_{\infty}=20$ m/s.}
\label{fig:aluminum}
\end{figure}
%%%%%%%%%%%%%%%%%%%%%%%%%%%%%%%%%%%%%%%%%%%%%%%%%%%%%%%%%%%%%%%%%%%%%%%%%%%%%%%%%%%%%%%%%%%%%%%%%%%%%%%%%%%%%%%%%%%%

Third, we mention the effect of heat conduction into a substrate beneath an ice sheet of finite thickness on the morphological instability. In the model herein, it was assumed that the ice region is semi-infinite and the undisturbed part of temperature gradient in the ice does not exist. We relax this assumption by including heat conduction into a planer aluminum substrate beneath the ice sheet. In Fig. \ref{fig:aluminum} (a), $b_{0}$ and $l_{\rm sub}$ are the thickness of ice and aluminum plate, respectively, $T_{\rm sub}$ is the temperature between the ice and aluminum plate and $T_{\rm sub0}$ is the temperature of other side of surface of the aluminum plate. Then, the undisturbed temperature gradient in the ice is 
$\bar{G}_{s}=(T_{sl}-T_{\rm sub})/b_{0}$, and 
Eq. (\ref{eq:amp}) is replaced by 
\begin{equation}
\sigma_{*}^{(r)}=-\frac{dH_{l}^{(r)}}{dy_{*}}\Big|_{y_{*}=0}
+K^{s}_{l}k_{l*}\frac{\cosh(k_{l*}b_{0}/\bar{h}_{0})}{\sinh(k_{l*}b_{0}/\bar{h}_{0})}
\left(-G^{s}_{l}+H_{l}^{(r)}|_{y_{*}=0}-1\right),
\label{eq:aluminum-amp}
\end{equation}
where $G^{s}_{l}\equiv \bar{G}_{s}/\bar{G}_{l}$ is the ratio of the undisturbed temperature gradient at the ice-water interface in ice to that in water. The ice thickness $b_{0}$ is determined by integrating the following equation subject to an initial condition of $b_{0}=0$ at $t=0$: \cite{UFYT10}
\begin{equation}
\frac{db_{0}}{dt}=\frac{K_{s}}{L}\frac{T_{sl}-T_{\rm sub}}{b_{0}}
+\frac{K_{a}}{L}\frac{T_{sl}-T_{\infty}}{b_{0}/\bar{G}_{a*}}.
\label{eq:b0}
\end{equation} 
If other side of surface of the aluminum plate is exposed to ambient cold air, i.e. assuming $T_{\rm sub0}=T_{\infty}$, $G^{s}_{l}$ can be expressed as \cite{UFYT10}
\begin{equation}
G^{s}_{l}=\frac{K_{l}}{K_{a}}\frac{\delta_{0}}{b_{0}}\Big/\left(1+\frac{K_{l}}{K_{\rm sub}}\frac{l_{\rm sub}}{b_{0}}\right),
\label{eq:Gsl}
\end{equation} 
where $K_{\rm sub}=237$ ${\rm J/(m\,K\,s)}$ is the thermal conductivity of the aluminum plate.

Figure \ref{fig:aluminum} (b) shows the dimensionless amplification rate $\sigma^{(r)}_{*}$ versus dimensionless wave number $k_{a*}$ for various ice thickness $b_{0}$. 
In the case of $b_{0}=\delta_{0}$, $\sigma^{(r)}_{*}$ is negative for all wave numbers, which means that the ice-water interface disturbances diminish with time.
On the other hand, in the case of $b_{0}=10\delta_{0}$, the ice-water interface disturbances in a finite range of wave numbers become unstable and $\sigma^{(r)}_{*}$ acquires a maximum value at a wave number. Noting that $G^{s}_{l}$ in Eq. (\ref{eq:Gsl}) is zero in the limit $b_{0} \gg \delta_{0}$ and $\bar{h}_{0}$ is the same order as $\delta_{0}$ as indicated in Table \ref{tab:tableI}, when $b_{0} \gg \delta_{0}$ Eq. (\ref{eq:aluminum-amp}) reduces to Eq. (\ref{eq:amp}) and the solid curve in Fig. \ref{fig:aluminum} (b) is the same as that for $u_{\infty}=20$ m/s in Fig. \ref{fig:ka-amp-lambda} (a).
When the ice thickness is small during the ice growth, the morphological instability of the ice surface is suppressed because the removal of the latent heat due to the conduction into the aluminum plate is dominant than that due to the convection by air and water flows. Even in the presence of the undisturbed part of temperature gradient $\bar{G}_{s}$ in ice, the morphological instability occurs when the ice thickness exceeds a critical value.

%*********************************************************************************************************************

Finally, we mention some limitations of the proposed model. 
First, freshwater icing sponginess containing non-negligible amount of liquid water is observed in icicles \cite{Makkonen88, Knight80} and aufeis. \cite{Streitz02} The spongy icing phenomenon is also well-known to the in-flight icing community. \cite{Blackmore03, Lozowski05} In the experiments of icing wind tunnel using a NACA 0012 airfoil, a considerable variation in sponginess (or liquid fraction) with air temperature, wind speed and liquid water content was observed. \cite{Lozowski05} The model herein cannot explain these experimental results. Therefore, it needs to modify an air-water-ice multi-phase system in Fig. \ref{fig:ice-water-air} to an air-water-spongy ice multi-phase system in Fig. \ref{fig:spongyice-water-air}, where a spongy layer is introduced in between a fully water region and a fully ice region. 
However, since the water region is thin liquid film and the latent heat transfer is strongly affected by the existence of the water-air interface, the configuration in Fig. \ref{fig:spongyice-water-air} is fundamentally different from the mathematical models of flow-induced morphological instability of mushy layers developed in previous studies, \cite{Worster00, Worster92, Feltham99, Chung01, Neufeld08} where the liquid region was taken to be semi-infinite. 

The conventional Stefan problem cannot describe the pattern formation observed in nature, because the dimensional information needed to set the scale of a crystal growth is absent. \cite{Langer80} 
In other words, if the temperature at the ice-water interface is kept at 0 $^{\circ}$C and neglecting surface energy, the morphological instabilities occur on arbitrarily small length scales given any amount of supercooling. \cite{Worster00} In practice, surface energy limits the instability at some scale and an ice surface under a supercooled water film results in dendritic growth. 
As a result, there can be a possibility of spongy ice formation, in which a portion of the surface liquid is incorporated into the dendritic ice matrix. \cite{Blackmore03} Hence the spongy layer is a mixture of ice and water and its temperature is a thermodynamic equilibrium one. \cite{Makkonen87} The use of this local equilibrium assumption is appropriate in the interior of the spongy layer at some distance from the tips of dendrites, where the increase in specific surface area of micro-scale phase boundaries promotes the release of latent heat into the interstices (pores) and hence the level of non-equilibrium can be kept very small. \cite{Worster92} 

%%%%%%% Fig. 9 %%%%%%%%%%%%%%%%%%%%%%%%%%%%%%%%%%%%%%%%%%%%%%%%%%%%%%%%%%%%%%%%%%%%%%%%%%%%%%%%%%%%%%%%%%%%%%%%%%%%
\begin{figure}[ht]
\begin{center}
\includegraphics[width=10cm,height=10cm,keepaspectratio,clip]{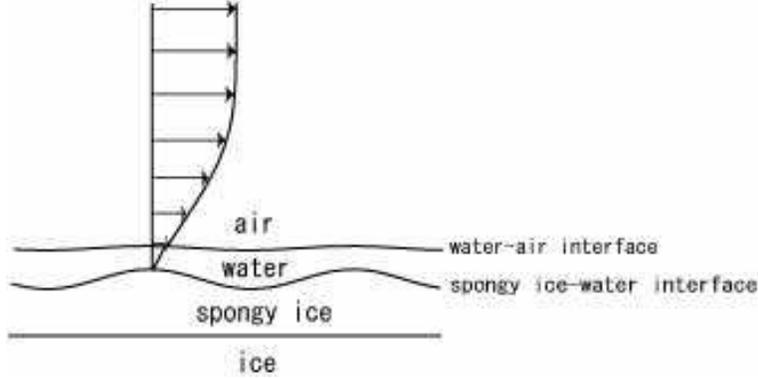} 
\end{center}
\caption{Schematic of air-water-spongy ice multi-phase system.}
\label{fig:spongyice-water-air}
\end{figure}
%%%%%%%%%%%%%%%%%%%%%%%%%%%%%%%%%%%%%%%%%%%%%%%%%%%%%%%%%%%%%%%%%%%%%%%%%%%%%%%%%%%%%%%%%%%%%%%%%%%%%%%%%%%%%%%%%%%

In Fig. \ref{fig:spongyice-water-air}, since the interface between the spongy ice region and water region does not have a well defined position on the micro scale of dendritic growth, the spongy ice-water interface is defined as the envelope (suitably smoothed) of the dendrite ice matrix. \cite{Worster00, Worster92} 
If the ice-water interface in Figs. \ref{fig:isotherms} (a) and (b) is replaced by the spongy ice-water interface, $T_{i}$ in Eq. (\ref{eq:Tsl}) is interpreted as the temperature at the spongy ice-water interface.
The local equilibrium assumption is likely to break down in the neighbourhood of the spongy ice-water interface, because if the spongy layer is postulated as a permeable material, \cite{Beavers67} a significant effect of the tangential and normal air shear stress disturbances found herein on the spongy layer through the thin water film can be expected and the latent heat advected by the water flow penetrates into near the spongy ice-water interface. The level of this non-equilibrium effect in the macro-scale is negligible at some distance from the spongy ice-water interface, where the temperature is 0 $^{\circ}$C, as shown in Fig. \ref{fig:isotherms} (b).
In order to gain a clear understanding of these viewpoints,  
it is necessary to add equations governing local liquid fraction and the internal evolution of the spongy layer to the model herein and the effects of the shear stress at a disturbed spongy ice-water interface on the distribution of liquid fraction, permeability and penetration depth should be investigated along with a study on non-equilibrium coexistence of crystal and shearing liquid flow. \cite{Butler02} Furthermore, the dependence of the liquid fraction on the icing parameter such as air temperature, wind speed and liquid water content in the experiments by Lozowski et al.\cite{Lozowski05} must be explained.

Second, in the linear stability analysis, the amplitude of the ice-water interface disturbance of the most unstable mode increases with time: 
$\delta_{b}(t_{*})=\exp(\sigma_{* \rm max}^{(r)}t_{*})\delta_{b}$,
which affects the magnitude of $h'_{x}/\bar{h}_{x}$ in Eq. (\ref{eq:h'x}). 
In order to evaluate the value of $\sigma_{*\rm max}^{(r)}$, it is necessary to determine the disturbed flow and temperature fields in the water film which are influenced by surrounding airflow and temperature fields.
However, the linear theory is unable to clarify further features related to the development of disturbance. 
We have to generalize the equation $d\delta_{b}(t_{*})/dt_{*}=\sigma^{(r)}_{* \rm max}\delta_{b}(t_{*})$ to a nonlinear amplitude evolution equation.

Third, the magnitude of $h'_{x}/\bar{h}_{x}$ depends on the shape of the ice-water interface. 
In the normal mode analysis presented here, the values of $h'_{x}/\bar{h}_{x}$ depend on the supposed sinusoidal form in Eq. (\ref{eq:h'x}).  
It is necessary to extend the current model to the problem of ice morphology produced on an arbitrary three-dimensional surface such as aircraft wings and overhead line cables, \cite{Myers02_2} taking into account water flow driven by both gravity and wind drag simultaneously. 
Removing restrictions mentioned above and further extension of the current model to practical icing problems
will be discussed in later papers.
  
%%%%%%%%%%%%%%%%%%%%%%%%%%%%%%%%%%%%%%%%%%%%%%%%%%%%%%%%%%%%%%%%%%%%%%%%%%%%%%%%%%%%%%%%%%%%%%%%%%%%%%%%%%%%%%%%%%%%%%
\begin{acknowledgements}
This study was carried out within the framework of the NSERC/Hydro-Qu$\acute{\rm e}$bec/UQAC Industrial Chair on Atmospheric Icing of Power Network Equipment (CIGELE) and the Canada Research Chair on Engineering of Power Network Atmospheric Icing (INGIVRE) at the Universit$\acute{\rm e}$ du Qu$\acute{\rm e}$bec $\grave{\rm a}$ Chicoutimi. 
The authors would like to thank all CIGELE partners (Hydro-Qu$\acute{\rm e}$bec, Hydro One, R$\acute{\rm e}$seau Transport d'$\acute{\rm E}$lectricit$\acute{\rm e}$ (RTE) and $\acute{\rm E}$lectricit$\acute{\rm e}$ de France (EDF), Alcan Cable, K-Line Insulators, Tyco Electronics, Dual-ADE, and FUQAC) whose financial support made this research possible. The authors would also like to thank H. Tsuji and anonymous referees for their useful comments. 
\end{acknowledgements}
%%%%%%%%%%%%%%%%%%%%%%%%%%%%%%%%%%%%%%%%%%%%%%%%%%%%%%%%%%%%%%%%%%%%%%%%%%%%%%%%%%%%%%%%%%%%%%%%%%%%%%%%%%%%%%%%%%%%%%

\end{document}